分类号________________ 密级________________

UDC ________________ 编号________________

# 中国科学院研究生院
# 博士学位论文

## 三维并行自适应有限元动态负载平衡及 $hp$ 自适应策略研究

刘 辉

| | | | |
|---|---|---|---|
| 指导教师 | 张林波 研究员、博士 | | |
| | 中国科学院数学与系统科学研究院 | | |
| 申请学位级别 | 博士 | 学科专业名称 | 计算数学 |
| 论文提交日期 | | 论文答辩日期 | |
| 培养单位 | 中国科学院数学与系统科学研究院 | | |
| 学位授予单位 | 中国科学院研究生院 | | |
| | | 答辩委员会主席 | |

# 摘 要

本文工作围绕着 PHG 平台展开. PHG 是“科学与工程计算国家重点实验室”正在发展的一个并行自适应有限元软件平台. PHG 适合分布式存储并行计算机, 其设计目标是支持使用 $h$, $p$ 及 $hp$ 自适应有限元方法和有限体积方法求解二维和三维偏微分方程组. PHG 支持多种网格及曲面单元, 支持多种网格输入格式, 提供常用的基函数, 支持多种线性方程组解法器和预条件子, 提供接口以调用外部线性方程组和特征值问题求解软件包, 同时提供前处理及后处理功能等. 本文的工作主要涉及两方面的内容. 第一方面内容是研究、实现 PHG 中的网格划分算法及动态负载平衡模块, 包括研究哈密尔顿路的构造和希尔伯特序的编码解码算法, 以及基于这些算法的网格划分和动态负载平衡方法. 第二方面内容是基于 PHG 平台研究 $hp$ 自适应有限元算法中的自适应策略并开展数值实验. 具体内容如下：

(1) 针对 PHG 所处理的协调四面体网格, 本文在网格的对偶图连通的条件下证明了过顶点哈密尔顿圈和任意两个单元之间的过顶点哈密尔顿路的存在性, 并在此基础上设计了线性复杂度的四面体网格上过顶点哈密尔顿圈和过顶点哈密尔顿路的构造算法. 文中还对相应并行算法的构造进行了探讨.

(2) 设计了任意维空间中具有线性复杂度的希尔伯特序编码解码算法并提出了希尔伯特空间填充曲线的一种变体. 本文同时对编码解码算法进行了改进，设计了复杂度更低的算法, 降低了计算量. 文中给出的希尔伯特空间填充曲线的变体保证曲线的编码顺序不随曲线阶数的改变而变化.

(3) 设计实现了加密树及空间填充曲线划分方法并设计了 PHG 的动态负载平衡模块, 同时实现并改进了 Zoltan 中的一个一维划分算法用于对排序后的单元进行一维划分. 此外, 本文还实现了 PLUM 中设计的子网格 – 进程映射算法, 用于减少动态负载平衡中的数据迁移量. 数值实验表明, 本文在 PHG 中所实现的空间填充曲线划分算法不论在划分时间还是划分质量上均优于其它一些软件包 (其中包括由美国 Sandia 国家实验室开发的著名的网格划分和数据迁移软件包 Zoltan) 中所提供的相应算法. 本文所实现的基于希尔伯特序的网格划分算法已经成为 PHG 中默认的网格划分方法, 经受了一些并行规模达数千子网格、计算规模超过十亿单元的大型应用的考验, 表现出很高的效率和很强的稳健性.

(4) 设计了一个基于误差下降预测的 $hp$ 自适应策略. 当前, $hp$ 自适应有限元算法

研究中的一个难点是自适应策略的设计, 目前尚没有成熟的、已被广泛接受的算法. 本文对 $hp$ 自适应策略进行了研究，在前人提出的几种基于误差下降预测的 $hp$ 自适应策略的基础上给出了一个新的 $hp$ 自适应加密策略, 该策略适用于二维三角形、四边形和三维四面体、六面体等不同类型的单元, 适用于正则加密、二分加密等不同自适应加密方式. 数值实验表明, 该策略可以达到最优的误差指数下降, 并在数值解的精度和计算效率上优于文献中的一些策略. 这部分工作的另外一个目的是验证 PHG 的 $hp$ 自适应模块的正确性和适用性.

## Researches on Dynamic Load Balancing Algorithms and *hp* Adaptivity in 3-D Parallel Adaptive Finite Element Computations

Liu Hui (majored in computational mathematics)
Directed by Prof. Zhang Lin-bo

This work is related to PHG (Parallel Hierarchical Grid). PHG is a toolbox for developing parallel adaptive finite element programs, which is under active development at the State Key Laboratory of Scientific and Engineering Computing. PHG is designed for distributed memory parallel computers, and its purpose is to support development of parallel algorithms and codes for solving real world application problems using $h$, $p$, or $hp$ adaptive finite element methods. PHG supports importing meshes from several mesh file formats. PHG provides solvers and preconditioners, as well as interfaces to many external packages, for solving linear systems of equations and eigenvalue problems resulted from adaptive finite element discretization.

This work is divided into two parts. The first part consists of design and realization of the dynamic load balancing module of PHG, including studies on mesh partitioning and data migration algorithms. The second part studies $hp$ adaptive strategies for finite element computations.

The main results of this work are as follows.

(1) For the tetrahedral meshes used in PHG, under reasonable assumptions, we proved the existence of through-vertex Hamiltonian paths between arbitrary two vertices, as well as the existence of through-vertex Hamiltonian cycles, and designed an efficient algorithm with linear complexity for constructing through-vertex Hamiltonian paths. The resulting algorithm has been implemented in PHG, and is used for ordering elements in the coarsest mesh for the refinement tree mesh partitioning algorithm.

(2) We designed encoding and decoding algorithms for high dimensional Hilbert order. Hilbert order has good locality, and it has wide applications in various fields in computer science, such as memory management, database, and dynamic load balancing. We analysed existing algorithms for computing 2D and 3D Hilbert order, and designed improved algorithms for computing Hilbert order in arbitrary space dimensions. We also proposed an alternate form of Hilbert space filling curve which has the advantage of preserving the ordering between different levels. The algorithms have been implemented in PHG and are used in mesh partitioning.

(3) We implemented refinement tree and space filling curve based mesh partitioning algorithms in PHG, and designed the dynamic load balancing module of PHG. The refinement tree based partitioning algorithm was originally proposed by Mitchell, the one implemented in PHG was improved in several aspects. The space filling curve based mesh partitioning function in PHG can use either Hilbert or Morton space filling curve. We also implemented a submesh to process mapping algorithm from the PLUM package in PHG, and use it to reduce the amount of data migration during mesh redistribution. Numerical experiments show that the dynamic load balancing functions work well on thousands of processes with more than one billion elements, and the space filling curve based mesh partitioning module of PHG is faster and yields better results than the corresponding function in other well known mesh partitioning packages.

(4) We studied existing *hp* adaptive strategies in the literature, and proposed a new strategy. Numerical experiments show that our new strategy achieves exponential convergence, and is superior, in both precision of the solutions and computation time, to the strategy compared. This part of the work also serves to validate the *hp* adaptivity module of PHG.

# 目　录

# 第一章　引言

偏微分方程出现在十八世纪, 由物理学家在求解实际问题时提出, 目前已广泛应用到物理、化学、生物以及经济等领域. 由于存在复杂的初始条件、边界条件等, 许多问题得不到解析解, 只可以用数值方法求出近似解, 常用方法有有限差分方法、有限元方法、有限体积方法等. 它们先将问题离散, 然后使用计算机求解. 随着科技的进步, 需要求解的问题规模越来越大, 对计算精度及运行时间的要求越来越高. 正是这些因素的存在, 促使着人们发展更有效的数值方法、计算机硬件和软件. 数值方法的发展和并行计算机软硬件的进步, 为我们求解挑战性问题带来了机遇.

## §1.1　自适应有限元方法

有限元方法是求解偏微分方程的一种数值方法, 此方法在 20 世纪 50 年代初由工程师们提出, 并用于求解简单的结构问题. 有限元方法的原始思想来自 Courant, 他在 1943 年提出在三角形网格上用分片线性函数去逼近 Dirichlet 问题. 有限元方法作为一种系统的数值方法, 并奠定数学基础, 则是在 60 年代中期, 由以冯康先生为代表的中国学者与西方学者独立并行地完成 [207].

有限元方法从物理问题出发, 利用变分原理, 整体地考虑物理问题, 得到方程的变分表达式, 同时对物理问题的计算区域做剖分形成计算网格, 然后在剖分区域 (单元) 上用简单函数去逼近原问题的解. 有限元方法将一个无穷维问题转化成有限维问题, 通过求解这个问题, 得到近似解. 有限元方法的优点之一是能方便地处理复杂区域.

来源于实际问题的偏微分方程往往存在局部奇性, 采用传统有限元方法求解时, 在解有奇性的区域需要使用非常细的网格才能保证有限元解的精度. 传统有限元由于使用一致网格, 会导致网格规模非常大. 而解光滑的区域并不需要这么细的网格, 从而会浪费计算资源. 更有效的做法是仅加密解有奇性的区域. 这是自适应有限元方法思想的由来.

自适应有限元方法根据解的性质动态地调整网格/基函数. 对自适应有限元方法 [199], 要解决的问题是如何确定有奇性的区域以及如何对数值解给出一个可信的误差估计. 标准的先验误差估计只能给出数值解的渐近误差行为, 并且需要知道解的正则性, 而解的正则性往往是不知道的. 自适应有限元方法需要从数值解及已知的数据导出误差估计 – 后验误差估计, 并且后验误差估计的计算代价需要远小于数值解求解的代价. 此外, 误差估计需要是局部的, 并且是真实误差的

上下界, 其中上界表明解的精度是可信的, 小于事先给定的误差, 下界表明误差估计是精确的, 网格/基函数将在需要调整的地方被调整. 自适应过程是一个计算数值解、计算误差、调整网格/基函数、重新计算的迭代过程, 具体步骤如图 1.1 所示 [204]. 常用的自适应方法有: $h$ 自适应方法, $p$ 自适应方法, $hp$ 自适应方法以及 $r$ 自适应方法.

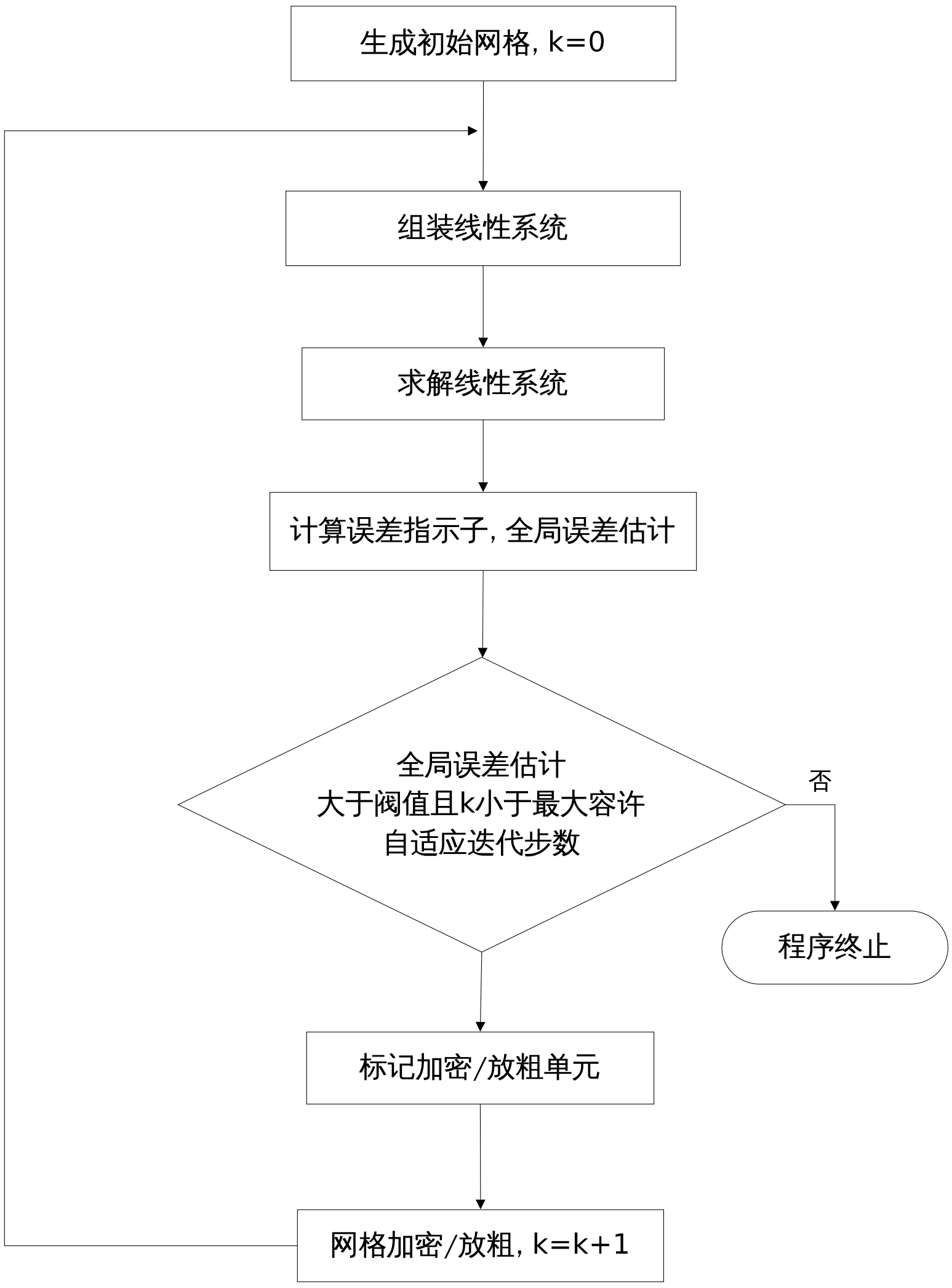

图 1.1 自适应有限元流程图

为了能让自适应流程 1.1 高效运行, 需要有正确的离散方法、求解器、误差指示子、网格标记策略和合适的加密策略, 同时也要选取适当的参数. 误差指示子是自适应有限元方法的关键之一, 常用的误差指示子主要有: 残量型误差指示

子, 对偶权重残量误差指示子, 多重基误差指示子, 平均化方法以及平衡残量误差指示子等.

在给定误差指示子的情况下, 如何选取单元进行加密/放粗同样会影响自适应有限元的收敛性和效率. 常用的单元标记策略主要有最大误差策略 (max stratedy), 误差等分布策略 (equidistribution stratedy), 保证误差下降策略 (guaranteed error reduction strategy) [39], MNS (MNS-refinement strategy) 策略 [40] 等. 后面章节将对标记策略做详细的介绍.

自适应有限元方法另一个关键模块是网格单元类型及加密策略的选取. 二维区域常选取四边形或者三角形作为基本单元. 四边形网格常采取正则加密, 加密部分单元会导致悬点的出现, 否则只能进行一致加密. 当执行这种加密时, 在一条边上可能会出现多个悬点, 这会对有限元空间及程序的数据结构有很多的限制. 为了克服这个问题, 开发人员常采用 Bank[41, 42] 设计的一悬点 (1-irregular) 和三邻居 (three-neighbors) 原则. 一悬点原则限制每条边上的悬点最多只有一个, 如果某条边由于邻居单元的加密引入新的悬点, 则该单元需要被加密以保证一悬点原则. 三邻居原则指出如果四边形的三条边都有悬点, 则该单元需要被加密. 对三角形单元, 常用的加密算法有正则加密和二分加密 [43]. 正则加密 [44] 是连接三角形的三个中点, 将一个三角形单元加密成四个子单元, 对于出现的悬点, 则是对邻居使用一个临时的二分加密以保证协调性. 二分加密是连接顶点和顶点对应的边, 将一个三角形加密成两个子单元. 二分加密算法有最长边二分加密 [45, 46, 47] 和最新顶点二分加密 [48]. 由 Rivara 提出的最长边二分加密算法是将最长边作为加密边, 将其中点与相对的顶点相连, 把单元分成两个子单元. 由 Sewell 提出的最新顶点加密算法是将三角形中最新产生的顶点作为加密顶点, 连接对应边的中点, 将三角形加密成两个子单元.

三维网格常选取四面体或者六面体作为单元. 六面体的自适应要复杂一些, 有的算法基于八叉树, 即将一个六面体剖分成八个子六面体, 由于会出现悬点, 接着对邻居做临时的剖分, 将这些邻居单元剖分成金字塔和棱柱, 当临时单元被标记为加密单元时, 先把它们恢复成六面体, 然后剖分 [49]. 有的做法是定义加密模板, 当某个单元被标记为加密单元时, 为该单元及邻居选择合适的模板, 生成一个没有悬点的六面体网格 [50]. 六面体网格的自适应是很复杂的, 程序的运行时间也较长, 而四面体网格的自适应则要方便得多. 四面体常采用二分加密算法, 和三角形一样, 可采用最长边加密和最新顶点加密. 四面体二分加密时选取一条边作为加密边将其一分为二, 产生两个子单元. 最长边二分加密算法由 Rivara 提出 [45, 46, 47]. 最新顶点二分加密算法首先由 Sewell 针对二维区域提出 [48], 然后由 Bänsch 扩展到三维区域 [51]. 最新的进展可以参考 Kossaczky[52], Liu 和 Joe [54],

及 Arnold [53] 等人的文章.

在网格二分加密的并行算法方面, 许多科研人员做出了有意义的工作. Rivara 等人 [55] 为四面体网格的全局加密设计了并行算法, 但不适合局部加密. Pébay 和 Thompson [57] 对基于边分割的四面体网格设计了并行算法. Jones 和 Plassmann [43] 为三角形网格设计了并行局部自适应加密算法. Barry、Jones 及 Plassmann [56] 同时给出了一个实现并行自适应有限元程序的框架. Castãnos 和 Savage [58] 为四面体网格设计了一种并行局部自适应加密算法, 并且应用在软件包 Pared 中. 最近, 张 [1] 为四面体最新顶点二分加密策略设计了并行算法, 并且应用在并行自适应有限元软件平台 PHG (Parallel Hierarchical Grid) [59] 中.

## §1.2 动态负载平衡

偏微分方程的并行数值求解的关键因素之一是处理好进程间的数据分布. 数据分布指根据进程数对数据进行划分, 将数据分布存储在各进程中. 计算过程中数据分布保持不变的程序, 如传统有限元程序, 在整个计算过程中仅需在开始做一次静态划分. 然而对自适应有限元程序, 由于每个进程的负载是在不断地变化的, 并且这个变化是不可以或很难预测的. 这种情况下, 就需要动态地调整数据分布以保持负载平衡. 好的数据划分方法既要使得每个进程拥有相同的计算负载, 又要减少进程间的通信.

并行自适应有限元计算中最常用的数据划分方法是划分计算网格, 每个进程负责一个子网格上的计算, 为保证可扩展性, 每个进程仅存储该子网格及定义在其上的数据, 不在本地的数据在需要时通过进程间通信获得. 这种计算模型包含三个模块: 数据划分、维护本地数据结构以方便计算和进程间通信、在需要的时候进行通信. 在自适应有限元程序中, 工作负载是与网格的单元、面、边及顶点相关的. 数据划分问题就是将这些对象划分到不同的进程中, 使每个进程拥有尽可能相同数目的对象. 在有限元计算中, 以单元划分数据是最常用的. 然而每个单元的负载可能并不相同, 例如在不同的单元上求解不同的物理问题或者像在 $p$ 自适应有限元中每个单元的自由度数可能并不相同, 可以给每个单元一个权重以表示计算负载. 数据划分的目标就是使得每个进程拥有相同的权重, 同时使得有限元计算中的通信量最小.

在计算过程中根据各进程中负载的变化调整数据的划分以保持负载平衡称为动态负载平衡. 并行自适应有限元计算中, 由于各进程的负载不断变化, 动态负载平衡是必需的. 动态负载平衡中的数据划分算法需满足一些条件: 1) 能在分布式的数据上并行执行; 2) 执行速度快, 因为它们是纯开销, 并且可能被频繁地

调用; 3) 最好是增量的, 即网格微小的变化仅导致划分的微小变化; 4) 内存使用是适度的; 5) 容易使用.

对粒子模拟, 粒子间的作用是由距离决定的, 通常距离近的相互影响较大, 按照距离划分数据会降低通信量. 而对于基于网格的计算, 如有限元、有限体积等方法, 通信一般发生在邻居之间, 邻接关系决定了通信模式. 动态负载平衡中的数据划分算法也分为两类: 基于几何性质的算法和基于图的算法. 基于几何性质的算法利用对象的几何信息, 如坐标信息, 主要算法有递归坐标二分算法 (RCB, Recursive Coordinate Bisection) [60] 、递归惯量二分算法 (RIB, Recursive Inertial Bisection) [61]、空间填充曲线算法 (SFC, Space-Filling Curve) [62] 等; 基于图的算法利用对象的拓扑关系, 如邻居关系或者顶点边的连接信息等, 主要方法有递归图二分算法 (RGB, Recursive Graph Bisection) [60]、贪婪算法 [63]、递归谱二分方法 (RSB, Recursive Spectral Bisection) [60]、K-L 方法 (Keinighnan-Lin Algorithm) [64] 等. 著名的网格划分软件包有: ParMETIS[14], Zoltan[19], JOSTLE[65] 等.

## §1.3 并行自适应有限元软件平台 PHG

PHG (Parallel Hierarchical Grid) 是一个并行自适应有限元软件包, 由科学与工程计算国家重点实验室研制. 该软件包主要用 C 语言开发 (部分使用 C++), 底层的通信由 MPI 完成, 面向分布式存储并行机, 具有很强的可移植性. PHG 对用户隐藏了通信接口, 无需用户干预通信, 顶层代码和串行代码一样. PHG 提供了一个可扩展的并行自适应有限元程序框架, 它提供了网格输入、网格划分、网格加密/放粗、自由度操作、线性系统组装和求解、特征值求解、可视化、性能统计等功能. PHG 采取了面向对象的设计, 方便用户添加新的功能. 下面简要介绍一下 PHG 的功能及特点:

(1) 网格对象. PHG 目前处理的网格对象为协调四面体网格, 输入文件可以为 ALBERTA [66], MEDIT [67], GAMBIT [68] 等格式, 可以处理周期网格及镜像网格. PHG 提供了丰富的边界类型及边界管理函数, 可以满足各种常规的需求. 网格局部自适应采用最新顶点二分加密算法 [1], 网格的加密/放粗层次结构被保留, 并分布在不同的进程中. 在周期网格中, PHG 自动维护网格的周期性.

(2) 动态负载平衡. PHG 提供动态负载平衡功能, 提供了加密树划分方法 [84]、空间填充曲线划分方法 [62] 等网格划分方法. PHG 同时提供了接口以调用外部函数库, 如 Zoltan, ParMETIS 等, 方便用户根据需要选取不同的算法.

(3) 自由度管理. 自由度对象是 PHG 的基本数据结构之一, 用于描述和处理分布在网格上的各种数据. 它可以用于定义有限元函数、空间, 也可以用于存储、处理任何其他分布在网格上的数据, 如几何信息等. PHG 提供了自由度的创建、插值、赋值、复制、基本算术运算等操作, 同时提供了有限元函数的微积分运算. PHG 还提供了一些特殊类型的自由度, 如常量自由度、解析自由度、几何量自由度等. PHG 提供了统一的创建自由度的模板用来创建各种自由度.

(4) 数值积分. 为方便有限元计算 PHG 提供了三角形和四面体上任意阶的数值积分公式 [69], 这些积分公式的权重都是非负的, 并且所有的积分点均不在单元外部. PHG 提供了基函数缓存机制, 极大减少重复计算, 加快积分过程.

(5) 向量和稀疏矩阵管理. PHG 提供了一组函数用于管理连续分块存储的分布式向量, 以及基于行划分、行压缩存储的分布式稀疏矩阵. 向量和矩阵的分布通过映射进行描述与管理. 这些模块的提供, 使得线性系统可以独立于网格而存在, 亦使得 PHG 可以独立作为线性方程组解法器使用.

(6) 线性解法器. PHG 提供了一套统一的接口用于求解线性方程组. PHG 本身实现了一些解法器, 如 PCG, GMRES 等, 和一些预条件子. PHG 同时提供了丰富的接口, 以供用户调用外部解法器软件包, 如 PETSc, HYPRE, SuperLU, MUMPS, SPC, LASPack, SPOOLES 等. PHG 定义了线性解法器对象 SOLVER, 用户可以通过该对象方便地定义、操作线性方程组的未知量、系数矩阵和右端项. PHG 定义的解法器对象独立于特定的方法, 方便用户按照自己的需要添加解法器及预条件子.

(7) 特征值解法器. PHG 目前没有实现自己的特征值求解器, 但提供了与 PARPACK, JDBSYM, LOBPCG, SLEPc, Trilinos/Anasazi 和 PRIMME 等软件包的接口, 用于计算广义特征值和特征向量. 特征值、特征向量的计算采用统一接口, 可在运行程序时通过命令行选项选择任何一个可用的特征值求解器.

(8) 可视化. PHG 目前实现了 OpenDX[70] 以及 VTK[71] 等格式可视化文件的输出, 可以输出自由度数据及网格数据.

(9) 参数控制. PHG 提供了灵活的命令行选项管理机制用于控制程序中的参数. PHG 的选项按照目录分类, 可读性很强.

(10) 性能统计. PHG 提供了获取程序运行时浮点性能及内存使用的函数, 方便用户获取程序的实际性能.

PHG 同时提供了内存管理、信息输出、输入等功能, 其中信息输出按照用户指定的输出等级, 满足不同的调试要求. 完整的 PHG 功能以及具体的函数接口在这里不再一一介绍, 用户可以在 PHG 网站 [59] 中获得.

## §1.4 论文主要工作

本论文的工作是围绕 PHG 平台开展的, 主要包括两部分内容. 第一部分内容涉及网格划分和动态负载平衡算法, 论文中研究了哈密尔顿路、希尔伯特序、网格划分算法等, 并在此基础上设计和实现了 PHG 的动态负载平衡模块. 第二部分内容涉及 $hp$ 自适应策略的研究, 论文针对二分加密网格设计了 $hp$ 自适应策略, 并与已有策略进行了比较. 论文所完成的主要工作具体如下:

(1) 证明了过顶点哈密尔顿路和哈密尔顿圈的存在性并设计实现了线性复杂度的算法. 针对 PHG 所处理的协调四面体网格, 本文在网格的对偶图连通的条件下证明了过顶点哈密尔顿圈和任意两个单元之间的过顶点哈密尔顿路的存在性, 并在此基础上设计了线性复杂度的四面体网格上过顶点哈密尔顿圈和过顶点哈密尔顿路的构造算法. 文中还对相应并行算法的构造进行了探讨. 本文在 PHG 平台中实现了该算法, 并将其应用于加密树网格划分方法中初始网格单元的排序. 将来如果需要, 该算法亦可作为单元排序方法直接应用于网格划分中.

(2) 设计了任意维空间中具有线性复杂度的希尔伯特序的编码解码算法并提出了希尔伯特空间填充曲线的一种变体. 希尔伯特序具有良好的空间局部性, 广泛应用于内存管理、数据库索引、动态负载平衡等领域. 本文总结、分析了文献中已有的生成二维和三维空间希尔伯特序的算法, 设计了任意维空间中具有线性复杂度的希尔伯特序编码、解码高效算法, 同时并对算法进行了改进, 设计了复杂度更低的算法, 降低了计算量. 文中还给出了希尔伯特空间填充曲线的一个变体, 它可保证曲线的编码顺序不随曲线阶数的改变而变化. 本文在 PHG 平台中实现了二维及三维算法, 并成功将其应用于网格划分中.

(3) 设计实现了加密树及空间填充曲线划分方法并设计了 PHG 的动态负载平衡模块. 加密树划分方法最初由 Mitchell 提出, 本文在实现该算法的同时亦对算法进行了改进, 降低了算法的计算复杂度, 提高了算法实现的效率. 同时实现了空间填充曲线划分框架, 可以使用希尔伯特空间填充曲线或 Morton 空间填充曲线, 并实现并改进了 Zoltan 中的一个一维划分算法用于对排序后的单元进行一维划分. 此外, 本文还实现了 PLUM 中设计的子网格 – 进程映射

算法, 用于减少动态负载平衡中的数据迁移量. 数值实验表明, 本文在 PHG 中所实现的空间填充曲线划分算法不论在划分时间还是划分质量上均优于其它一些软件包 (其中包括由美国 Sandia 国家实验室开发的著名的网格划分和数据迁移软件包 Zoltan) 中所提供的相应算法. 本文所实现的基于希尔伯特序的网格划分算法已经成为 PHG 中默认的网格划分方法, 经受了一些并行规模达数千子网格、计算规模超过十亿单元的大型应用的考验, 表现出很高的效率和很强的稳健性.

(4) 设计了一个基于误差下降预测的 $hp$ 自适应策略. 当前, $hp$ 自适应有限元算法研究中的一个难点是自适应策略的设计, 目前尚没有成熟的、已被广泛接受的算法. 本文对 $hp$ 自适应策略进行了研究，在前人提出的几种基于误差下降预测的 $hp$ 自适应策略的基础上给出了一个新的 $hp$ 自适应加密策略, 该策略适用于二维三角形、四边形和三维四面体、六面体等不同类型的单元, 适用于正则加密、二分加密等不同自适应加密方式. 数值实验表明, 该策略可以达到最优的误差指数下降, 并在数值解的精度和计算效率上优于文献中的一些策略. 这部分工作的另外一个目的是验证 PHG 的 $hp$ 自适应模块的正确性和适用性.

## §1.5 论文内容安排

论文分为六章. 第一章为引言, 本章对自适应有限元及并行自适应有限元软件平台 PHG 进行简要介绍. 第二、三、四章研究动态负载平衡方法, 其中第二章研究哈密尔顿路和哈密尔顿圈. 第三章研究希尔伯特序的编码解码算法. 第四章介绍 PHG 的动态负载平衡及数据迁移算法. 第五章研究 $hp$ 自适应策略. 最后一章给出总结及未来工作.

# 第二章 哈密尔顿路和哈密尔顿圈

在三维自适应有限元计算中, 协调四面体网格是最常用的网格类型之一. 四面体单元适合非结构网格, 适合处理复杂边界, 其数据结构也简单. 在并行自适应有限元计算中, 一种常用的网格划分方法是先将网格线性化, 即将单元进行排序, 接着根据进程数将单元按照线性化后的顺序分成几段, 每一段映射给不同的进程. 哈密尔顿路是常用的单元排序方法之一. 本章研究 PHG 中的非结构协调四面体网格上哈密尔顿路的基本性质及构造方法. 目前在 PHG 中, 哈密尔顿路被用于 RTK 网格划分方法 (加密树划分方法) 中初始单元顺序的确定，将来如果需要亦可将它直接用于网格划分。

## §2.1 协调四面体网格

本章中, 单元是一个四面体, 它被看作 $\mathbb{R}^3$ 中的闭区域. 一个四面体有四个顶点、六条边和四个三角形面, 如图 2.1 所示. 本章中涉及到的四面体都是非退化的, 即四个顶点不在一个平面上.

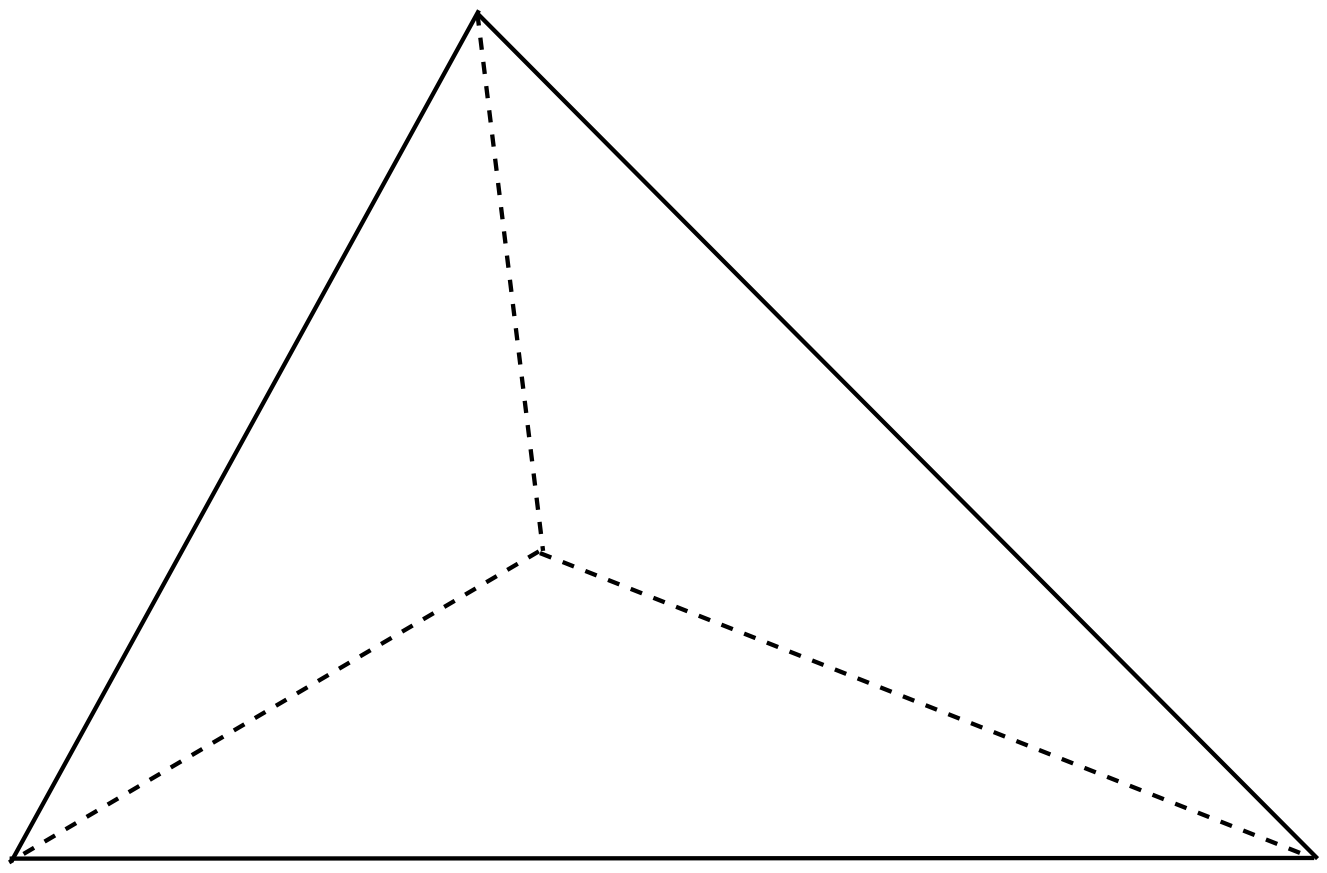

图 2.1 四面体单元

**定义 2.1.1.** *令 $\Omega$ 为 $\mathbb{R}^3$ 中一个连通的开区域, 一个四面体网格 $\mathcal{M}$ 是四面体单元的集合, 记作 $\{T_i\}$, 这些单元满足如下条件: $\overline{\Omega} = \cup T_i$, $\overset{o}{T_i} \cap \overset{o}{T_j} = \emptyset$ $(i \neq j)$, 其中 $\overset{o}{T}$ 表示单元 $T$ 的内部, $\emptyset$ 为空集.*

一个顶点如果位于区域 $\Omega$ 的边界上, 则被称作边界点, 否则是内部点. $|\mathcal{M}|$ 表示网格 $\mathcal{M}$ 的单元数目. 两个单元 $T_i$ 和 $T_j$ $(i \neq j)$ 如果有一个公共面, 则它们

被称为邻居.

**定义 2.1.2.** 网格 $\mathcal{M}$ 被称为协调的, 如果对 $\mathcal{M}$ 中任意两个不同的单元 $T_i$ 和 $T_j$, $T_i \cap T_j$ 是一个顶点、一条边、一个面或者空集.

协调四面体网格的生成方法主要有八叉树方法 [76, 77]、Delaunay 方法 [78, 79] 等, 常用的四面体网格生成软件有 Gmsh [80]、Netgen [81]、TetGen [82] 等. 协调四面体网格局部自适应加密算法主要有最长边二分加密算法 [45, 46, 47]、最新顶点加密算法 [1, 48, 51, 53, 52, 54] 等, 其质量度量标准主要有最小 – 最大立体角、最小平面角、半径比、边长比等.

## §2.2 哈密尔顿路与哈密尔顿圈

自适应有限元方法根据数值解调整计算网格/基函数. 在分布式存储并行计算机中, 自适应会导致负载不平衡, 网格划分对并行程序而言是一个重要的模块. 它被用来对网格进行初始划分, 并且当网格变化时, 对网格进行重划分, 以保证并行程序在计算过程中有均衡的负载. 一类重要的网格划分方法是: 先把网格单元排序, 形成一个一维的列表, 然后把这个列表划分成段, 每段具有相等的权重. 哈密尔顿路 (Hamiltonian path) 是这类方法中的排序方法之一 [62, 125, 129].

哈密尔顿路是图论中的概念, 它是一条通过所有结点的路, 且每个结点仅通过一次. 在四面体网格中, 哈密尔顿路被定义成一个单元序列, 每个单元仅出现一次. Heber 等人 [85] 对二维三角形单元网格证明了哈密尔顿路的存在性, 其中哈密尔顿路通过一个顶点或者一条边进入下一个单元, 同样通过一个顶点或者一条边离开一个单元, 定理的证明需要如下条件: 三角形网格是协调的并且不能存在局部割点 (local cut vertex) [83]. Mitchell [83] 针对二维及三维空间研究了三角形、四边形、四面体、六面体网格的哈密尔顿路和哈密尔顿圈的存在性. 对于四面体网格, 由于仅通过面的哈密尔顿路不一定存在, 因此 Mitchell 的文章中引入了过顶点哈密尔顿路 (through-vertex Hamiltonian path) 以及过顶点哈密尔顿圈 (through-vertex Hamiltonian cycle) 的概念. Mitchell 证明了: 如果协调四面体网格没有局部割点和局部割边, 那么过顶点哈密尔顿路和哈密尔顿圈存在. 证明是构造性的, 可以直接得到构造算法.

本文将在一个假设下, 即网格的对偶图连通, 证明过顶点哈密尔顿路和哈密尔顿圈的存在性. 证明同样是构造性的, 可以直接导出线性复杂度的算法.

## §2.3 存在性定理

本节将研究四面体网格上过顶点哈密尔顿路和哈密尔顿圈的存在性, 内容分为两部分, 第一部分先给出基本定义, 介绍一些记号和概念; 第二部分给出存在性定理.

### §2.3.1 基本定义

为了叙述的方便, 在这里将介绍所用到的概念及记号, 其中部分概念来源于 Mitchell 的文章 [83].

**定义 2.3.1.** 协调四面体网格 $\mathcal{M}$ 的对偶图定义为一个无向图, 在这个图中, 网格 $\mathcal{M}$ 中每一个单元对应图中的一个顶点, $\mathcal{M}$ 中的两个邻居单元则对应图中的一条边.

**定义 2.3.2.** 在一个图中, 如果任意两个顶点之间存在一条通路, 则称图是连通的.

**定义 2.3.3.** 在协调四面体网格 $\mathcal{M}$ 中, 一条长度为 $n$ 的路 (path), 是一个单元序列, 记作 $T_1T_2\ldots T_n$, $T_i \in \mathcal{M}$, $i = 1,\ldots,n$, 其中 $T_i \cap T_{i+1} \neq \emptyset$, 并且 $T_i \neq T_{i+1}$, $i = 1, n-1$.

**定义 2.3.4.** 在协调四面体网格 $\mathcal{M}$ 中, 一个长度为 $n$ 的圈 (cycle), 是一条长度为 $n+1$ 的路, 其中 $T_1 = T_{n+1}$.

**定义 2.3.5.** 协调四面体网格 $\mathcal{M}$ 上的一条哈密尔顿路 (Hamiltonian path) 是一条路, 其中网格 $\mathcal{M}$ 的每个单元正好出现一次.

**定义 2.3.6.** 协调四面体网格 $\mathcal{M}$ 上的一条部分哈密尔顿路 (partial Hamiltonian path) 是一条路, 其中每个单元至多出现一次.

**定义 2.3.7.** 协调四面体网格 $\mathcal{M}$ 上的一个哈密尔顿圈 (Hamiltonian cycle) 是一个圈, 其中网格 $\mathcal{M}$ 的每个单元正好出现一次.

**定义 2.3.8.** 一条过顶点路 (through-vertex path) 是一条路, 其中总是通过顶点进入或离开一个单元, 并且对同一个单元, 进入或者离开此单元的顶点不是同一个.

在上述定义下, 可以引申出过顶点哈密尔顿路.

**定义 2.3.9.** 一条(部分) 过顶点哈密尔顿路 (through-vertex Hamiltonian path) 是一条过顶点路, 并且是一条(部分) 哈密尔顿路.

过顶点哈密尔顿路可以表示为 $H = T_1 v_1 T_2 \cdots v_{n-1} T_n$, 其中 $T_i \neq T_j$ $(i \neq j)$ 且 $v_i \neq v_{i+1}$ $(1 \leq i \leq n-2)$, $v_i$ 表示连接 $T_i$ 和 $T_{i+1}$ 的顶点. 过顶点哈密尔顿圈可以类似表示.

令 $\mathcal{M}$ 为区域 $\Omega$ 上的一个网格, $v$ 为 $\mathcal{M}$ 中的一个顶点, 如果 $\overline{\Omega} \backslash v$ 是非连通的, 则称 $v$ 为 $\mathcal{M}$ 中的一个割点. 如果 $v$ 满足如下条件: $\exists R > 0$ 使得 $\forall\ r$ $(0 < r \leq R)$, $(B(v,r) \cap \overline{\Omega}) \setminus v$ 是非连通的, 则称 $v$ 为 $\mathcal{M}$ 的局部割点 [83], 其中 $B(v,r)$ 表示以 $v$ 中心的半径为 $r$ 的球. 割点和局部割点如图 2.2 和 2.3 所示. 图 2.2 中顶点 $v$ 是割点, 图 2.3 中顶点 $v$ 是局部割点. 为了简便, 我们使用了二维图形. 类似的, 可以定义割边及局部割边.

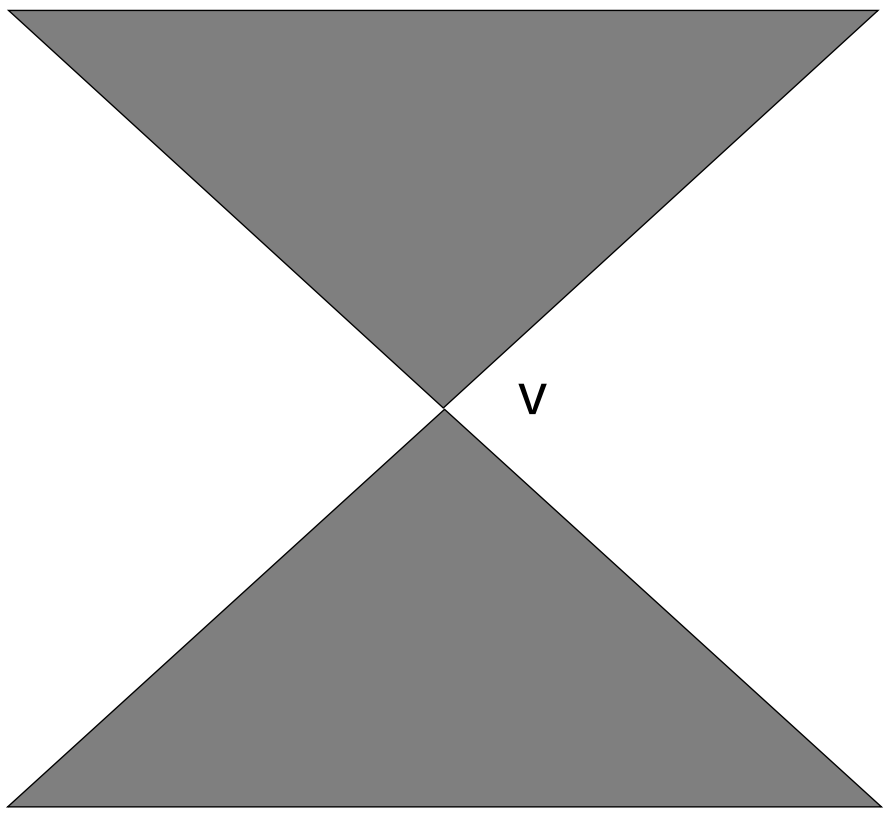

图 2.2 含有割点 $v$ 的网格

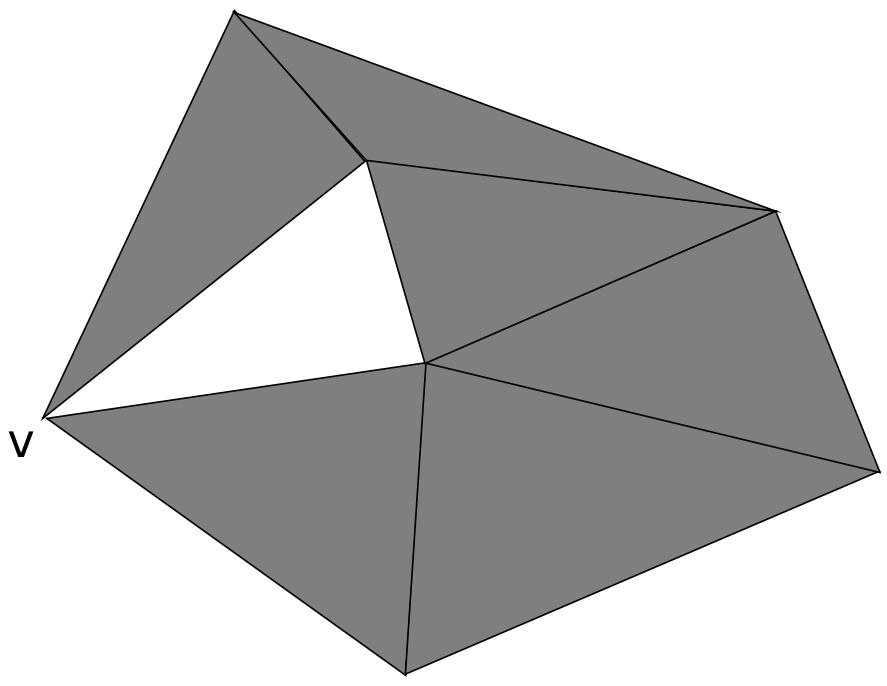

图 2.3 含有局部割点 $v$ 的网格

从定义可以很容易看出, 对偶图连通的四面体网格是没有割点的, 但可能会有局部割点. 没有割点及局部割点的四面体网格由于可能存在割边或者局部割

边, 对偶图亦不一定连通.

### §2.3.2 存在性定理

对协调四面体网格, 本节将证明过顶点哈密尔顿路和哈密尔顿圈的存在性. 本节先给出三个引理, 它们将在后面的证明过程中被用到. 第一个引理给出邻居间共享顶点的个数. 它是我们构造和扩充哈密尔顿路的基础.

**引理 2.1.** *假设网格 $\mathcal{M}$ 是协调四面体网格, 其对偶图是连通的并且 $|\mathcal{M}| \geq 2$, 则有如下结论成立:*

*(1) 若 $\mathcal{M}_1$ 和 $\mathcal{M}_2$ 是 $\mathcal{M}$ 的非空子集, 且 $\mathcal{M}_1 \cap \mathcal{M}_2 = \emptyset$, $\mathcal{M}_1 \cup \mathcal{M}_2 = \mathcal{M}$, 则存在单元 $T_1 \in \mathcal{M}_1$ 及 $T_2 \in \mathcal{M}_2$, 使得它们的交集 $T_1 \cap T_2$ 是一个公共面, 即它们是邻居.*

*(2) 如果 $T, T_1$ 和 $T_2$ 是 $\mathcal{M}$ 中不同的单元, 并且 $T_1$ 和 $T_2$ 均是 $T$ 的邻居, 则这三个单元拥有两个公共顶点.*

*(3) 如果 $T, T_1, T_2$ 和 $T_3$ 是 $\mathcal{M}$ 中不同的单元, 并且 $T_1, T_2$ 和 $T_3$ 均是 $T$ 的邻居, 则这四个单元拥有一个公共顶点.*

*证明.* 结论 (1) 是对偶图连通的直接推论.

结论 (2): 因为单元 $T_1$ 和 $T_2$ 都是 $T$ 的邻居, 它们中的每一个都分别和 $T$ 共享一个面, 并且很容易验证这两个面共享一条边, 所以 $T_1, T_2$ 及 $T$ 三个单元共享一条边, 拥有两个相同的顶点, 结论成立.

结论 (3): 因为 $T_1, T_2$ 和 $T_3$ 都是 $T$ 的邻居, 它们中的每一个都和 $T$ 共享一个面, 在四面体中, 任意三个面共享一个顶点, 结论成立. □

**引理 2.2.** *若协调四面体网格 $\mathcal{M}$ 的对偶图是连通的, $H = T_1 v_1 T_2 v_2 \cdots T_{k-1} v_{k-1} T_k$ $(k \geq 2)$ 为 $\mathcal{M}$ 的一个部分过顶点哈密尔顿路. 那么,*

*(1) 如果 $T_1$ 和 $T_2$ 是邻居, 那么 $T_1$ 的任何不属于 $H$ 的邻居可以插入到 $H$ 中, 并且位于 $T_1$ 和 $T_2$ 之间.*

*(2) 如果 $T_{k-1}$ 和 $T_k$ 是邻居, 那么 $T_k$ 的任何不属于 $H$ 的邻居可以插入到 $H$ 中, 并且位于 $T_{k-1}$ 和 $T_k$ 之间.*

*证明.* 由于两个结论是相似的, 仅给出结论 (1) 的证明过程.

令 $T$ 为 $T_1$ 的不属于 $H$ 的邻居. 根据引理 2.1 可知, $T_1, T_2$ 和 $T$ 有两个共同的顶点, 记作 $u_1$、$u_2$.

如果 $H$ 仅包含 $T_1$ 和 $T_2$, 即 $k=2$, 那么 $T$ 可以按照如下的方式插入到 $H$ 中: $H=T_1u_1Tu_2T_2$.

现在假设 $k>2$. 如果 $v_1\in\{u_1,u_2\}$, 不失一般性, 假设 $v_1=u_2$, 那么 $T$ 可以按照如下方式插入到 $H$ 中: $H=T_1u_1Tv_1T_2v_2\cdots T_k$. 如果 $v_1\notin\{u_1,u_2\}$, 那么 $u_1$ 和 $u_2$ 中必定有一个顶点和 $v_2$ 是不同的, 记作 $u_1$, 在这种情况下 $T$ 可以按照如下方式插入到 $H$ 中: $H=T_1v_1Tu_1T_2v_2\cdots T_k$. 重复这个过程, $T_1$ 的任意一个邻居都可以插入到 $T_1$ 和 $T_2$ 之间.

□

由第二个引理可以知道, 如果创建合适的初始路, 那么我们可以保持该哈密尔顿路的头部和尾部不变, 把它们所有的邻居单元插入到这两个单元之间. 接下来介绍第三个引理. 这个引理指出, 如果某个部分过顶点哈密尔顿路没有包含所有的单元, 则可以通过插入新的单元来扩展这条路, 直至包含所有的单元.

**引理 2.3.** *若协调四面体网格 $\mathcal{M}$ 的对偶图是连通的, $H=T_1v_1\cdots v_{n-1}T_n$ 是 $\mathcal{M}$ 中一条非空的部分过顶点哈密尔顿路, 它包含了 $T_1$ 及 $T_n$ 的所有邻居. 令 $\mathcal{M}_1=\{T_1,\cdots,T_n\}$, 如果 $\mathcal{M}_1\neq\mathcal{M}$, 则 $H$ 可以通过插入新单元来扩展, 并且插入的单元位于 $T_1$ 和 $T_n$ 之间.*

*证明.* 记 $\mathcal{M}_2=\mathcal{M}\backslash\mathcal{M}_1$. 如果 $\mathcal{M}_2$ 是非空的, 根据引理 2.1, 存在 $T\in\mathcal{M}_2$ 及 $T_i\in\mathcal{M}_1$ $(1<i<n)$ , 使得 $T$ 和 $T_i$ 是邻居.

由于 $T_i$ 有四个顶点, 那么 $T_i$ 和 $T$ 共有的三个顶点之一必定是 $T_i$ 的入口顶点 $v_{i-1}$ 或者出口顶点 $v_i$, 不失一般性, 假定它是 $v_i$ (否则的话, 可以反转 $H$, 在完成插入之后, 再重新反转), 并且 $T_i$ 和 $T$ 的三个共有顶点之一必定不是 $v_{i-1}$ 或者 $v_i$, 记作 $v$. 那么新的部分过顶点哈密尔顿路可以构造为: $H=T_1v_1\ldots T_ivTv_iT_{i+1}\ldots T_n$. 定理得证. □

*注记:* 引理 2.3 的一个条件: 部分过顶点哈密尔顿路包含 $T_1$ 及 $T_n$ 的全部邻居是必需的. 反例, 假定 $H=T_1vT_n$, 其中 $T_1$ 和 $T_2$ 仅共享一个顶点 $v$, 如果 $T_1$ 有不含有顶点 $v$ 的邻居 $T$, 那么不可能将 $T$ 插入到 $T_1$ 和 $T_n$ 之间.

**定理 2.4.** *假设协调四面体网格 $\mathcal{M}$ 的对偶图是连通的, 且 $|\mathcal{M}|\geq 2$. 对网格 $\mathcal{M}$ 中任意两个单元 $T_1$ 和 $T_n$, $T_1\neq T_n$, 存在一条 $T_1$ 到 $T_n$ 的过顶点哈密尔顿路.*

*证明.* 由于 $\mathcal{M}$ 的对偶图是连通的, 可以在网格 $\mathcal{M}$ 中找到单元序列 $T_1T_2\cdots T_n$, 使得序列中任意两个相邻单元是邻居, 并且可以创建初始的过顶点哈密尔顿路 $H=T_1v_1T_2v_2\cdots v_{n-1}T_n$ (采用归纳法, 假设已经创建 $H=T_1v_1\cdots T_{i-1}v_{i-1}T_i$. 对

单元 $T_{i+1}$, 它和 $T_i$ 有三个公共顶点, 其中之一和 $v_{i-1}$ 是不同的, 记作 $v_i$, 然后 $H$ 可以扩展成 $H = T_1v_1 \cdots T_{i-1}v_{i-1}T_iv_iT_{i+1}$. 重复这个过程, 便可构造出包含 $T_1, \cdots, T_n$ 的初始部分过顶点哈密尔顿路). 利用引理 2.2, 可以把 $T_1$ 和 $T_n$ 的所有邻居插入到 $H$ 中, 并且全部位于 $T_1$ 和 $T_n$ 之间. 接着利用引理 2.3, 可将所有剩余的单元插入到 $H$ 中. 证明完毕. □

下面一个定理是定理 2.4 的直接推论, 它给出协调四面体网格中过顶点哈密尔顿圈的存在性.

**定理 2.5.** *如果协调四面体网格 $\mathcal{M}$ 的对偶图是连通的, 并且 $|\mathcal{M}| \geq 2$, 那么在网格 $\mathcal{M}$ 中存在过顶点哈密尔顿圈.*

证明. 选取两个邻居单元 $T_1$ 和 $T_n$. 根据定理 2.4, 存在一个哈密尔顿路 $H = T_1v_1T_2v_2 \ldots v_{n-1}T_n$. 由于 $T_1$ 和 $T_n$ 是邻居, 它们有三个公共顶点, 其中一个是和 $v_1$ 及 $v_{n-1}$ 不同的, 记作 $v_n$. 那么可以按照如下方式创建哈密尔顿圈, $H = T_1v_1T_2v_2 \ldots v_{n-1}T_nv_nT_1$. □

最后一个定理, 定理 2.6, 对于将来设计哈密尔顿路及哈密尔顿圈的并行算法是有用的. 由于并行程序中网格是分布在不同进程中的, 这个定理指出, 每个进程可以独立地创建自己的哈密尔顿路和哈密尔顿圈, 然后将这些部分哈密尔顿路和哈密尔顿圈组合起来, 形成一个完整的哈密尔顿路和哈密尔顿圈.

**定理 2.6.** *若四面体网格 $\mathcal{M}$ 的对偶图是连通的, $\mathcal{M}_1 \subset \mathcal{M}$ 及 $\mathcal{M}_2 \subset \mathcal{M}$ 是网格 $\mathcal{M}$ 的子网格, 满足条件: $|\mathcal{M}_1| \geq 1$, $\mathcal{M}_2 \geq 1$, $\mathcal{M}_1 \cap \mathcal{M}_2 = \emptyset$ 及 $\mathcal{M}_1 \cup \mathcal{M}_2 = \mathcal{M}$. 如果 $\mathcal{M}_1$ 和 $\mathcal{M}_2$ 的对偶图也是连通的, 那么可以分别对两个子网格构建哈密尔顿路, 记作 $H_1$ 和 $H_2$, 使得 $H = H_1vH_2$ 是 $\mathcal{M}$ 的一个过顶点哈密尔顿路, 其中 $v$ 是 $\mathcal{M}_1$ 和 $\mathcal{M}_2$ 的一个公共顶点.*

证明. 由于 $\mathcal{M}$ 的对偶图是连通的, 根据引理 2.1, 存在 $T_1 \in \mathcal{M}_1$ 及 $T_2 \in \mathcal{M}_2$ 使得 $T_1$ 和 $T_2$ 是邻居. 由于两个子网格的对偶图均是连通的, 根据定理 2.4, 在两个子网格中分别存在过顶点哈密尔顿路 $H_1 = T_0 \cdots v_1T_1$ 及 $H_2 = T_2v_2 \cdots T_n$. 因为 $T_1$ 和 $T_2$ 是邻居, 它们有三个公共顶点, 其中必然有一个顶点和顶点 $v_1$ 及 $v_2$ 是不同的, 记作 $v$, 从而 $\mathcal{M}$ 上的过顶点哈密尔顿路可以按照如下的方式创建 $H = H_1vH_2 = T_0 \cdots v_1T_1vT_2v_2 \cdots T_n$. □

## §2.4 算法及数值算例

上一节中定理的证明是构造性的, 它们可以直接转化成构建过顶点哈密尔顿

路和哈密尔顿圈的高效算法. 由证明过程也可以看出, 几个定理的证明是很相似的, 算法同样也是类似的. 为简便起见, 本节仅给出构造过顶点哈密尔顿路的算法. 由于证明本身是构造性的, 在此仅给出主要步骤, 具体细节可以在定理的证明过程中得到. 算法共有三个步骤. 第一步是创建初始的哈密尔顿路; 第二步是插入头部和尾部单元所有的邻居; 第三步是扩展哈密尔顿路使得这条路包含网格所有的单元.

**算法 1**: 哈密尔顿路构造算法

假设 $\mathcal{M}$ 是一个协调四面体网格, 其对偶图是连通的, $n = |\mathcal{M}| \geq 2$. 对任意给定的四面体 $T_1$ 和 $T_n$, 一个过顶点哈密尔顿路 $H = T_1 v_1 T_2 \cdots v_{n-1} T_n$ 可以按如下三个步骤构造:

***Step 1.*** *创建一条初始路. 这条路中, 任意两个相邻的单元都有一个公共面. 具体的创建办法见定理 2.4.*

***Step 2.*** *把 $T_1$ 和 $T_n$ 的所有邻居插入到初始哈密尔顿路中. 具体做法参看引理 2.2 的证明.*

***Step 3.*** *将所有剩余的单元插入到哈密尔顿路中. 具体做法参看引理 2.3 的证明.*

下面分析算法 1 的计算复杂度. 第一步是一个图论问题, 这个问题可以由宽度优先搜索 (breadth fist search) 算法实现, 复杂度为 $O(n)$, 其中 $n$ 为网格的单元数目. 由于每个单元的邻居是有限的, 被 $O(n)$ 控制, 因此第二步插入邻居过程的复杂度是 $O(n)$. 在第三步时, 插入每个单元的代价是 $O(1)$, 而剩余单元的数目是有界的, 不超过 $n$, 因此第三步的计算复杂度是 $O(n)$. 综合以上分析, 算法的计算复杂度是 $O(n)$.

我们在自适应有限元工具包 PHG (Parallel Hierarchical Grid) [1, 59] 中实现了算法 1. 下面给出一些测试结果, 测试平台是一台 DELL PowerEdge 2950 服务器 (2 个四核 1.60GHz Intel E5310 CPU, 4MB 二级缓存). 我们选取了五组不同的网格, 单元数目从十多万到几百万不等. 这些网格均由 J. Schöberl 开发的网格生成软件 NETGEN [81] 生成. 表 2.1 给出了程序的计算时间统计, 可以看出即使单元数达到三百万的规模, 用于生成过顶点哈密尔顿路的时间仍然少于四秒. 同时, 程序的运行时间是与网格规模成比例增长的, 表明算法的计算复杂度是线性的.

## §2.5 本章小结

本章证明了协调四面体网格上过顶点哈密尔顿路和哈密尔顿圈在一定条件下的存在性, 并实现了线性复杂度的算法, 该算法可用于网格划分中单元的排序.

**表 2.1 程序运行时间**

| | 单元数 | 面数 | 顶点数 | 时间 (seconds) | 单元/秒 |
|---|---|---|---|---|---|
| Mesh 1 | 155,456 | 323,264 | 32,238 | 0.127770 | 1,216,686 |
| Mesh 2 | 239,616 | 480,704 | 40,781 | 0.240720 | 995,413 |
| Mesh 3 | 453,736 | 923,336 | 85,338 | 0.515140 | 880,801 |
| Mesh 4 | 1,243,648 | 2,536,704 | 232,288 | 1.265506 | 982,727 |
| Mesh 5 | 3,211,264 | 6,498,304 | 573,628 | 3.736147 | 859,512 |

# 第三章　空间填充曲线

空间填充曲线提供了一类有效的网格划分方法. 这类曲线出现于 19 世纪, 由于其特殊的性质一直吸引数学家进行研究, 研究主要集中在空间填充曲线的数学性质上, 如连续性、可微性等. 空间填充曲线具有良好的空间局部性, 近几十年, 人们将其广泛应用在内存管理、数据库索引、动态负载平衡、图像处理等领域, 在空间填充曲线的算法研究方面取得了长足的进步. 本章对高维希尔伯特空间填充曲线的编码解码算法进行研究.

## §3.1　基础知识

George Cantor 在 1878 年指出任意两个有限维光滑流形具有相同的势 [86]. 该结果暗示从单位区间 $[0,1]$ 到单位正方形区域 $[0,1]^2$ 存在双射. 问题是这种类型的映射是不是连续的. 数学家 Netto 在 1879 年给出一个著名结果 [86], 指出这样的双射是间断的.

**定理 3.1.** *(Netto 定理) 如果 $f$ 是一个从 $m$ 维光滑流形到 $n$ 维光滑流形的双射, 其中 $m \neq n$, 那么 $f$ 是间断的.*

有了这个定理之后, 问题变成: 如果去掉双射的限制, 是否存在一个满射将区间 $[0,1]$ 映射到正方形区域 $[0,1]^2$ 上. Giuseppe Peano (1858–1932) 在 1890 年发现了第一个这样的曲线. 因此, 具有这样性质的曲线不仅被称为空间填充曲线, 又被称为 Peano 曲线. 曲线的广义定义如下:

**定义 3.1.1.** *一个 $n$ 维空间填充曲线是一个连续的满射, 将单位区间 $[0,1]$ 映射到 $n$ 维超立方 $[0,1]^n$.*

在 Peano 之后, 其他数学家也对这个问题进行了深入研究, 进一步的例子分别被 D. Hilbert (1891), E.H. Moore (1900), W. F. Osgood (1903), H. Lebesgue (1904), W. Sierpiński (1912), G. Pólya (1913), K. Knopp (1917) 等人给出 [86]. 一般情况下, 空间填充曲线由线段序列迭代, 当迭代次数趋于无穷时生成. 用希尔伯特空间填充曲线做例子, 如图 3.1, 3.2 及 3.3 所示: 图 3.1 是经过一次迭代产生的曲线; 图 3.2 是经过二次迭代产生的曲线; 图 3.3 是经过六次迭代产生的曲线.

空间填充曲线主要有: Peano 空间填充曲线、希尔伯特空间填充曲线、Moore 空间填充曲线、Sierpiński 空间填充曲线、Morton 空间填充曲线、Lebesgue 空间填充曲线、Osgood 空间填充曲线等. 其中 Sierpiński 空间填充曲线 [105] 如图 3.4 所示, Morton 空间填充曲线 [88] 如图 3.5 所示 (高维曲线同样存在, 出于简洁的考

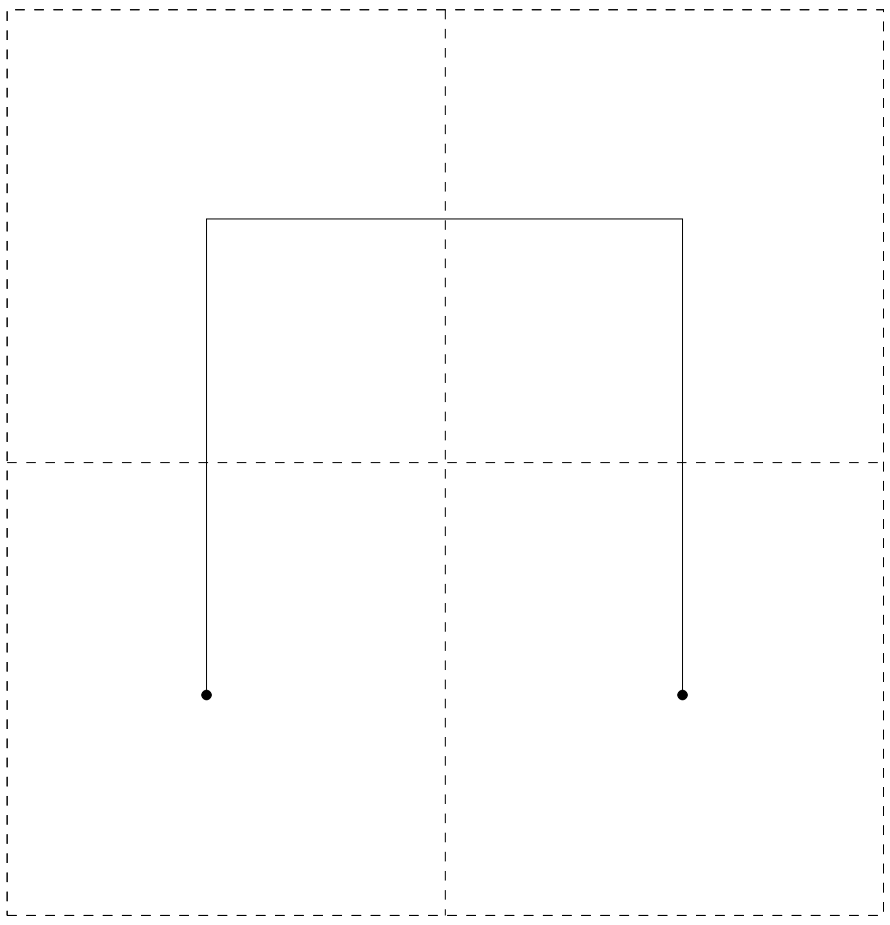

图 3.1 一阶希尔伯特空间填充曲线

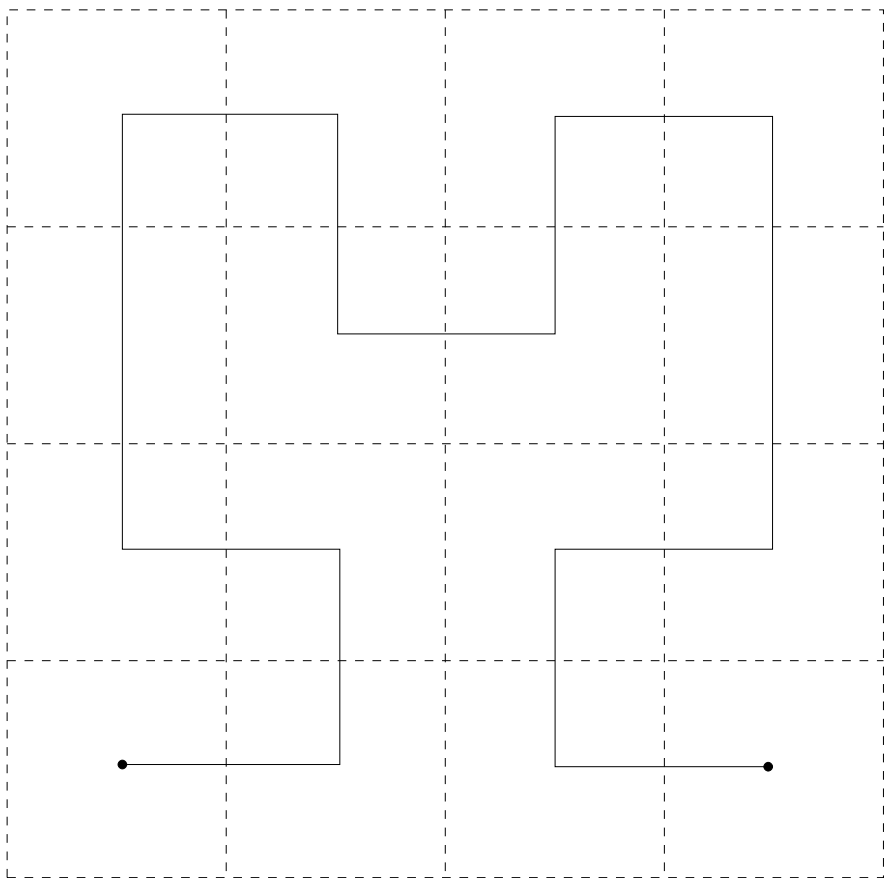

图 3.2 二阶希尔伯特空间填充曲线

虑, 仅给出了二维图). 这些曲线大多是在一个世纪前发现的, 但它们的具体应用却是在近几十年. 空间填充曲线具有自相似性、空间局部性等特点, 被广泛用在数学、算法、地理信息系统、图像处理、数据库、电路设计、加密、内存管理、数据压缩、科学计算及动态负载平衡等领域 [89, 90, 91, 92, 93, 94, 95, 96, 97, 98, 100, 101, 102, 103, 104]. Velho 等人利用空间填充曲线设计了图像抖动算法, 应用于激光打印机等 [87, 104, 118]. Salmon 等人利用空间填充曲线为 $N$- 体模拟程序进行数据划分及重排序以改进程序的空间局部性 [103]. Challacombe 等人利用空间填充曲线在并行机上对稀疏矩阵进行划分 [95]. Mellor-Crummey 等人 [100, 101] 以

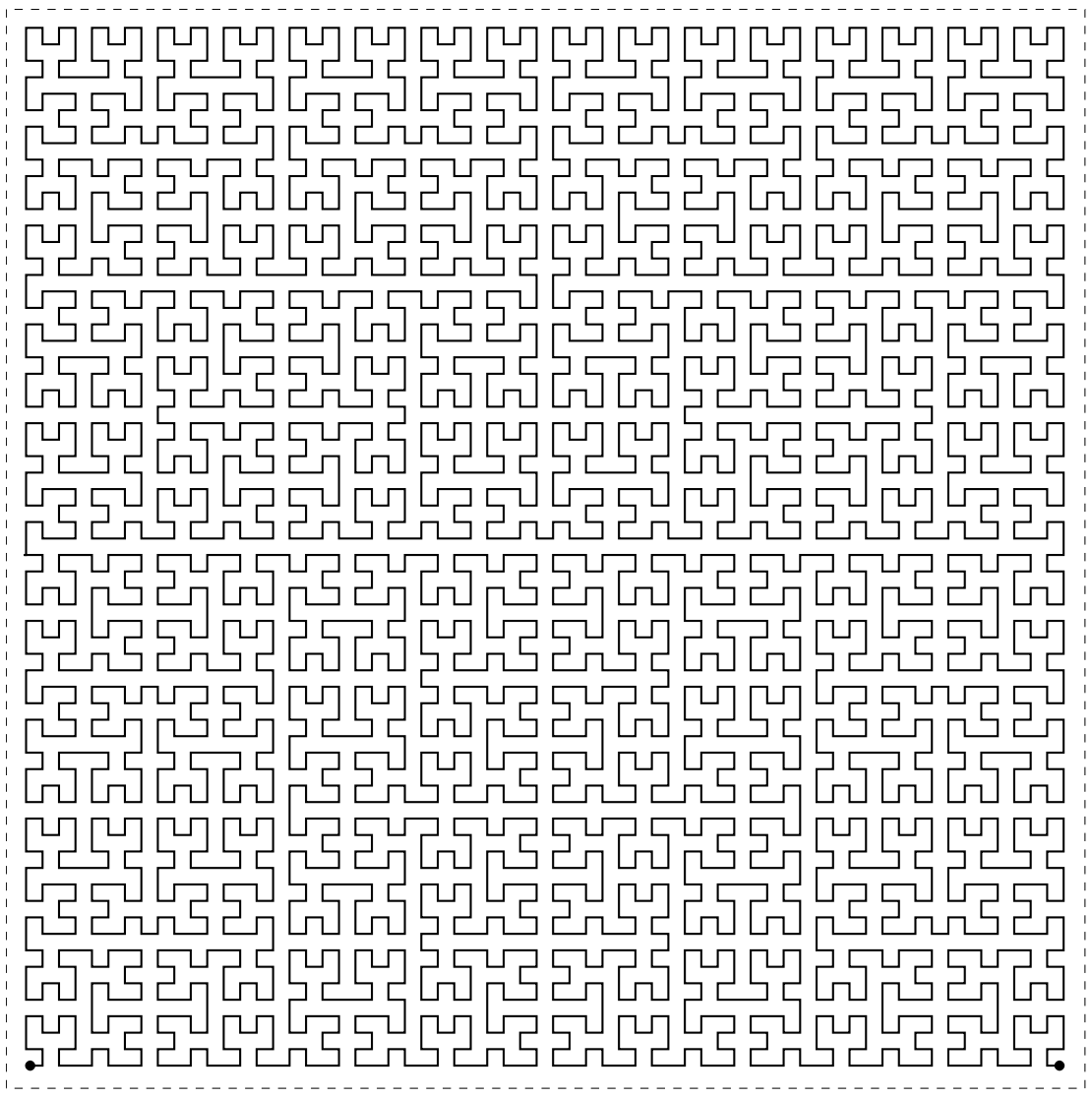

图 3.3 六阶希尔伯特空间填充曲线

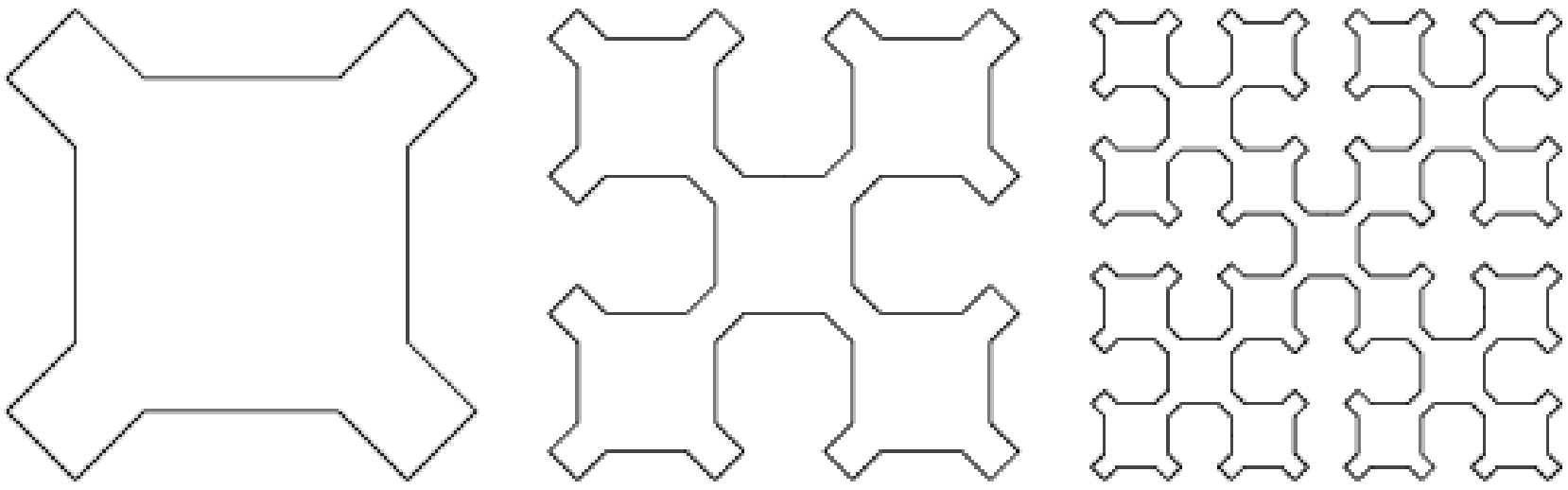

图 3.4 Sierpiński 空间填充曲线 (一阶, 二阶, 三阶)

及 Hu 等人 [96] 利用空间填充曲线对数据及计算重排序以改善内存使用. Matias 等人 [98] 利用空间填充曲线对视频进行处理. Hwansoo Han[99] 等人将空间填充曲线用在 CFD 计算等科学计算领域中. 空间填充曲线有良好的空间局部性, 也被用在数据压缩 [106]、索引 [107] 等方面. 在并行计算方面, 空间填充曲线划分算法也是动态负载平衡常用的算法 [18, 19, 62, 123], 其中希尔伯特空间填充曲线, Morton 空间填充曲线, Sierpiński 空间填充曲线具有最好的空间局部性, 应用也是最广泛的. 由于 Morton 空间填充曲线的算法简单, 人们首先使用这种曲线, 应用领域相当广泛, 但由图 3.5 可以看出, 该曲线具有跳跃性, 其空间局部性相对于另外两种曲线要差一些. 随着希尔伯特空间填充曲线和 Sierpiński 空间填充曲线的编码解码算法日趋成熟, 这两种空间填充曲线越来越占有主导地位, 尤其是希尔

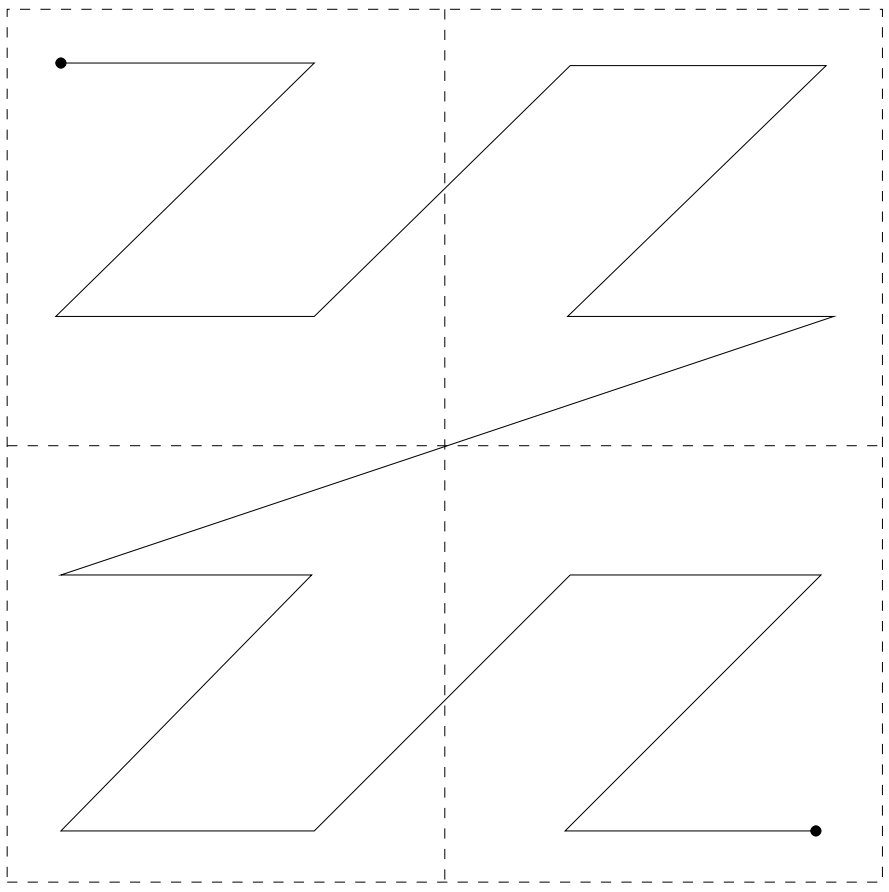

图 3.5 Morton 空间填充曲线 (二阶)

伯特空间填充曲线.

## §3.2 希尔伯特序

G. Peano 第一个发现了空间填充曲线, 但第一个给出空间填充曲线一般化生成方法的人却是希尔伯特 (D. Hilbert, 1891) [86, 118]. 在这里, 用二维希尔伯特空间填充曲线作为例子来阐述空间填充曲线的生成过程, $I$ 表示区间 $[0,1]$, $Q$ 表示正方形 $[0,1]^2$. 希尔伯特空间填充曲线的生成方式是: 将区间 $I$ 剖分成四个子区间, 将 $Q$ 剖分成四个子正方形, 第一次迭代的结果如图 3.1 所示; 接着将 $I$ 的每个子区间剖分成四个子区间, $I$ 子区间的个数为 16, 将 $Q$ 的每个子正方形剖分成四个子正方形, 子正方形的个数同样为 16, 第二次迭代的结果如图 3.2; 经过六次迭代产生的曲线如图 3.3 所示. 该过程一直持续, 当趋于无穷时, 产生一个充满整个正方形的曲线. 这些子正方形的安排按照如下方式: 相邻子区间对应的子正方形有一条公共边.

假设迭代次数为 $m$, 迭代过程会将单位区间 $I$ 及单位正方形 $Q$ 剖分成 $2^{2m}$ 个子区间和 $2^{2m}$ 个子正方形. 对区间 $I$ 的每个子区间编号, 编号取值为整数, 其范围为 $I_2^m = \{0, 1, \cdots, 2^{2m} - 1\}$ , 单位正方形 $Q$ 被分成了 $2^{2m}$ 个子正方形, 每个子正方形赋予一个整数坐标, 其取值范围为 $Q_2^m = \{(x, y) : 0 \leq x, y < 2^m - 1\}$. 迭代次数 $m$ 又称为阶 (order)、分辨率 (resolution) 或深度 (depth). 在下面的讨论中, 阶这个术语将会被一直使用. 我们用 $H_n^m$ 表示 $m$- 阶 $n$- 维希尔伯特空间填充曲线. 仍用二维希尔伯特空间填充曲线做例子, 取 $m$ 为 2, 集合 $Q_2^2$ 中每一个整数

坐标均对应集合 $I_2^2$ 中的一个编号, $H_2^2$ 如图 3.6 所示, 曲线的起始点为 0, 终点为 15, 用黑色实心圆表示. 我们可以看出 $H_2^2$ 是由 $H_2^1$ 生成的: $H_2^2$ 包含了四个经过空间变换的 $H_2^1$, 如图 3.6 所示; $H_2^3$ 也是由四个经过空间变换的 $H_2^2$ 得到. 从这个观察可以得到一般化的希尔伯特空间填充曲线生成方法: $H_2^m$ 可以由四个 $H_2^{m-1}$ 拼接得到, 只是这四个 $H_2^{m-1}$ 分别需要做一次空间变换. 具体的变换方式将在接下来的内容中仔细分析.

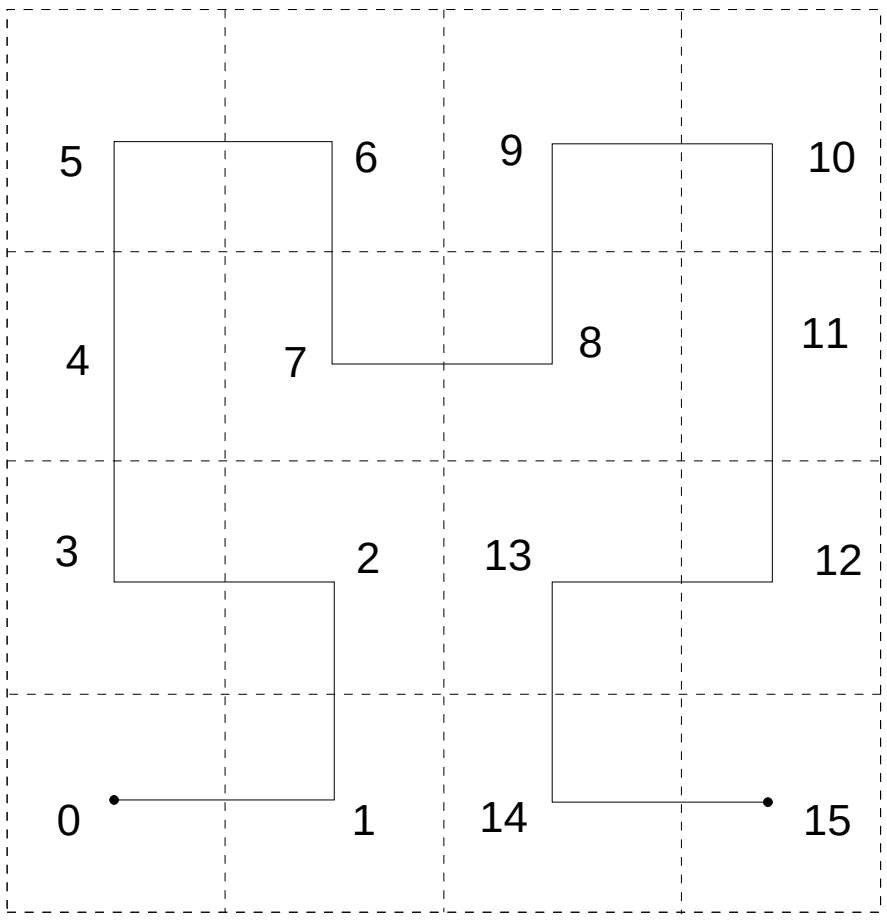

图 3.6 希尔伯特序 ($H_2^2$, 二阶)

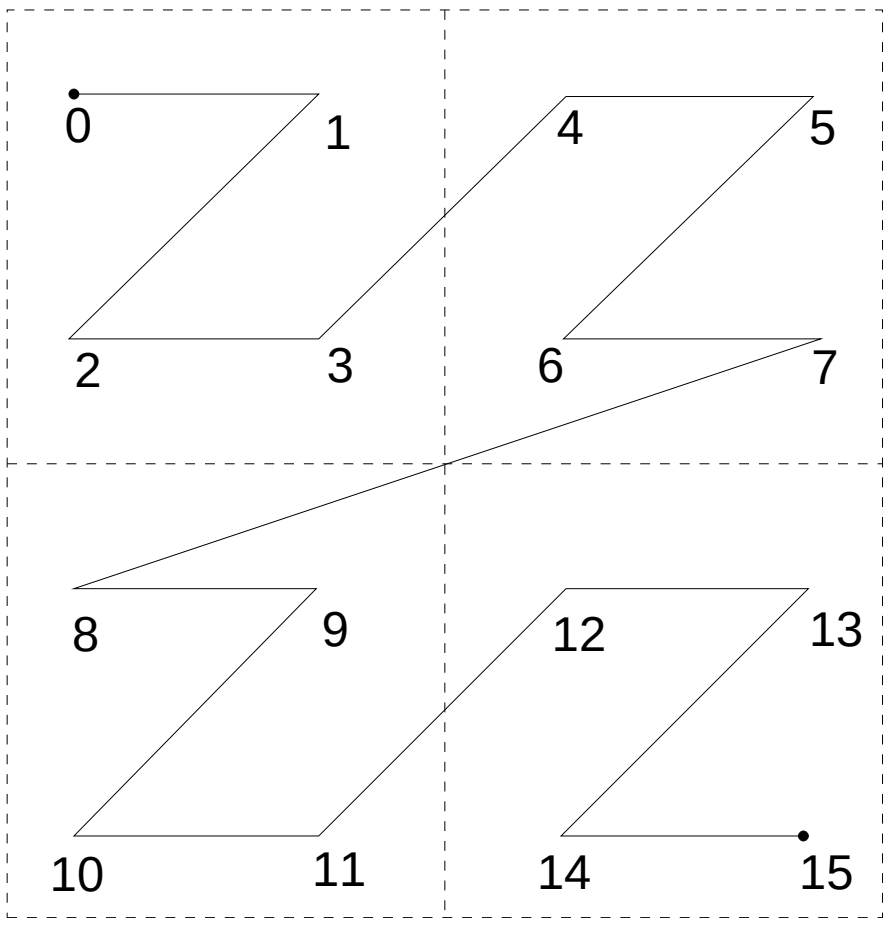

图 3.7 Z-order (Morton order, 二阶)

对任意的正整数 $m$, 上述方法为集合 $Q_2^m$ 定义了一种顺序 (order). 由于该顺序是由希尔伯特空间填充曲线定义的, 又被称为希尔伯特序 (Hilbert order, Hilbert code). 对于不同类型的空间填充曲线, 可以定义不同的排序方法 (ordering), 如 Morton 空间填充曲线, 由于其曲线与 $Z$ 很相似, 又称为 $Z$- 序 (Z-order, Morton order, Morton code) [88], 图 3.7 是一个二阶 Morton 空间填充曲线定义的 $Z$- 序. $Q_2^m$ 与 $I_2^m$ 是一个一一对应关系, 将 $Q_2^m$ 中一个坐标映射成 $I_2^m$ 中一个整数的过程称为编码 (enconding), 编码过程为每一个坐标建立一个索引 (index); 将 $I_2^m$ 中一个整数映射成 $Q_2^m$ 中一个坐标的过程称为解码 (decoding), 解码过程是一个曲线遍历的过程. 高维空间填充曲线的编码解码定义和二维是一样的. 希尔伯特空间填充曲线和希尔伯特序是同一个内容, 只是名称不一样.

接下来的内容集中在希尔伯特序 (希尔伯特空间填充曲线) 上. 很多研究人员对希尔伯特空间填充曲线的计算方法进行了研究, 提出了许多有效的算法, 它们可以划分为两类: 递归算法 [109, 110, 111, 112] 和迭代算法 [62, 113, 114, 115, 116, 118]. 对递归算法, Butz, Goldschlager[109, 110] 等人设计了简洁的算法, 这些算法是空间填充曲线递归算法的基础. Breinholt 和 Schierz[108] 设计了效率更高的递归算法. Witten 等人使用 GROPER 语言描述了一种递归技术 [111]. Cole 在一个图形系统上实现了递归过程, 并为 Peano 及 Sierpiński 空间填充曲线描述了递归算法 [112]. 在通常情况下, 迭代算法的速度要快于递归算法, 尤其是表驱动 (table-driven) 的迭代算法 [62, 113, 114, 115, 116, 118]. Fisher[113], Liu 和 Schrack[115, 116] 为二维和三维问题设计了基于二进制操作的迭代算法. Faloutsos 等人 [117] 通过分析希尔伯特空间填充曲线和 $Z$- 序之间的关系, 实现了编码解码算法. 在文 [62] 中, 作者展示了基于表的算法, 其中使用了两个表分别存储了各个象限的变换信息及状态信息, 由于仅需要查表而不需要重新计算, 极大提高了生成速度. 以上提到的算法一般都被限制在二维和三维空间, 对高维空间填充曲线, 由于复杂性很高, 这方面的工作比较少. Butz[109] 的递归算法可以对高维曲线进行计算, 但复杂度较高, 速度慢, 不适合大规模计算. Lawder[119] 重新对 Butz 的算法进行研究, 并给出了改进. 最近 Kamata 等人 [120] 提出了基于硬件的高维算法. Li 等人 [121] 分析了高维希尔伯特空间填充曲线不同象限的变换关系.

下面介绍本文作者在希尔伯特序编码和解码算法方面的工作. 本文的工作基于文章 [118, 121]. Li 等人设计了算法分析高维希尔伯特序的演化规则, 这些规则包含了坐标变换信息, 由于问题的复杂性, 作者并没有给出编码解码算法. 本文对这些演化规则进行了细致地分析和理解, 并给出这些变换的数学表达式和解释, 在此基础上, 设计了编码和解码算法. 对于编码和解码算法, 分别给出了基于位运算的算法和基于算术运算的算法. 基于位运算的算法是基于算术运算算法

的改进, 计算效率更高. 算法的时间计算复杂度为 $O(mn)$. 对于给定的编码解码问题, 算法可以在 $m$ 步给出结果. Chen 等人 [118] 利用希尔伯特空间填充曲线特有的性质, 针对二维问题设计了复杂度更低的编码解码算法, 计算复杂度为 $O(k)$, 其中 $k$ 定义为 $k = \log_2(\max(x, y)) + 1$, 与 $m$ $(k \leq m)$ 的大小无关. 利用类似的性质, 可以将三维希尔伯特序的编码解码算法的复杂度降为 $O(k)$ [123], 其中 $k$ 定义为 $k = \log_2(\max(x, y, z)) + 1$. 本章给出了这种思想的理论依据, 并推广到高维希尔伯特空间填充曲线, 得到一系列复杂度更低的算法. 同一个坐标, 在不同阶数的希尔伯特空间填充曲线中的序并不一定相同, 我们对传统的希尔伯特空间填充曲线做了改造, 得到希尔伯特曲线新的变体, 该曲线满足性质: 同时属于不同阶曲线的一组坐标在这些曲线中的顺序是一样的. 本章为该曲线变体设计了编码解码算法.

## §3.3 希尔伯特序编码解码算法

本节将介绍编码算法和解码算法. 我们先给出预备知识, 然后分析文章 [121] 中介绍的算法, 分析高维希尔伯特空间填充曲线的性质, 给出理论结果, 在此基础上介绍编码和解码算法, 其中每类算法将使用四种不同的方式表达, 两种使用算术运算表达, 两种使用位运算表达, 并且具有不同的计算复杂度. 这里介绍的算法对任意维数 $(n \geq 2)$ 的希尔伯特空间填充曲线均适用. 本文给出的是算法框架, 可以作为模板应用到不同维数的问题中.

### §3.3.1 预备知识

本节所有的操作均在非负整数上进行. 需要声明的是操作同样可以在实数域上进行, 只需将输入及输出做一次转换即可. 令 $Q_n^m$ 为整数坐标集合, $Q_n^m = \{(x_n, \cdots, x_2, x_1) | 0 \leq x_i < 2^m, 1 \leq i \leq n\}$, $(x_n, \cdots, x_1)$ $(\in Q_n^m)$ 是希尔伯特空间填充曲线的坐标, $x_i$ 为坐标的第 $i$ 个分量. 记 $I_n^m = \{0, 1, \cdots, 2^{nm} - 1\}$. 当 $n$ 为 2 时, 即通常的平面 (此时有 $x_2 = x, x_1 = y$); 当 $n$ 为 3 时, 即通常的三维空间 (此时有 $x_3 = x, x_2 = y, x_1 = z$). 当所有的 $x_i$ 和 $y_i$ $(1 \leq i \leq n)$ 均为 0 或者 1 时, 定义如下逻辑运算 $\wedge$

$$(x_n, \cdots, x_2, x_1) \wedge (y_n, \cdots, y_2, y_1) = (x_n \wedge y_n, \cdots, x_2 \wedge y_2, x_1 \wedge y_1), \tag{3.1}$$

在这里 $\wedge$ 代表异或运算符 XOR.

引入符号 $(a_1a_2\cdots a_k)_d$ 表示 $d$ 进制数字, 其中 $0\le a_i<d$, $1\le i\le k$, 且 $k$ 可以为任意的正整数. 当 $d$ 为 2 时, 表示二进制; 当 $d$ 为 10 时, 表示通常的十进制. 各种不同的数字表示方式可以互相转换. 当 $d$ 为 2 时, 定义取反运算符 $\mathrm{Re}_k$ 如下:

$$\mathrm{Re}_k(j)=\mathrm{Re}_k((a_1a_2\cdots a_k)_2)=(b_1b_2\cdots b_k)_2,\quad b_i=1-a_i,\ 1\le i\le k, \tag{3.2}$$

其中 $j$ 是一个十进制数, $j=(a_1a_2\cdots a_k)_2$ 为 $j$ 的二进制表示, $j$ 满足条件 $0\le j<2^k$. 对二进制数, 可以引入通常的按位与 (and), 右移, 左移运算符, 分别记作 &, $\gg$ 和 $\ll$. 引入定义 [122]

$$p_n^i(a_1,a_2,\cdots,a_n)=(\sum_{j=1}^{i}a_j) \bmod 2, \tag{3.3}$$

其中 $a_i$ 等于 0 或者 1 ($1\le i\le n$). 借助 $p_n^i$, 定义映射 $f_n$

$$f_n(a_1,a_2,\cdots,a_n)=(b_1b_2\cdots b_n)_2=j,\quad b_1=a_1,\quad b_i=\begin{cases} a_i, \text{ if } p_n^{i-1}=0\\ 1-a_i, \text{ if } p_n^{i-1}=1\end{cases}, \tag{3.4}$$

其中 $a_i$ 等于 0 或者 1, 并且 $j$ 为十进制数. 该函数将一个向量 $(a_1,a_2,\cdots,a_n)$ 映射成一个整数. 同时定义 $f_n$ 的逆映射 $b_n$

$$b_n(j)=b_n((a_1a_2\cdots a_n)_2)=(b_1,b_2,\cdots,b_n),\quad b_1=a_1,\quad b_i=\begin{cases} a_i, \text{ if } a_{i-1}=0\\ 1-a_i, \text{ if } a_{i-1}=1\end{cases}, \tag{3.5}$$

其中 $j$ 为一个十进制数且等于 $(a_1a_2\cdots a_n)_2$.

*注记:* 在实际应用中, 对特定的 $n$, 映射 $b_n$ 及 $f_n$ 可以事先计算好存储在表中, 需要的时候直接查询. 对于 $b_n$, 可以采用如下方式快速计算

$$b_n((a_1\cdots a_n)_2)=(a_1,\cdots,a_n)\wedge(0,a_1,\cdots,a_{n-1}). \tag{3.6}$$

由于 $f_n$ 和 $b_n$ 互为逆映射, 计算 $b_n$ 之后, 可以直接得到 $f_n$ 的结果.

下面部分概念来自文章 [121]. $n$- 维希尔伯特单元 ($n$-dimensional cell) 是一个 1 阶的 $n$- 维希尔伯特空间填充曲线. $n$- 维希尔伯特因子列表 (Hilbert gene list) 是一系列坐标变换指令, 它们给出了从 $(m-1)$- 阶希尔伯特空间填充曲线到 $m$- 阶曲线的变换指令. 坐标的变换指令分为两类: 交换指令 (exchange command) 和反转指令 (reverse command). 它们可以解释为坐标关于某个超平面的对称变换. 希尔

伯特空间填充曲线是有方向的, 它有一个入口和一个出口. 当对一个希尔伯特单元做坐标变换时, 入口和出口的空间坐标可能会改变. 对一个 $n$- 维空间, 它有 $2^n$ 个象限, 例如二维平面有四个象限, 三维空间有八个象限, 编号均从 0 开始, 编号方式由 $f_n$ 直接计算. 每个象限均有不同的变换指令. 引入 $H_n^{i,0}$ 和 $H_n^{i,1}$ $(0 \le i < 2^n)$ 表示单元经过 $i$- 象限的指令变换后的入口和出口坐标. 在文章 [121] 中, 作者介绍了如何生成希尔伯特单元和希尔伯特因子列表.

希尔伯特单元在文章 [121] 中是按照递归的方式生成的, 如果要生成 $n$- 维希尔伯特单元, 需先生成 $(n-1)$- 维希尔伯特单元. 本文分析了其生成方式, 基于理论分析, 设计了映射 $b_n$ 直接生成单元, 同时引入了 $b_n$ 的逆映射 $f_n$. $b_n$ 将在解码算法中被使用, $f_n$ 在编码算法中被使用.

下面我们对希尔伯特因子列表进行分析, 给出其数学解释并给出具体的变换表达式. 希尔伯特因子列表包含两种指令: 交换指令和反转指令. 这两组指令均可以解释为关于某个超平面的对称变换. 对于交换指令, 如果变换发生在分量 $x_i$ 与 $x_j$ $(i \neq j)$ 之间, 则其对称面为超平面 $x_i - x_j = 0$, 坐标 $(x_n, \cdots, x_i, \cdots, x_j, \cdots, x_1)$ 经过变换之后则变为 $(x_n, \cdots, x_j, \cdots, x_i, \cdots, x_1)$, 该结果可以由简单的数学演算直接推导出来, 简单说就是交换第 $i$ 个与第 $j$ 个位置的分量, 这是称其为交换指令的原因. 对反转指令, 假设变换的位置为 $i$ 并且当前希尔伯特空间填充曲线的阶数为 $m$, 则其对称面为超平面 $x_i = \frac{2^m-1}{2}$, 坐标 $(x_n, \cdots, x_i, \cdots, x_1)$ 经过变换后变为 $(x_n, \cdots, 2^m - 1 - x_i, \cdots, x_1)$. 该操作与映射 $\mathrm{Re}_m(x_i)$ 是等价的. 反转变换改变的仅是 $x_i$ 位置的坐标值, 结果与指令的执行顺序是无关的. 这两种变换的逆变换均是自身, 它们均是保长度的, 并且保持曲线上各个点的相对位置不变, 即对给定的曲线, 变换只会改变曲线在空间中的位置信息, 但不会改变各个坐标对应的序的相对编号. 总结如下:

**性质 3.2.** *交换变换的本质是镜面对称变换, 如果变换发生在分量 $x_i$ 与 $x_j$ $(i \neq j)$ 之间, 其对称面为超平面 $x_i - x_j = 0$; 反转变换的本质也是镜面对称变换, 如果变换发生在 $x_i$ 分量且当前曲线的阶数为 $m$, 则对称面为超平面 $x_i = \frac{2^m-1}{2}$, 如果有多个反转变换, 结果与执行顺序无关. 交换和反转变换的逆变换均为自身.*

文章 [121] 给出了生成这些指令的算法, 为了表达的方便, 本文对生成算法进行了重写, 表达式如下:

$$G_n^{i,0} = (b_n(0) \wedge b_n(2^n - 1)) \wedge (H_n^{i,0} \wedge H_n^{i,1}), \quad (0 \le i < 2^n), \tag{3.7}$$

$$G_n^{i,1} = b_n(0) \wedge H_n^{i,0}, \quad (0 \le i < 2^n). \tag{3.8}$$

$G_n^{i,0}$ 和 $G_n^{i,1}$ 表示第 $i$ 象限的交换指令和反转指令. 根据 $b_n$ 的定义, 可以将 $b_n(0)$ 及 $b_n(2^n-1)$ 分别写成如下形式 $(0,\cdots,0)$ 和 $(1,0,0,\cdots,0)$, 有关系 $b_n(0)\wedge b_n(2^n-1)=(1,0,0,\cdots,0)$ 成立. 希尔伯特单元的出口和入口坐标仅有一个分量不同, 并且两种变换是保持曲线上各点的相对位置不变, 经过变换后的单元的入口坐标和出口坐标仅有一个分量的值是不同的, 所以入口坐标 $H_n^{i,0}$ 与出口坐标 $H_n^{i,1}$ 做异或运算, 得到的结果中除了一个分量为 1 外全部为 0. 根据式 (3.7), $G_n^{i,0}$ 的分量要么全部为 0 要么仅有两个不为 0. 这意味着, 每个象限上要么不执行交换指令要么仅需要执行一次. 如果 $G_n^{i,0}$ 的第 $i$ 个与第 $j$ 个分量为 1, 则在这两个分量之间执行交换指令. 反转指令则可能需要执行多次.

在此介绍希尔伯特曲线的一个特殊性质. 分析象限 0 的性质, 根据定义可以把 $b_n(0)$, $b_n(2^n-1)$, $H_n^{0,0}$ 及 $H_n^{0,1}$ 分别写成如下形式: $(0,\cdots,0)$, $(1,0,\cdots,0)$, $(0,\cdots,0)$ 和 $(0,\cdots,0,1)$. 根据式 (3.7) 及 (3.8), 有 $G_n^{0,0}=(1,0,\cdots,0,1)$ 和 $G_n^{0,1}=(0,\cdots,0)$. 这意味着在第 0 象限, 有且仅有交换变换, 变换发生在第一个和最后一个分量之间. 这是设计计算复杂度更低的算法的理论基础.

**性质 3.3.** *在每个象限中, 最多执行一次交换变换, 反转变换可以被执行多次. 在 0 象限中, $G_n^{0,0}=(1,0,\cdots,0,1)$, $G_n^{0,1}=(0,\cdots,0)$, 有且仅有交换变换.*

### §3.3.2 编码算法

本文给出 $n$ 维空间 $(n\geq 2)$ 希尔伯特空间填充曲线的编码算法. 本小节仅给出算法模板, 当给定具体的维数 $n$ 后, 可以根据这个模板导出相应的算法.

前面已经提到, 希尔伯特空间填充曲线是自相似的, 曲线的生成是一个迭代 (递归) 过程, $m$- 阶希尔伯特空间填充曲线 $H_n^m$ 的生成需要 $(m-1)$- 阶曲线 $H_n^{m-1}$. 具体过程可以描述为: $n$ 维空间有 $2^n$ 个象限, 其每一个象限包含一个 $(m-1)$- 阶希尔伯特空间填充曲线, 对每个象限的 $(m-1)$- 阶希尔伯特空间填充曲线实施该象限对应的交换指令和反转指令, 然后连接相邻象限中曲线的入口和出口, 即可得到一个 $m$- 阶的希尔伯特空间填充曲线. 如图 3.1 及 3.2 所示, 可以看出 $H_2^2$ 的四个子正方形中的每个均包含一个 $H_2^1$. 从曲线的入口出发, 遍历曲线, 可以为曲线中每个点建立一个序, 这是编码过程.

下面详细分析编码过程。对任意给定的坐标 $P$，它的序等于从入口开始遍历曲线到指定位置所经过的点的个数. 假设当前曲线的阶数为 $m$, 它由 $2^n$ 个 $(m-1)$- 阶曲线生成, $P$ 必定位于某个 $(m-1)$- 阶曲线中, 假设位于第 $i$ 象限中, 遍历此曲线, 在它到达所处的 $(m-1)$- 阶曲线之前, 要经过 $i$ 个 $(m-1)$- 阶曲线. 使用记号 $X_m(P)$ 表示点 $P$ 在 $m$- 阶曲线中的希尔伯特序, 那么它的希尔伯特序可以由下式表示:

$$X_m(P) = i \times 2^{n(m-1)} + \widetilde{X}_{m-1}(P),$$

其中 $\widetilde{X}_{m-1}(P)$ 表示点 $P$ 在经过变换的 $(m-1)$- 阶希尔伯特曲线中的序. 前面已经讲过, 交换指令和反转指令是保持长度的, 不会改变曲线的序, 所以我们将 $P$ 所处的某个 $(m-1)$- 阶曲线做逆变换, 变换成一个没有经过改变的 $(m-1)$- 阶曲线, 上式相应地变为

$$X_m(P) = i \times 2^{n(m-1)} + X_{m-1}(P). \tag{3.9}$$

该式是设计迭代算法的出发点, 它保证我们可以使用统一的算法框架计算希尔伯特序.

假设当前希尔伯特空间填充曲线的阶数为 $m$, 任意一个点 $(x_n, \cdots, x_1)$ $(\in Q_n^m)$, 其每个分量 $x_i$ $(1 \le i \le n)$ 可以写成二进制表示 $x_i = (x_i^m x_i^{m-1} \cdots x_i^1)_2$. 希尔伯特序保存在整数 $(r_m r_{m-1} \cdots r_1)_{2^n}$ $(\in I_n^m)$ 中. 算法中用到的变换信息 $G_n^{i,0}$ 及 $G_n^{i,1}$ $(0 \le i < 2^n)$ 需事先计算好, 具体的过程可以从文章 [121] 得到.

希尔伯特空间填充曲线的生成是从低阶到高阶的递归过程, 这样的方式可以生成一个完整的曲线. 由于编码解码问题仅需计算一个点, 这样的方式是浪费时间和空间的, 根据式 (3.9), 本文采取从高阶到低阶的一个回溯过程, 以迭代的方式计算. 变换指令有两类, 在编码过程中, 反转指令先执行, 然后是交换指令. 下面给出第一个算法, 它基于算术运算, 具体过程如下:

**编码算法 1.1.**

(1) 如果 $m = 0$, 程序终止. 否则, $r_m = f_n(x_n^m, x_{n-1}^m, \cdots, x_1^m)$.

(2) 对每个整数 $i$ $(1 \le i \le n)$, 如果 $x_i^m = 1$, 那么 $x_i = x_i - 2^{m-1}$.

(3) 反转. 对每个整数 $i$ $(1 \le i \le n)$, 如果 $G_n^{r_m,1}$ 的第 $i$ 个分量是 1, 那么 $x_i = 2^{m-1} - 1 - x_i$.

(4) 交换. 如果 $G_n^{r_m,0}$ 存在两个非零分量, 其位置为 $i$ 和 $j$, 则交换 $x_i$ 和 $x_j$.

(5) $m = m - 1$, 转至 (1).

在*编码算法 1.1* 中, 如果将整数写成二进制数, 会发现在步骤 (2) 中, $x_i = (x_i^m \cdots x_i^1)_2$ 是被 $(x_i^{m-1} \cdots x_i^1)_2$ 所代替, 仅丢掉一个高位有效数字, 该运算可以由简单的按位与操作 (&) 完成. 步骤 (3) 中的算术运算等价于一个按位取反运算,

等价于 $\mathrm{Re}_{m-1}$ 操作. 在实际应用中, 位运算的效率要高于算术运算, 上述算法可以按照位运算的方式重写:

**编码算法 2.1.**

(1) 如果 $m=0$, 程序终止. 否则, $r_m = f_n(x_n^m, x_{n-1}^m, \cdots, x_1^m)$.

(2) 对每个整数 $i$ $(1 \le i \le n)$, $x_i = (x_i^{m-1} \cdots x_i^1)_2 = (x_i^m \cdots x_i^1)_2 \ \& \ (011 \cdots 1)_2$.

(3) 反转. 对每个整数 $i$ $(1 \le i \le n)$, 如果 $G_n^{r_m,1}$ 的第 $i$ 个分量是 1, 那么 $x_i = \mathrm{Re}_{m-1}(x_i) = (x_i^{m-1} \cdots x_i^1)_2 \wedge (1 \cdots 1)$.

(4) 交换. 如果 $G_n^{r_m,0}$ 存在两个非零分量, 其位置为 $i$ 和 $j$, 则交换 $x_i$ 和 $x_j$.

(5) $m = m-1$, 转至 (1).

*编码算法 1.1* 和 *编码算法 2.1* 只是表达不同, 它们具有相同的复杂度. 下面分析它们的计算复杂度. 算法在步骤 (2) 和 (3) 需要更新 $n$ 个坐标, 复杂度为 $O(n)$, 步骤 (4) 的计算复杂度为 $O(1)$, 算法会在 $m$ 步内终止, 因此算法的时间复杂度为 $O(nm)$. 对于空间复杂度, 算法需要存储 $2^n$ 个象限的变换信息, 空间复杂度为 $O(2^n)$. 问题本身的复杂度是按照指数增长的, 这是前人的工作大多集中在二维及三维的原因. 当 $n$ 较小时, 存储变换信息所占的内存是可以接受的, 例如维数 $n$ 为 20 时, 所占内存大概为 2M. 当维数较高时, 高效实现仍然是挑战性的.

根据 $f_n$ 的定义, 有 $r = f_n(0, 0, \cdots, 0) = 0$. 由性质 3.3 知道在象限 0 中仅有交换指令. 观察 *编码算法 1.1*, 会发现仅有步骤 (4) 被执行. 这个过程将一直执行到 $r$ 是非零值, 在这个过程仅有坐标分量 $x_n$ 与 $x_1$ 交换, 并且交换两次的结果是保持坐标不变. 这个过程可以跳过, 如果跳过的次数为偶数, 保持坐标不变; 如果次数为奇数, 则交换 $x_n$ 与 $x_1$. 通过这个观察, 可以设计计算复杂度更低的算法. 定义 $k = \mathrm{floor}(\log_2(\max\{x_n, \cdots, x_1\})) + 1$, 并令 $\log_2(0)$ 等于 0. 上述两个算法可以分别改写成如下形式:

**编码算法 1.2.**

(1) 设定 $(r_m \cdots r_1)_{2^n}$ 为 0. 如果 $m$ 与 $k$ 具有不同的奇偶性, 则交换 $x_1$ 与 $x_n$. $m = k$.

(2) 如果 $m=0$, 程序终止. 否则 $r_m = f_n(x_n^m, x_{n-1}^m, \cdots, x_1^m)$.

(3) 对每个整数 $i$ $(1 \le i \le n)$, 如果 $x_i^m$ 为 1, 则 $x_i = x_i - 2^{m-1}$.

(4) 反转. 对每个整数 $i$ $(1 \le i \le n)$, 如果 $G_n^{r_m,1}$ 的第 $i$ 个分量是 1, 那么 $x_i = 2^{m-1} - 1 - x_i$.

(5) 交换. 如果 $G_n^{r_m,0}$ 存在两个非零分量, 其位置为 $i$ 和 $j$, 则交换 $x_i$ 和 $x_j$.

(6) $m = m - 1$, 转至 (2).

**编码算法 2.2.**
(1) 设定 $(r_m \cdots r_1)_{2^n}$ 为 0. 如果 $m$ 与 $k$ 具有不同的奇偶性, 则交换 $x_1$ 与 $x_n$. $m = k$.
(2) 如果 $m = 0$, 程序终止. 否则 $r_m = f_n(x_n^m, x_{n-1}^m, \cdots, x_1^m)$.
(3) 对每个整数 $i$ $(1 \le i \le n)$, $x_i = (x_i^{m-1} \cdots x_i^1)_2 = (x_i^m \cdots x_i^1)_2$ & $(011 \cdots 1)_2$.
(4) 反转. 对每个整数 $i$ $(1 \le i \le n)$, 如果 $G_n^{r_m,1}$ 的第 $i$ 个分量是 1, 那么 $x_i = \mathrm{Re}_{m-1}(x_i)$.
(5) 交换. 如果 $G_n^{r_m,0}$ 存在两个非零分量, 其位置为 $i$ 和 $j$, 则交换 $x_i$ 和 $x_j$.
(6) $m = m - 1$, 转至 (2).

*编码算法 1.2* 和 *编码算法 2.2* 的时间计算复杂度为 $O(kn)$, 对某个给定的坐标, 其时间复杂度是与曲线的阶数 $m$ 无关的. 对阶数 $m$ 较高的应用, 这两个算法更有优势. 它们的空间复杂度与 *编码算法 1.1* 及 *编码算法 2.1* 是一样的, 都是 $O(2^n)$.

### §3.3.3 解码算法

解码过程是给定整数 $z$ $(\in I_n^m)$ 及希尔伯特空间填充曲线的阶 $m$, 求出整数对应的空间坐标 $(x_n, \cdots, x_2, x_1)$ $(\in Q_n^m)$ 的过程. 解码算法所需要的理论基础均已在上面给出, 此处仅给出每个编码算法对应的解码算法, 不再对解码算法做详细的讨论.

记 $m$ 为希尔伯特空间填充曲线的阶, $(r_m r_{m-1} \cdots r_1)_{2^n}$ 为整数 $z$ 的 $2^n$- 进制数表示. 引入 $k$, 它表示 $(r_m r_{m-1} \cdots r_1)_{2^n}$ 中第一个非零元素的位置, 即 $r_k > 0$ 且 $r_i = 0$ $(i > k)$. 同时假定当 $z$ 为 0 时, $k$ 为 1. 假定 $\log_{2^n}(0)$ 为 0, $k$ 的定义等价于 $k = \mathrm{floor}(\log_{2^n}(z)) + 1$. 计算结果保存在坐标 $(x_n, \cdots, x_2, x_1)$ $(\in Q_n^m)$ 中.

在解码算法中, 交换指令先执行, 然后是反转指令. 下面分别给出上述四个编码算法对应的解码算法.

**解码算法 1.1.**
(1) 初始化 $(x_n, \cdots, x_1)$, $(x_n, \cdots, x_2, x_1) = b_n(r_1)$. $v = 2$.
(2) 如果 $v > m$, 程序终止. 否则 $(s_n, \cdots, s_2, s_1) = b_n(r_v)$.
(3) 交换. 如果 $G_n^{r_v,0}$ 在位置 $i$ 及 $j$ 有非零值, 则交换 $x_i$ 与 $x_j$.
(4) 反转. 对每个整数 $i$ $(1 \le i \le n)$, 如果 $G_n^{r_v,1}$ 的第 $i$ 分量为 1, 则 $x_i = 2^{v-1} - 1 - x_i$.

(5) 对每个整数 $i$ $(1 \le i \le n)$, 如果 $s_i$ 等于 1, 则 $x_i = x_i + 2^{v-1}$.
(6) $v = v + 1$, 转至 (2).

**解码算法 2.1.**
(1) 初始化 $(x_n, \cdots, x_1)$, $(x_n, \cdots, x_2, x_1) = b_n(r_1)$. $v = 2$.
(2) 如果 $v > m$, 程序终止. 否则 $(s_n, \cdots, s_2, s_1) = b_n(r_v)$.
(3) 交换. 如果 $G_n^{r_v,0}$ 在位置 $i$ 及 $j$ 有非零值, 则交换 $x_i$ 与 $x_j$.
(4) 反转. 对每个整数 $i$ $(1 \le i \le n)$, 如果 $G_n^{r_v,1}$ 的第 $i$ 分量为 1, 则 $x_i = \mathrm{Re}_{v-1}(x_i)$.
(5) 对每个整数 $i$ $(1 \le i \le n)$, 如果 $s_i$ 等于 1, 则 $x_i = (x_i) \wedge (1 \ll (v-1))$.
(6) $v = v + 1$, 转至 (2).

**解码算法 1.2.**
(1) 初始化 $(x_n, \cdots, x_1)$, $(x_n, \cdots, x_2, x_1) = b_n(r_1)$. $v = 2$.
(2) 如果 $v > k$, 程序终止. 否则 $(s_n, \cdots, s_2, s_1) = b_n(r_v)$.
(3) 交换. 如果 $G_n^{r_v,0}$ 在位置 $i$ 及 $j$ 有非零值, 则交换 $x_i$ 与 $x_j$.
(4) 反转. 对每个整数 $i$ $(1 \le i \le n)$, 如果 $G_n^{r_v,1}$ 的第 $i$ 分量为 1, 则 $x_i = 2^{v-1} - 1 - x_i$.
(5) 对每个整数 $i$ $(1 \le i \le n)$, 如果 $s_i$ 等于 1, 则 $x_i = x_i + 2^{v-1}$.
(6) $v = v + 1$, 转至 (2).
(7) 如果 $m$ 与 $k$ 具有不同的奇偶性, 则交换 $x_n$ 与 $x_1$. 最终的坐标为 $(x_n, \cdots, x_1)$.

**解码算法 2.2.**
(1) 初始化 $(x_n, \cdots, x_1)$, $(x_n, \cdots, x_2, x_1) = b_n(r_1)$. $v = 2$.
(2) 如果 $v > k$, 程序终止. 否则 $(s_n, \cdots, s_2, s_1) = b_n(r_v)$.
(3) 交换. 如果 $G_n^{r_v,0}$ 在位置 $i$ 及 $j$ 有非零值, 则交换 $x_i$ 与 $x_j$.
(4) 反转. 对每个整数 $i$ $(1 \le i \le n)$, 如果 $G_n^{r_v,1}$ 的第 $i$ 分量为 1, 则 $x_i = \mathrm{Re}_{v-1}(x_i)$.
(5) 对每个整数 $i$ $(1 \le i \le n)$, 如果 $s_i$ 等于 1, 则 $x_i = (x_i) \wedge (1 \ll (v-1))$.
(6) $v = v + 1$, 转至 (2).
(7) 如果 $m$ 与 $k$ 具有不同的奇偶性, 则交换 $x_n$ 与 $x_1$. 最终的坐标为 $(x_n, \cdots, x_1)$.

这四个算法的时间计算复杂度与对应的编码算法的复杂度是一样的. *解码算法 1.1* 和 *解码算法 2.1* 的时间计算复杂度为 $O(nm)$, 其余两个的时间计算复杂度为 $O(nk)$. 它们具有相同的空间复杂度, 均为 $O(2^n)$.

## §3.4 希尔伯特曲线的新变体

由前面的分析可知, $H_n^m$ 可以由 $H_n^{m-1}$ 得到, $H_n^m$ 的第 0 象限包含的并不是 $H_n^{m-1}$ 本身, 而是经过交换指令变换得到的曲线. 用二维曲线做例子, 如图 3.8 所示, $H_2^2$ 的第 0 象限中的曲线并不是 $H_2^1$, $H_2^3$ 的第 0 象限的曲线并不是 $H_2^2$. 这就说明, 同一个坐标在不同阶的曲线中的序可能是不同的. 当曲线的阶数改变时, 需要重新计算希尔伯特序.

$H_n^m$ 第 0 象限的曲线由 $H_n^{m-1}$ 经过一次交换指令得到, 而交换变换的逆变换是自身，即同一个交换指令执行两次保持曲线不变. 从图 3.8 可以看出, $H_1^1$ 恰巧是 $H_2^3$ 的起始曲线. 这说明只要对奇数或者偶数阶的曲线做一次 $x_n$ 与 $x_1$ 间的交换变换, 就能保证变体曲线 $\widetilde{H}_n^m$ 第 0 象限的曲线是 $\widetilde{H}_n^{m-1}$ 自身. 本文选择对偶数阶曲线进行一次 $x_n$ 与 $x_1$ 之间交换变换, 得到了希尔伯特空间填充曲线的一个变体 $\widetilde{H}_n^m$. 在曲线 $\widetilde{H}_n^m$ 中, 同一坐标在不同阶曲线中对应的序都是相同的, 同一个序对应的坐标值也是一样的. 同样取二维空间做例子, 如图 3.9 示, 可以看到此时 $\widetilde{H}_2^3$ 第 0 象限的曲线是 $\widetilde{H}_2^2$ , $\widetilde{H}_2^2$ 第 0 象限的曲线是 $\widetilde{H}_2^1$.

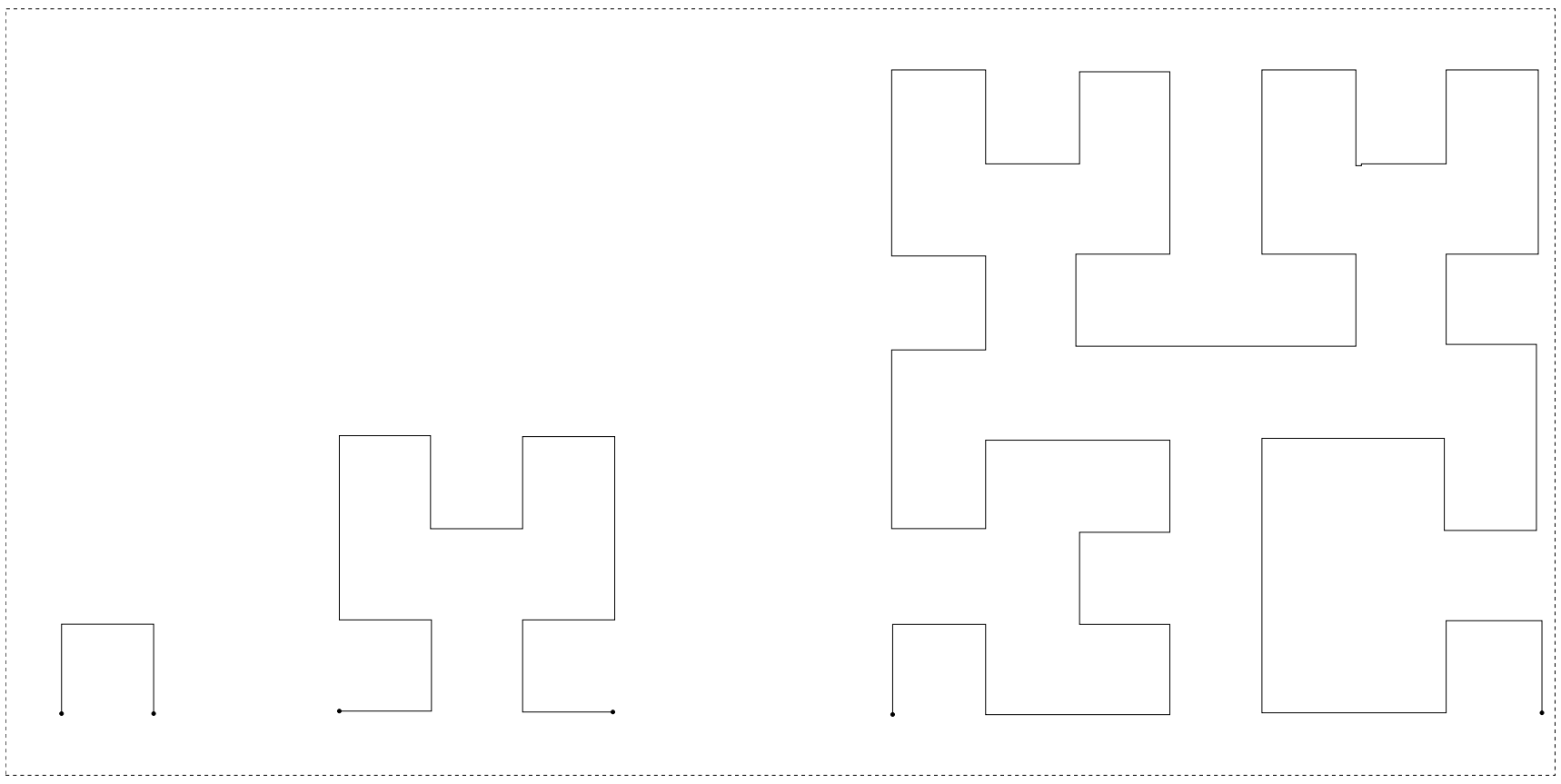

图 3.8 希尔伯特曲线 $(H_2^1, H_2^2, H_2^3)$

下面给出该变体曲线的编码及解码算法. 由于仅偶数阶希尔伯特曲线被实施了一次交换变换, 奇数阶曲线没有改变, 可以这么做: 奇数阶的曲线没有被改变, 编码解码算法和原来的算法相同; 对偶数阶的曲线, 可以先实施一次交换变换回到原来的希尔伯特曲线, 然后调用原来的算法. 这样的好处是不用重写各种算法, 不用重新计算不同象限的变换指令.

由于变体曲线的算法与传统的算法非常相似, 本文仅给出一个编码算法和解

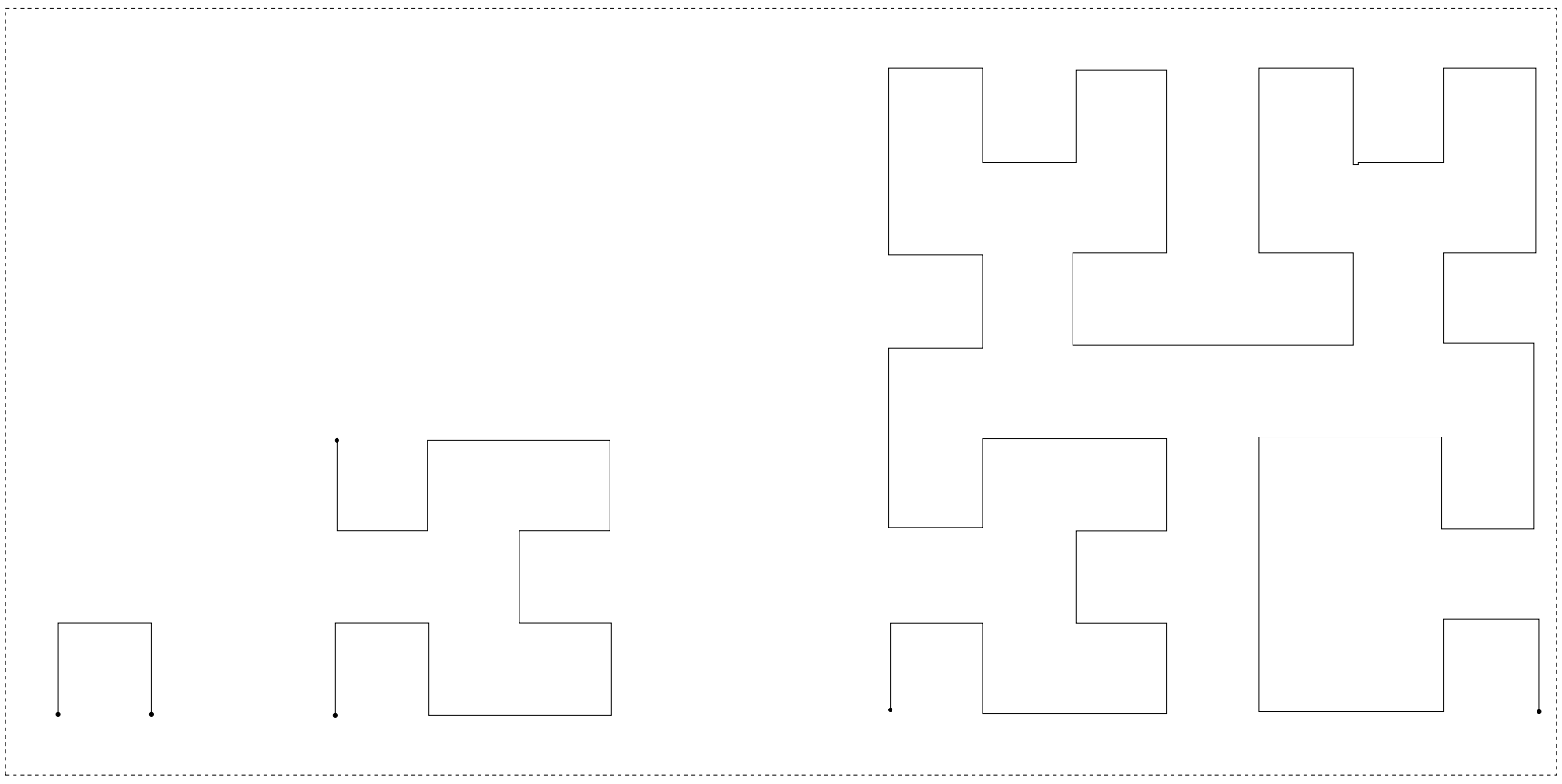

图 3.9 希尔伯特曲线变体 ($\widetilde{H}_2^1$, $\widetilde{H}_2^2$, $\widetilde{H}_2^3$)

码算法作为代表, 其他基于位运算或者复杂度更低的算法, 都可以类似得到. 算法描述如下:

**编码算法 3.1.**
(1) 如果 $m$ 偶数, 交换 $x_n$ 与 $x_1$.
(2) 如果 $m=0$, 程序终止. 否则, $r_m = f_n(x_n^m, x_{n-1}^m, \cdots, x_1^m)$.
(3) 对每个整数 $i$ $(1 \le i \le n)$, 如果 $x_i^m = 1$, 那么 $x_i = x_i - 2^{m-1}$.
(4) 反转. 对每个整数 $i$ $(1 \le i \le n)$, 如果 $G_n^{r_m,1}$ 的第 $i$ 个分量是 1, 那么 $x_i = 2^{m-1} - 1 - x_i$.
(5) 交换. 如果 $G_n^{r_m,0}$ 存在两个非零分量, 其位置为 $i$ 和 $j$, 则交换 $x_i$ 和 $x_j$.
(6) $m = m-1$, 转至 (2).

**解码算法 3.1.**
(1) 初始化 $(x_n, \cdots, x_1)$, $(x_n, \cdots, x_2, x_1) = b_n(r_1)$. $v = 2$.
(2) 如果 $v > m$, 程序终止. 否则 $(s_n, \cdots, s_2, s_1) = b_n(r_v)$.
(3) 交换. 如果 $G_n^{r_v,0}$ 在位置 $i$ 及 $j$ 有非零值, 则交换 $x_i$ 与 $x_j$.
(4) 反转. 对每个整数 $i$ $(1 \le i \le n)$, 如果 $G_n^{r_v,1}$ 的第 $i$ 分量为 1, 则 $x_i = 2^{v-1} - 1 - x_i$.
(5) 对每个整数 $i$ $(1 \le i \le n)$, 如果 $s_i$ 等于 1, 则 $x_i = x_i + 2^{v-1}$.
(6) $v = v+1$, 转至 (2).
(7) 如果 $m$ 偶数, 交换 $x_n$ 与 $x_1$.

这两个算法的时间计算复杂度均为 $O(nm)$, 它们的空间复杂度为 $O(2^n)$. 这两个算法需要的空间变换信息和前一节的是一样的. 复杂度更低的算法亦可以类似得到, 这里不再给出.

## §3.5 算法的具体化

前几节对希尔伯特空间填充曲线做了理论分析并给出了算法模板, 适用于任意维数的希尔伯特空间填充曲线. 下面对算法做具体化, 同时提供一些程序优化建议. 我们仅选取两个算法模板做例子, 出于简单的考虑, 将选择二维希尔伯特空间填充曲线, 高维空间是完全类似的. 选取*编码算法 1.1* 和 *解码算法 1.1*.

现代计算机内存容量较大, 一个常见的编程思想就是通过增加内存使用而加快程序运行速度. 上述所有算法都需要计算 $b_n$, $f_n$, $G_n^{i,0}$ 及 $G_n^{i,1}$ $(0 \le i < 2^n)$, 对特定的维数 $n$, 这些信息是保持不变的, 可以把这些信息事先计算好存储在源文件中, 在计算时直接查询, 这种技术会大大提高程序的运行速度.

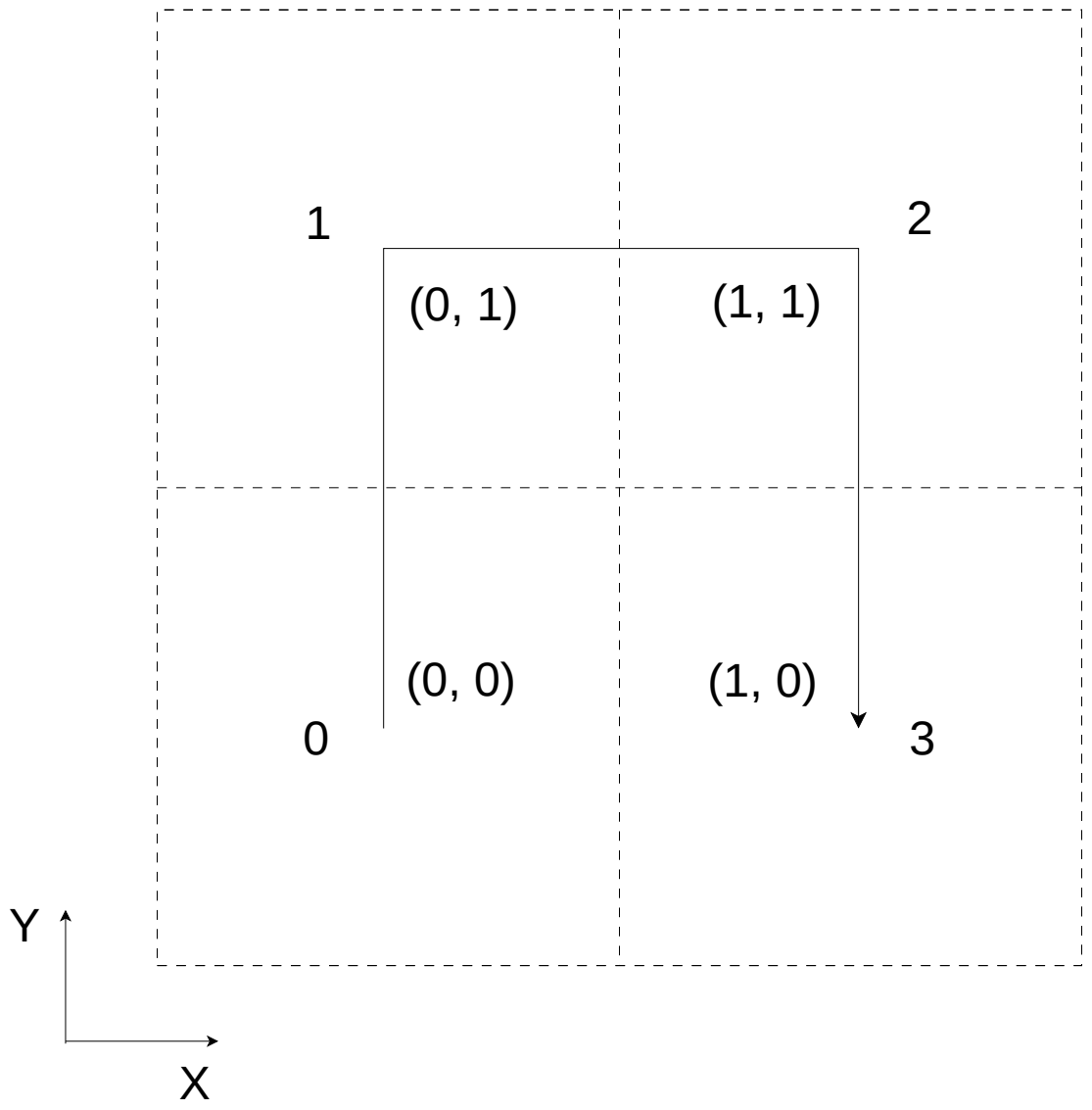

图 3.10 希尔伯特单元 $(H_2^1)$

在上节的算法描述中使用 $(x_2, x_1)$ 表示坐标, 本文用更符合习惯的 $(x, y)$ $(x = x_2, y = x_1)$ 代替, 先给出二维希尔伯特单元 $H_2^1$, 如图 3.10 所示, 并给出了曲线方向及每个点的整数坐标. 映射 $b_2$ 及 $f_2$ 根据定义可以直接求得, 它们存储在表 3.1 中, 同时标注在图 3.10 中. 变换指令和反转指令存储在表 3.2 中. 直接观察可以

看出, $f_2$, $b_2$, $G_2^{i,0}$ 及 $G_2^{i,1}(0 \le i < 4)$ 按照数组的方式存储有较好的查询效率.

**表 3.1 左边表格为映射 $f_2$, 右边表格为映射 $b_2$**

| | |
|---|---|
| $f_2(0,0)$ | 0 |
| $f_2(0,1)$ | 1 |
| $f_2(1,1)$ | 2 |
| $f_2(1,0)$ | 3 |
| $b_2(0)$ | (0, 0) |
| $b_2(1)$ | (0, 1) |
| $b_2(2)$ | (1, 1) |
| $b_2(3)$ | (1, 0) |

**表 3.2 左边的表格为交换指令, 右边的表格为反转指令**

| | |
|---|---|
| $G_2^{0,0}$ | (1, 1) |
| $G_2^{1,0}$ | (0, 0) |
| $G_2^{2,0}$ | (0, 0) |
| $G_2^{3,0}$ | (1, 1) |
| $G_2^{0,1}$ | (0, 0) |
| $G_2^{1,1}$ | (0, 0) |
| $G_2^{2,1}$ | (0, 0) |
| $G_2^{3,1}$ | (1, 1) |

观察 *编码算法 1.1*, 坐标 $(x_n, \cdots, x_1)$ 在步骤 (2) $\sim$ (4) 中被更新, 其中用了三个步骤描述使算法清晰, 而在实际应用中, 可以把三个步骤组合到一块, 用一个表达式来表示, 将成倍加快程序的速度. 记当前曲线的阶为 $v$, $x_{new}$, $y_{new}$ 分别为更新后的坐标. 取第三象限做例子, 在这个象限中, 先执行反转指令, 然后是交换指令. 步骤 (2) 是一个消去 $x$, $y$ 的高位有效位的过程, 有 $(x^v, y^v) = (1,0)$, 所以 $(x_{new}, y_{new}) = (x - 2^{v-1}, y)$. 步骤 (3) 是反转变换, 有 $G_2^{3,1} = (1,1)$, 需要对两个坐标做反转, 有 $(x_{new}, y_{new}) = (2^v - 1 - x, 2^{v-1} - 1 - y)$. 步骤 (4) 是一个交换的过程, 由于 $G_2^{3,0} = (1,1)$, 最终有 $(x_{new}, y_{new}) = (2^{v-1} - 1 - y, 2^v - 1 - x)$. 类似地可以推导出另外三个象限的更新规则. 我们把这四个象限的更新规则总结到表 3.3 中. 对解码算法可做类似的优化, 我们把解码算法的坐标更新规则总结在表 3.4 中. 在实际计算中, 将 *编码算法 1.1*的步骤 (2) $\sim$ (4) 用一个坐标更新步骤代替, 将 *解码算法 1.1*的步骤 (3) $\sim$ (5) 同样用一个坐标更新步骤代替. 至此便完成了二维空间填充曲线编码解码算法的具体化。

## §3.6 本章小结

本文简要地介绍了空间填充曲线的知识, 并在这些知识的基础上给出高维曲线的编码解码算法. 对编码问题, 给出了四种算法, 两种基于算术运算, 两种基于位运算, 它们的时间计算复杂度分别为 $O(nm)$ 和 $O(kn)$, 其中复杂度 $O(kn)$ 与曲

**表 3.3 编码过程的坐标更新规则**

| Quadrant | $x_{new}$ | $y_{new}$ |
|---|---|---|
| 0 | $y$ | $x$ |
| 1 | $x$ | $y-2^{v-1}$ |
| 2 | $x-2^{v-1}$ | $y-2^{v-1}$ |
| 3 | $2^{v-1}-1-y$ | $2^{v}-1-x$ |

**表 3.4 解码过程的坐标更新规则**

| Quadrant | $x_{new}$ | $y_{new}$ |
|---|---|---|
| 0 | $y$ | $x$ |
| 1 | $x$ | $y+2^{v-1}$ |
| 2 | $x+2^{v-1}$ | $y+2^{v-1}$ |
| 3 | $2^{v}-1-y$ | $2^{v-1}-1-x$ |

线的阶数 $m$ 无关, 在高阶编码解码应用中有优势. 解码问题是编码问题的逆问题, 同样给出了四个算法.

此外, 本文还给出了希尔伯特曲线的一个变体, 该变体相对于传统的希尔伯特曲线有一个优势, 即任意一个坐标或者序, 它对应的序或者坐标, 不随曲线的阶数改变而变化. 该性质有助于在实际应用中减少计算时间. 相应的, 给出了这个变体的编码解码算法.

最后通过二维的例子展示了将本章给出的算法具体化的过程, 并给出了程序优化的建议.

# 第四章　PHG 动态负载平衡模块的设计与实现

偏微分方程的并行求解, 关键问题之一是网格划分, 它不仅要求每个进程拥有相等的计算负载, 同时要求有良好的划分质量, 以减少进程间通信. 在自适应有限元计算过程中, 网格/基函数不断调整, 会导致负载不平衡, 必须动态地调整网格分布, 从而实现动态负载平衡. 动态负载平衡算法需要速度快、网格划分质量高、具有增量性质并且是并行运行的. 本章介绍 PHG 中动态负载平衡的算法及实现.

## §4.1　PHG 的层次网格结构与动态负载平衡

PHG 的计算网格为协调四面体网格. PHG 采用的是最新顶点二分加密算法, 存储了四面体单元所有的前辈, 以二叉树的形式表示. PHG 将网格分布到不同进程中, 每个进程拥有一个子网格, 子网格的单元 (叶子) 是互不重叠的, 同时进程中存储了单元所有的前辈, 形成一个分布式层次网格结构. 本节介绍 PHG 的层次网格结构、网格划分及分布式层次网格.

### §4.1.1　二分自适应网格加密、放粗与 PHG 的层次网格结构

PHG 的自适应网格加密方法为最新顶点二分加密算法 [1], 也称为“边加密”算法. 对一个单元加密时, 将它一条边 (称为加密边) 的中点与与之相对的两个顶点相连, 将单元一分为二, 如图 4.1 所示. 一个单元被加密后, 加密产生的单元称为它的子单元, 该单元称为加密所产生的子单元的父单元.

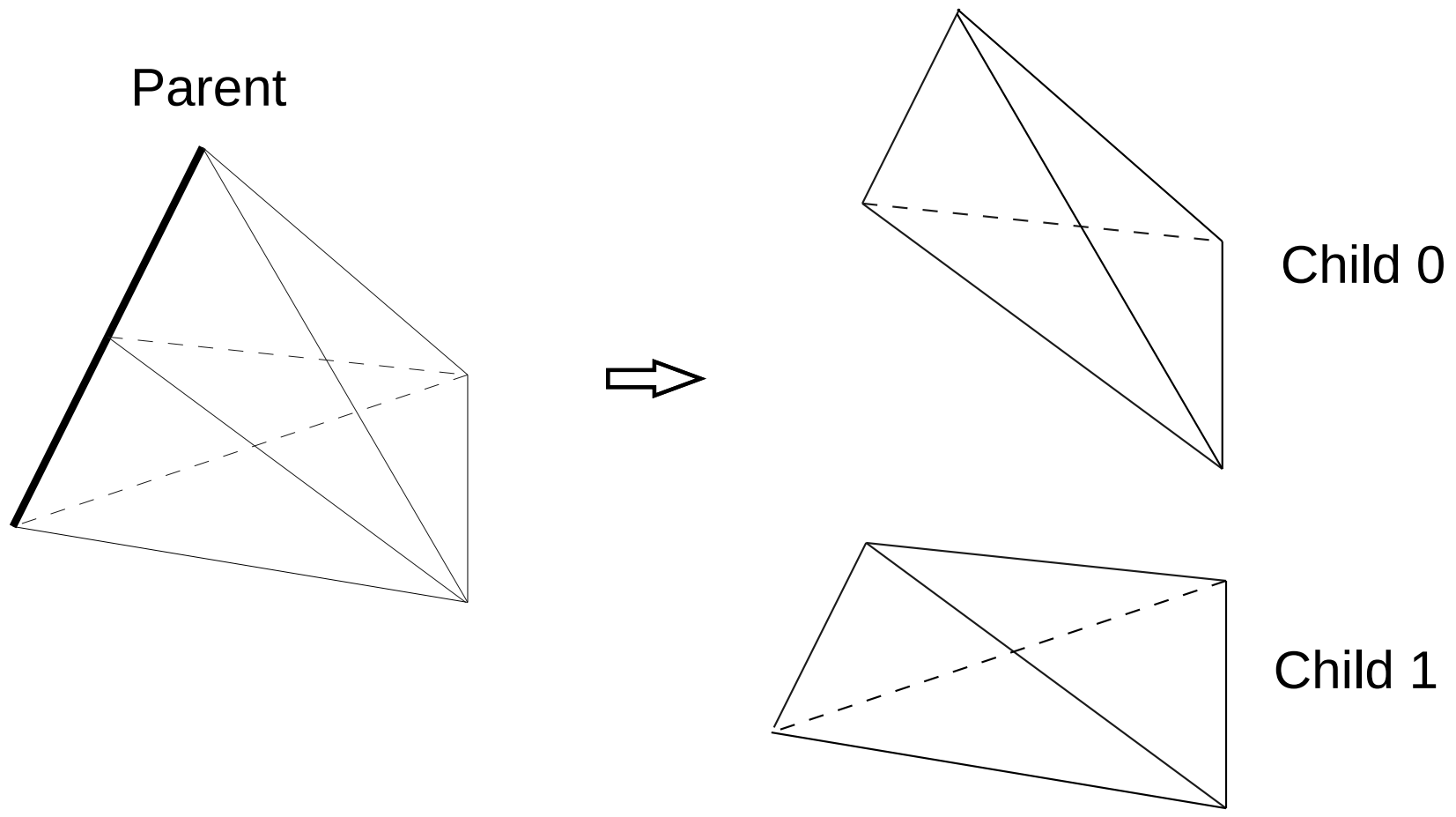

图 4.1　四面体二分加密

PHG 的四面体网格经过多次二分加密后, 形成一个以单元为结点的二叉树结构, 一个单元经过加密后产生两条分枝, 分别对应它的两个子单元. 所有叶子单元构成当前网格层, PHG 保证当前网格层总是协调的. 下面用二维协调三角形网格做例子 (三维协调四面体网格的二叉树结构与二维协调三角形网格相同), 如图 4.2 所示. 在这个例子中, 初始网格包含一个单元 $e_0$, 将它加密一次得到两个新单元 $e_1$ 和 $e_2$, 再将 $e_1$ 和 $e_2$ 各加密一次得到四个新单元 $e_3$, $e_4$, $e_5$, $e_6$, 最后将 $e_3$, $e_4$ 和 $e_5$ 各加密一次得到新单元 $e_7$, $e_8$, $e_9$, $e_{10}$, $e_{11}$ 和 $e_{12}$, 最终网格由 $e_6$, $e_7$, $e_8$, $e_9$, $e_{10}$, $e_{11}$ 和 $e_{12}$ 构成.

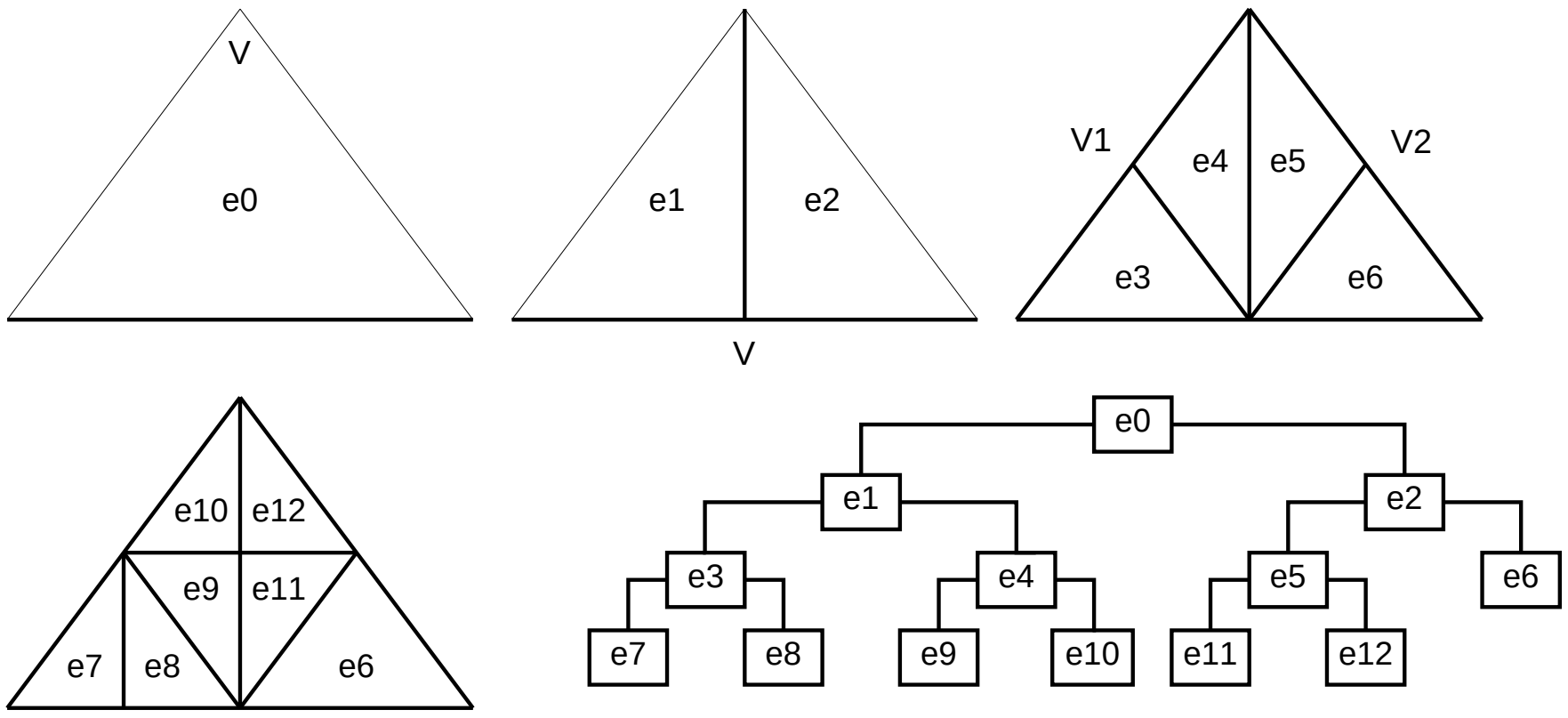

图 4.2 层次网格结构的二叉树表示

层次网格结构简单, 数据组织直观, 直接反映了 PHG 的内存管理模式. PHG 设计了单元数据结构 `SIMPLEX`, 提供了子单元成员分别指向两个子单元, 通过这个成员实现了二叉树. PHG 实现了深度优先搜索算法遍历整个加密树, 同时提供了函数接口访问所有网格单元; PHG 提供了邻居成员指向各个面对应的邻居单元, 通过指针实现邻居单元的快速访问; PHG 提供了 `parent` 成员指向父单元, 对根单元而言 `parent` 为空指针, 在该成员的帮助下, 可以实现自下而上的回溯访问.

### §4.1.2 网格划分

PHG 采用的划分方法是以单元为单位将网格划分为子网格, 每个进程负责一个子网格, 每个进程仅存储它所拥有的子网格上的数据, 不在本地的数据通过进程间通信获得. 对于传统并行有限元, 由于计算过程中网格保持不变, 仅需做一次网格划分, 划分质量要求很高, 但对划分时间的限制较少. 而对并行自适应有限元或者涉及时间的问题, 网格/基函数根据数值解调整, 会导致负载不平衡, 程序需要根据计算负载对网格分布做动态调整. 在这类应用中, 网格划分方法需要

满足以下条件: 1) 能在分布式数据上并行执行; 2) 执行速度快, 因为它们可能会被频繁调用; 3) 能产生高质量的划分; 4) 最好是增量的, 即网格的微小变化仅导致划分的微小变化; 5) 内存使用是适量的.

并行计算中的数据划分方法可以分为两类: 基于几何性质的划分方法和基于图的划分方法 [125, 126, 127, 128, 129]. 基于几何性质的划分方法是利用对象 (粒子、单元等) 的几何信息划分数据, 如坐标, 常用的方法有递归坐标二分方法 RCB (Recursive Coordinate Bisection) [60, 140, 141] 、递归惯量二分方法 RIB (Recursive Inertial Bisection) [61, 142]、空间填充曲线划分方法 SFC (Space-Filling Curve) [62, 130, 131, 145, 146] 等. 基于图的划分方法是利用对象的拓扑信息, 如网格中单元的邻居关系, 常用方法有递归图二分方法 (Recursive Graph Bisection) [60]、贪婪算法 [63]、递归谱二分方法 RSB (Recursive Spectral Bisection) [60, 142, 143, 144]、K-L 算法 (Keinighnan-Lin Algorithm) [64]、多水平方法 (multilevel methods) [16, 135, 136, 137, 132, 133, 134] 和扩散方法 (diffusive methods) [18, 138, 139] 等. 图方法和几何方法各有优缺点. 图方法运行时间长, 由于显式地控制通信量, 划分质量高. 几何方法在空间局部性很重要或者拓扑结构不存在的情况下是非常有效的, 几何方法没有显式控制通信量, 它只是根据空间位置信息划分, 可能会导致通信量很大, 但几何划分方法简单, 它们容易实现并且速度快. 同时, 像递归坐标二分方法、空间填充曲线划分方法等, 它们是隐式增量的 [125], 数据迁移的代价比较小. 在并行自适应有限元计算中, 最终目标是极小化总体计算时间, 包括划分时间、数据迁移时间及有限元求解时间. 它们是相关的: 质量高的划分会减少有限元计算时间, 但可能会需要更多的划分时间; 质量差的划分需要时间少, 但可能增加有限元计算时间. 在实际应用中, 需要根据实际情况综合考虑这些因素. 下面给出几种常用划分方法的细节:

- 递归坐标二分方法 (RCB) [60, 140, 141]. 交替地在每个空间方向上用一个平面 (3D) 或者一条直线 (2D) 将对象分成两个子集, 使得两个子集中的对象数目 (或权重) 是近似相等的. 划分的平面或者直线垂直于给定的方向. 对于一个待划分的集合, RCB 选择最长方向进行分割. RCB 方法没有考虑集合的形状及长宽比. Jones 和 Plassmann[141] 对这个算法做了改进, 他们将新的方法叫做 URB (Unbalanced Recursive Bisection). 由于通信量与长宽比有很大的关系, URB 将待分集合分成两个长宽比接近的集合. 假设有 $n$ 个进程, URB 会从子集大小比例为 1 : (n - 1), 2 : (n - 2), 3 : (n - 3) 等划分中选择满足上述条件的划分. Jones 和 Plassmann 在文章中指出该划分会降低通信量. RCB 及 URB 是隐式增量的.

- 递归惯量二分方法 (RIB) [61, 142]. RIB 也是递归地选择一个平面或者直线将对象集合分成两个大小相等的子集, 但它选择的平面或者直线不是垂直于坐标轴的, 而是根据对象的分布, 在空间中选择一个最长方向, 用一个垂直于这个方向的平面或者直线将对象分成两个子集. 该方法对惯量轴敏感, 因而不是增量的划分方法.

- 空间填充曲线划分方法 (SFC) [62, 130, 131, 145, 146]. 该方法由 Patra 和 Oden 首先应用在网格自适应加密问题中 [145, 146]. SFC 方法是根据空间填充曲线将 $n$- 维问题映射成 1- 维问题, 然后对 1- 维问题进行划分. 对于映射, 可以选择希尔伯特序, Morton 序等. 该方法是增量的, 方法简单, 容易实现. SFC 方法划分质量不如图划分质量高, 但运行速度快, 在动态负载平衡中很有竞争力.

- 多水平方法 (multilevel method) [16, 132, 133, 134, 135, 136, 137]. 多水平方法是图划分中最流行的方法. 多水平方法类似 V- 循环多重网格方法, 通过将一些顶点缩成一个顶点, 将图放粗成一个更小的图, 形成一个放粗序列, 然后在最小的图上进行划分, 最后通过加密过程得到划分结果.

- 递归谱二分方法 (RSB) [60, 142, 143, 144]. 谱划分方法先构造图的 `Laplacian` 矩阵, 然后求该矩阵的第二小的特征值对应的特征向量 (Fiedler vector), 每个顶点给定一个对应于这个特征向量的值, 从而将图划分问题变成一维划分问题. 谱方法划分质量高, 但特征值特征向量的求解代价高, 一般用在静态划分中.

网格划分是并行计算的核心算法之一. 常用的图划分及几何划分软件如下:

- PaToh[151]. PaToH (Partitioning Tools for Hypergraph) 是一款串行多水平超图划分软件, 号称是最快的串行超图划分软件. 它可以划分普通的超图, 也可以划分带有多个限制条件的超图.

- Chaco[152]. Chaco 是一个串行多水平图划分软件. 主要功能有: 扩展了谱方法, 允许使用两个或者三个 Laplacian 特征向量将图四等分或者八等分, 并为谱方法提供了高效健壮的特征值特征向量解法器; 提供了通用的 Kernighan-Lin/Fiduccia-Mattheyses 算法, 可以处理带权图; 改进了图和并行体系结构之间的映射算法; 提供了各种后处理功能以改进划分质量.

- SCOTCH[154] 及 PT-SCOTCH[153]. SCOTCH 是一个串行软件, 提供了图划

分、静态映射、稀疏矩阵排序、网格和超图划分; PT-SCOTCH 是 SCOTCH 的并行化版本.

- Parkway[150]. Parkway 是一个并行超图划分软件, 它基于 C++ 和 MPI 实现了多水平超图划分算法.

- METIS[17], hMETIS[155] 及 ParMETIS[14]. METIS (Serial Graph Partitioning and Fill-reducing Matrix Ordering) 是一个串行程序集合, 提供了图划分、有限元网格划分以及稀疏矩阵填充约化 (fill-reducing) 排序功能. METIS 中实现的算法是基于多水平二分递归算法、k- 路多水平算法及多限制划分算法等. ParMETIS (Parallel Graph Partitioning and Fill-reducing Matrix Ordering) 是 METIS 的并行版本, 基于 MPI, 适合并行自适应计算及大规模并行处理. hMETIS (Hypergraph & Circuit Partitioning) 是一个串行超图划分软件, 它基于多水平超图划分算法实现.

- Zoltan [19, 18]. Zoltan 是美国 Sandia 国家实验室开发的一个软件, 它包含一系列并行工具以简化并行程序的编程. 它提供了几何划分、图划分、超图划分功能, 同时提供了数据迁移工具、并行图染色、非结构通信服务及动态内存管理等. 它为程序在不同划分算法之间切换提供了方便.

- JOSTLE [65]. JOSTLE 是为在分布式并行机上划分非结构网格 (有限元网格、有限体积网格等) 设计的, 同时可以对网格进行重划分. JOSTLE 将网格看成非结构图, 使用最新的图划分算法实现, 提供了多水平图划分算法及扩散算法, 运行速度很快, 拥有串行和并行两个版本.

### §4.1.3 PHG 的分布式二叉树

PHG 的网格是分布式存储的, 网格划分基于单元 (二叉树中的叶子) 进行, 当用 $p$ 个进程并行处理时, PHG 将网格划分成 $p$ 个子网格, 即将网格中单元集合分解成 $p$ 个互不相交的子集, 每个子集构成一个子网格. 相应地, 描述网格层次结构的二叉树被划分成 $p$ 棵子树, 每棵子树包含子网格中的叶子单元和它们所有的前辈 (上层单元). 图 4.3 是将图 4.2 中的网格划分为两个子网格 ($p = 2$) 时的子网格和子树示意图, 其中第一个子网格包含叶子单元 $e_7$, $e_8$ 和 $e_9$, 第二个子网格包含叶子单元 $e_6$, $e_{10}$, $e_{11}$ 和 $e_{12}$. 在不同子树中, 叶子单元互不相同, 但非叶子单元则可能有重复. 这种子树结构既方便在网格层次间进行遍历, 又具有并行可扩展性. 缺点是动态负载平衡时需要迁移子树及进行子树的重组, 大大增加了动态负载平衡实现的复杂度.

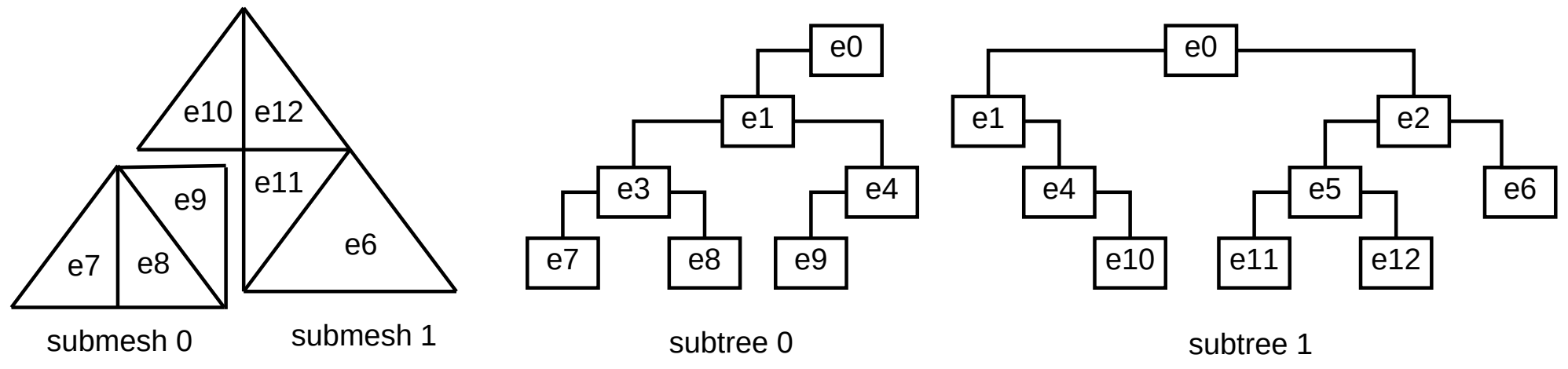

图 4.3 分布式层次网格

### §4.1.4 动态负载平衡

在自适应有限元计算中, 网格/基函数调整后, 如果进程的计算负载不平衡, 需要重新调整网格和数据分布, 使得各个进程保持相等量的计算负载, 保证每个进程都处于繁忙状态, 同时尽可能减少它们之间的通信. PHG 中动态负载平衡与有限元计算的关系如图 4.4 所示. PHG 中, 负载不平衡因子定义如下,

$$\text{LIF} = \frac{\sum_{i=0}^{p-1} W_i}{p \times \max\{W_0, \cdots, W_{p-1}\}}, \tag{4.1}$$

其中 $W_i$ 表示进程 $i$ 的计算负载. PHG 的动态负载平衡主要由如下几个步骤构成:

(1) 检查网格在各进程中的分布是否平衡, 即判断 LIF 是否大于事先设定的阈值. 如果 LIF 大于事先设定的阈值, 则调用相关的函数重新进行网格划分; 否则, 则继续使用当前的网格划分;

(2) 网格划分: 调用划分程序重新划分网格, 使得每个进程拥有相等的计算负载, 同时使得进程间的通信量最少.

(3) 子网格 – 进程映射: 对新的划分, 计算子网格到进程的映射, 使得从老划分到新划分的数据迁移量最少.

(4) 网格单元迁移和二叉树重组: 网格分布发生了改变, 将单元按新划分发送到所属的进程, 重组子网格和二叉树.

(5) 自由度数据迁移: 将单元上的自由度数据迁移到对应的进程上, 重建有限元空间, 维护数据的正确性和完整性.

本节仅给出动态负载平衡的几个主要步骤, 后面内容将详细介绍网格划分算法、子网格 – 进程映射、数据迁移等.

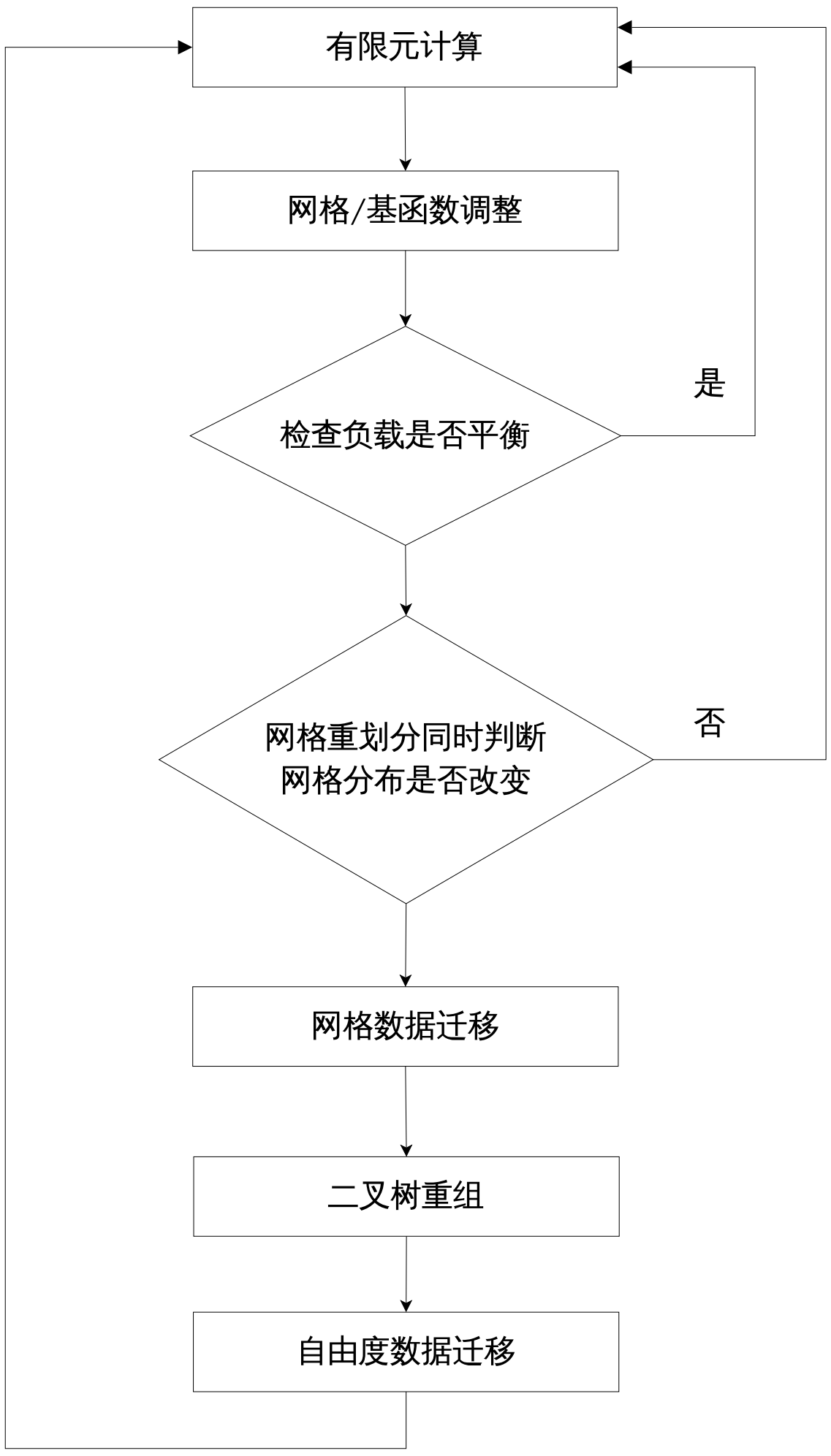

图 4.4 PHG 中的动态负载平衡

## §4.2 网格划分方法

PHG 提供了与软件包 ParMETIS[14, 16] 和 Zoltan[18, 19] 的接口, 前者提供了多水平图划分方法, 后者提供了超图划分方法以及常用的几何划分方法, 如递归坐标二分方法、递归惯量二分方法、希尔伯特空间填充曲线划分方法. 除此之外, 我们在 PHG 中实现两种网格划分方法: 加密树划分方法 [83, 84] 和空间填充曲线划分方法 [62, 125, 126, 127, 128]. 本节下面部分对这两个方法及实现进行介绍.

### §4.2.1 加密树划分方法

PHG 存储自适应过程中生成的加密树, 利用加密树划分方法实施网格划分是一个很自然的想法. 加密树划分方法 (refinement-tree partition method) 由 William Mitchell[84] 提出, 该方法基于加密/放粗过程中产生的二叉树, 它调用深度优先遍历算法访问二叉树, 按照访问顺序对网格单元排序. 二叉树的遍历需要满足条件: 先访问左孩子, 然后是右孩子. 由于要求当前叶子结点与下一个被访问的叶子结点有一个共享面, 加密树划分方法有良好的划分质量. 取二维区域做例子, 如图 4.5 所示, 这棵加密树与图 4.2 中的加密树是相同的, 只是调整了孩子结点的顺序, 对这棵加密树采用常规的深度优先遍历, 按照访问顺序相邻两个叶子单元之间有一个公共边 (三维为面). 如果加密树不能使上述条件成立, 则需要动态地调整孩子结点的访问次序使条件成立.

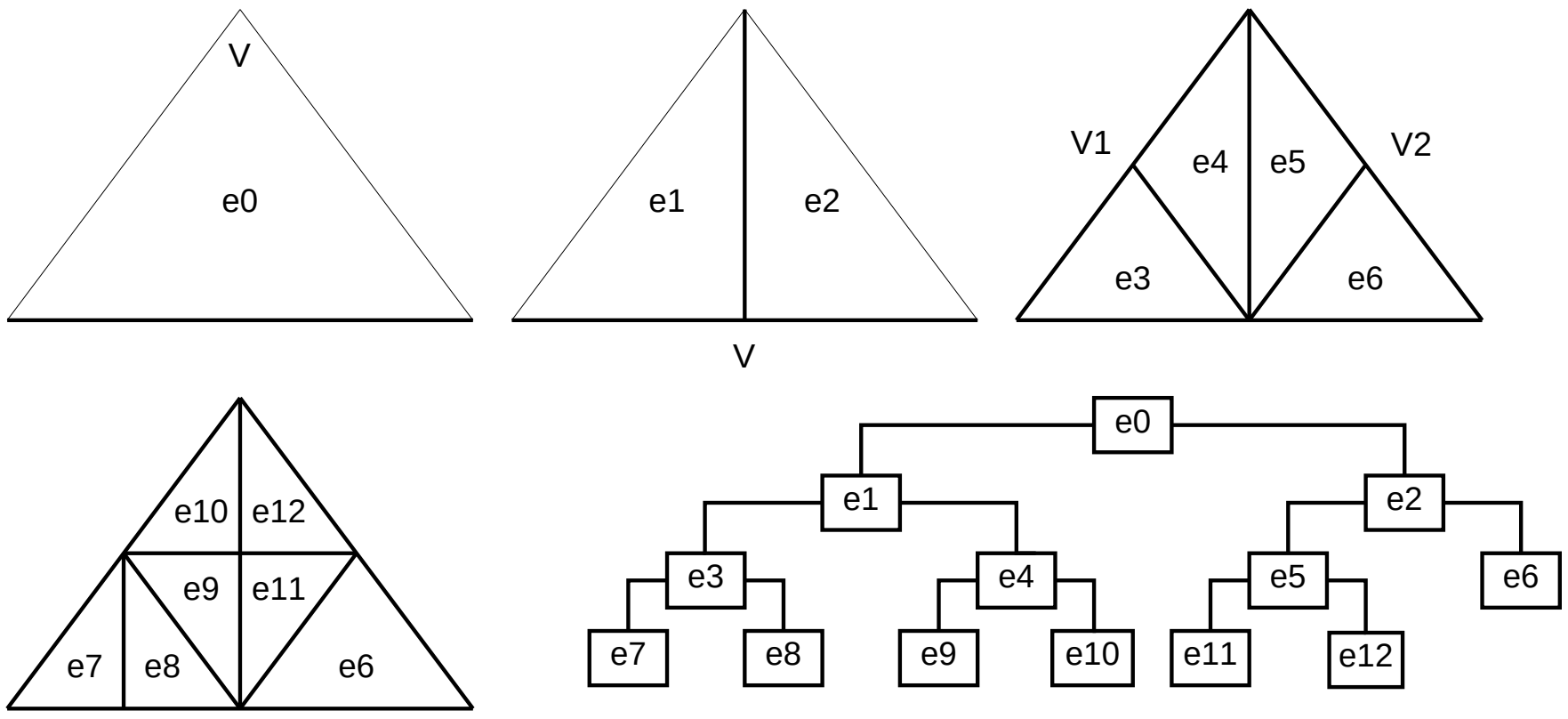

图 4.5 二分加密产生的加密树

在加密树中按下面方法给每一个结点赋予一个权重 $w$: 叶子结点的权重由用户给定, 非叶子结点的权重为以该结点为根的子树的所有叶子结点的权重和. Mitchell 的算法由两步组成, 第一步先计算每一个结点的权重, 第二步将加密树二分, 使得每个集合 (set) 中的叶子数目相等, 递归调用二分算法来完成网格划分. 按照 Mitchell 的分析, 算法的计算复杂度为 $O(N\log p + p\log N)$, 其中 $N$ 为叶子结点数目. 细节如算法 2 所示.

上述算法需要计算每一个结点的权重, 由于父单元在多个进程中同时存在, 在计算父单元的权重时, 通信比较复杂. 我们重新设计了算法, 为每一个叶子结点定义前缀和 (prefix sum). 按照叶子结点的访问顺序, 叶子结点的前缀和定义为: 在到达该叶子结点之前访问的所有叶子结点的权重之和. 下面给出形式化定义,

**算法 2**: 二分加密树划分算法

```
algorithm bisect
    compute subtree weights
    bisect subtree(root)
end algorithm bisect

algorithm bisect subtree(node)
    if node is a leaf then
        assign node to the smaller set
    elseif node has one child then
        bisect subtree(child)
    else node has two children
        select a set for each child
        for each child examine the sum of the subtree weight with
          the accumulated weight of the selected set
        for the smaller of the two sums, assign the subtree rooted
          at that child to the selected set and add the subtree weight
          to the weight of the set
        bisect subtree(other child)
    endif
end algorithm bisect subtree
```

按照叶子结点的访问顺序, 为每一个叶子结点赋予一个位置编号, 从 0 开始, 则第 i 个叶子结点的前缀和为

$$S_i = \sum_{j=0}^{i-1} w_j \tag{4.2}$$

其中 $w_j$ 为叶子结点 $j$ 的权重, 式 (4.2) 可以改写为

$$S_i = \sum_{j=0}^{i-1} w_j = S_q + \sum_{j=q}^{i-1} w_j,\ (q < i). \tag{4.3}$$

假设所有叶子结点的权重之和为 $W$, 进程数为 $p$, 那么所有前缀和属于区间 [W i/p, W (i+1)/p) 的叶子单元分配给子网格 $i$ $(0 \le i < p)$. 通过分析可以看出, 只要计算出叶子结点的前缀和, 就可以通过计算叶子结点所属区间来完成划分.

当网格位于一个进程中时, 仅需遍历一次加密树即可完成划分, 算法复杂度为 $O(N)$. 当网格分布在不同进程中时, 假设进程数为 $p$, 每个进程含有的叶子结点数目分别为 $n_i$ $(0 \le i < p)$, 进程 $i$ 上叶子结点权重之和为 $W_i$ $(0 \le i < p)$. 定义叶子单元在本进程中的局部位置编号, 即叶子结点按照在本进程中访问顺序从 0 开始的编号. 定义 $S_{i,j}$ 为进程 i 上局部编号为 j 的叶子单元的前缀和, 那么由式 (4.3) 可以推导出

$$S_{i,j} = \sum_{k=0}^{i-1} W_k + \sum_{k=0}^{j-1} w_j = S_{i,j-1} + w_{j-1}, \tag{4.4}$$

上式表明, 只要知道了 $W_i$ $(0 \le i < p)$ 就可以在每一个进程中计算出所有叶子结点的前缀和. 在每个进程上, 遍历一次加密树就可以知道本进程所有叶子结点的权重和 $W_i$ $(0 \le i < p)$, 然后通过第二次遍历就可以同时完成计算前缀和及网格划分. 算法如下所示.

**算法 3**: PHG 实现的加密树划分算法

***Step 1.*** 每个进程访问本地的子树, 计算所有本地叶子结点的权重和 $W_i$.

***Step 2.*** 调用 `MPI_Scan` 操作, 为每个进程收集其需要的 $W_i$ $(0 \le i < p)$.

***Step 3.*** 访问本地子树, 根据式 (4.4) 计算每一个叶子结点的前缀和, 并同时完成划分.

本文设计的算法简洁, 仅需遍历两次加密树以及一次 `MPI_Scan` 通信, 算法的计算复杂度为 $O(N)$. 对加密树划分算法, 由于初始网格会包含许多单元, 因而也会有多个加密树, 加密树的访问顺序是按照根结点的顺序, 只需对初始网格 (根结点) 排序, 并在整个自适应过程中维护这个序, 以保证在整个自适应过程中访问子树的顺序是相同的. PHG 提供了三种单元排序方式: 哈密尔顿路排序, 希尔伯特序和 Morton 序. 哈密尔顿路排序是将单元按照第二章设计的哈密尔顿路算法排序; 希尔伯特序和 Morton 序方法是将单元的重心坐标按照这两种序的定义映射到一维区间, 在一维区间上按照数值大小排序.

函数 `phgPartitionRTK` 实现了加密树划分算法, 接口如下:

```
BOOLEAN phgPartitionRTK(GRID *g, int np, DOF *w, FLOAT p),
```

其中 `np` 给出子网格数量, 自由度对象 `w` 给出单元的权重, `p` 为权重的指数, 单元 `e` 实际的权重为 $w_e^p$.

PHG 设计了如下辅助函数用于对初始网格排序:

(1) 函数 `phgHamiltonPath` 实现了哈密尔顿路排序, 接口如下:

```
int phgHamiltonPath(GRID *g),
```

其中返回 0 表示成功, 非 0 表示程序失败;

(2) 函数 `phgHilbertOrder` 实现了希尔伯特序排序, 接口如下:

```
  int phgHilbertOrder(GRID *g),
```

其中返回 0 表示成功, 非 0 表示程序失败;

(3) 函数 `phgMortonOrder` 实现了 Morton 序排序, 接口如下:

```
  int phgMortonOrder(GRID *g),
```

其中返回 0 表示成功, 非 0 表示程序失败;

(4) 函数 `phgGridInitOrder` 负责调用上面三个函数之一完成对初始网格的排序, 接口如下:

```
  int phgGridInitOrder(GRID *g, BOOLEAN dist),
```

其中 `dist` 变量表明网格是否为分布式的, 返回 0 表示成功, 非 0 表示程序失败.

目前上述几个排序程序都是串行的.

### §4.2.2 空间填充曲线划分方法

网格划分中利用空间填充曲线将高维空间映射成一维空间, 完成单元的排序, 从而将高维划分问题转化为一维划分问题. Morton 空间填充曲线和希尔伯特空间填充曲线均具有较好的空间局部性, 划分质量比较好, 是最常用的两种曲线 [62, 130, 131, 145, 146].

空间填充曲线划分方法分为三个步骤. 第一个步骤是计算空间填充曲线, 首先将计算区域映射到立方体 $(0,1)^3$, 然后调用空间填充曲线生成算法将立方体映射到区间 (0, 1); 第二个步骤是划分区间 (0, 1), 使得属于每个区间单元的权重和相等; 第三个步骤是调用子网格 – 进程映射算法, 对子网格重新编号以减少数据迁移.

对计算区域 $\Omega$, 存在一个长方体 (三维, 在二维情况下为长方形) 包围盒 (bounding box) 包含这个区域, 假设包围盒在 $x, y, z$ 三个方向的长度分别为 $\mathrm{len}_x, \mathrm{len}_y, \mathrm{len}_z$,

将区域映射到 $(0,1)^3$ 的通常的做法是

$$x_1 = (x - x_0)\ /\ \mathrm{len}_x,\ \ y_1 = (y - y_0)\ /\ \mathrm{len}_y,\ \ z_1 = (z - z_0)\ /\ \mathrm{len}_z,$$

其中 $(x_0, y_0, z_0)$ 是包围盒坐标最小的顶点, $(x_1, y_1, z_1)$ 为坐标 $(x, y, z)$ 经过变换后在立方体 $(0,1)^3$ 中的坐标. 上述变换改变了区域的长宽比, 使得变换后的长宽比为 $1:1:1$, 破坏了区域的空间局部性. PHG 中取 $\mathrm{len} = \max(\mathrm{len}_x, \mathrm{len}_y, \mathrm{len}_z)$, 采用变换

$$x_1 = (x - x_0)\ /\ \mathrm{len},\ \ y_1 = (y - y_0)\ /\ \mathrm{len},\ \ z_1 = (z - z_0)\ /\ \mathrm{len},$$

该变换保持区域的空间局部性, 改善了划分质量, 这一点将在后面用数值算例加以验证.

对空间填充曲线划分算法, 我们提供了两种空间填充曲线生成程序: Morton 空间填充曲线和希尔伯特空间填充曲线. 前者的算法简单, 但曲线本身有较大的跳跃, 使得空间局部性略差. 希尔伯特空间填充曲线的空间局部性要好很多, 但生成算法复杂.

利用空间填充曲线对单元排序, 然后调用 §4.2.3 中介绍的一维划分算法及 §4.2.4 中介绍的子网格 – 进程映射算法便可完成网格划分. 我们在 PHG 中实现了下述空间填充曲线生成及网格划分函数:

(1) 函数 `phgSFCInvHilbert2D` 及 `phgSFCInvHilbert3D` 实现了二维和三维希尔伯特空间填充曲线向一维区间 $[0,1]$ 的映射, 接口如下:

```
BOOLEAN phgSFCInvHilbert2D(SFC *hsfc, INT n),
BOOLEAN phgSFCInvHilbert3D(SFC *hsfc, INT n),
```

其中数组 `hsfc` 存储需要转换的坐标 ($\in [0,1]^3$) 和映射, `n` 为数组的长度, 数据结构 `SFC` 定义如下:

```
typedef struct {
    SFC_FLOAT co[3];
    SFC_FLOAT sfc;
    INT index.
} SFC;
```

其中成员 `co[3]` 给出空间点的坐标 ($\in [0,1]^3$), `sfc` 返回映射结果, 辅助成员 `index` 存储着单元的本地索引.

(2) 函数 phgSFCInvMorton2D 及 phgSFCInvMorton2D 实现了二维和三维 Morton 空间填充曲线向一维区间 [0, 1] 的映射, 接口如下:

```
BOOLEAN phgSFCInvMorton2D(SFC *msfc, INT n);
BOOLEAN phgSFCInvMorton3D(SFC *msfc, INT n);
```

其中参数 msfc 和 n 的意义与希尔伯特空间填充曲线相同. 这两种曲线的阶数都是 25, 在源文件中直接修改 MAXLEVEL 的数值可以获得阶数 20 到 30 之间的空间填充曲线.

(3) 函数 phgPartitionSFC 实现了空间填充曲线划分方法, 接口如下:

```
BOOLEAN phgPartitionSFC(GRID *g, int np, DOF *w, FLOAT p);
```

其中 g 为网格信息, 划分编号直接存储在单元 e 的成员 mark 中, np 为子网格数量, 自由度对象 w 为权重, p 为权重的指数, e 的实际权重取为 $w_e^p$.

### §4.2.3 一维划分算法

上节的空间填充曲线划分方法及其他一些网格划分方法最终均将问题转化为一维划分问题, 因此实现一个高效健壮的一维划分程序是必要的.

假设当前进程数为 $p$ 以及给定区间 [a, b), 这里的一维划分问题为: 如何将 [a, b) 划分成 $p$ 个子区间 $[a, a_1)$, $[a_1, a_2)$, $\cdots$, $[a_{p-2}, a_{p-1})$, $[a_{p-1}, b)$, 使得每个区间上包含的对象的权重和相等.

从问题的描述可以看出, 一维划分需要计算 $p-1$ 个分割点 $a_1$, $a_2$, $\cdots$, $a_{p-2}$, $a_{p-1}$. 我们在 PHG 中实现了一个一维划分算法求解该问题, 其基本算法来源于 Zoltan[18], 它是二分搜索算法的推广. 在二分搜索算法中, 只需求解一个点 $a_1$, 将区间分成两个子区间. 如果将区间 $k$ ($k \geq 2$) 分而不是二分, 那么 $a_1$ 将位于某个子区间中, 接着将这个子区间 $k$ 分, 重新得到 [a, b) 的一个划分, $a_1$ 将位于上述 $k$ 个更小的子区间中的一个. 重复这个过程, 可以在给定的精度下搜索到 $a_1$. 将这个算法推广, 将区间 [a, b) 划分成 $N$ ($N \gg p$) 个子区间, $a_i$ ($0 < i < p$) 将位于某个子区间, 然后将子区间细分, 最终在给定的精度下可以搜索到 $a_1$, $a_2$, $\cdots$, $a_{p-2}$, $a_{p-1}$.

具体实现时, 我们取子区间数为 $N = (p-1) * k + 1$, 程序为每一个划分点 $a_i$ ($0 < i < p$) 维护一个包围盒 (bounding box). 在每一次划分之前先更新这些包围盒, 缩小 $a_i$ ($0 < i < p$) 的取值范围, 重新划分区间 [0, 1) 的时候, 直接划分 $a_i$ 所在的包围盒, 而不是 $a_i$ ($0 < i < p$) 所在的某个子区间, 这样会加速求解 $a_i$ ($0 < i < p$) 的过程. 对该一维划分算法, 我们实现了两个不同的版本: 第一个版

本 phgPartition1DV1 的子区间数目始终为 $N$; 第二个版本 phgPartition1DV2 是一个自适应版本, 其子区间数目不是固定的, 当某个 $a_i$ $(0 < i < p)$ 达到精度要求之后, 其所在的包围盒将不再细分, 子区间数目是一个递减的过程, 通过减少不必要的划分, 加快划分速度. 两个函数的接口如下:

```
FLOAT phgPartition1DV1(DOTS *x, INT lenx, int np, MPI_Comm comm),
FLOAT phgPartition1DV2(DOTS *x, INT lenx, int np, MPI_Comm comm).
```

其中 x 是对象数组, 包含当前进程所有对象的权重、坐标 $(\in [0, 1)$, lenx 是数组的长度, np 是子网格数, comm 是当前的通信器. DOTS 的定义如下:

```
typedef struct {
    FLOAT key;
    FLOAT w;
    int pn;
} DOTS;
```

其中 key $(\in (0, 1))$ 是一维坐标, w $(\geq 0)$ 是对象的权重, 通常情况下为正数, pn 是对象所在划分的编号, 用于返回结果.

### §4.2.4 子网格－进程映射

网格划分完成后, 需要将子网格映射到进程上, 该映射尽量使得从老划分到新划分的数据迁移量最小 [148]. Oliker 和 Biswas 设计了一个启发式算法处理这个问题, 他们的算法可以得到次优解 [72, 147, 148]. 我们在 PHG 中实现了该算法.

Oliker 和 Biswas 的算法先对数据迁移建模. 用相似矩阵 S (similarity matrix) 表示数据在所有进程中的分布情况. S 在 PHG 中是一个 $p_{\text{old}} \times p$ 的矩阵, $p_{\text{old}}$ 是当前子网格数, $p$ 为新子网格数 (在通信器不变的情况下, $p_{\text{old}}$ 与 $p$ 是相等的), 相似矩阵 S 的元素 $\mathtt{S}_{i,j}$ 表示编号为 $i$ 的子网格中需要迁移到编号为 $j$ 的子网格中的数据量. 实际计算中, 每个进程并发地计算相似矩阵的一行, 然后通过一个主进程收集这些信息创建一个相似矩阵, 根据这个相似矩阵计算子网格 – 进程映射关系, 最后通过一个广播操作将映射关系发送给所有进程.

优化子网格 – 进程映射时, 需要针对不同的体系结构上需要建立不同的代价方程 (cost function) 以优化数据迁移开销. 常用的两种度量是: TotalV 和 MaxV. TotalV 极小化所有进程数据迁移量之和; MaxV 极小化单个进程的最大数据迁移量. Oliker 和 Biswas 的算法中使用的度量是 TotalV. 给定一个相似矩阵 S, 其数据总和是确定的, 极小化数据迁移与极大化保持数据不动是等价的, 目标是寻找一个映射 $i \to p_i$ $(i = 0, \cdots, p-1)$ 极大化如下代价方程

$$F = \sum_{\substack{p_j=i \\ 0\le j<p}} S_{i,j},$$

其中 $(p_0, p_1, \cdots, p_{p-1})$ 是 (0, 1, $\cdots$, p-1) 的一个置换.

Oliker 和 Biswas 设计了一个启发式贪婪算法解决该问题, 所得到的映射是次优的, 算法的伪代码如下所示, 本文对算法稍微做了修改以适应 PHG 中的应用情况.

**算法 4**: 启发式子网格—进程映射算法

```
for (j = 0; j < p; j++) part_map[j] = unassigned;
for (i = 0; i < p; i++) proc_unmap[i] = 1;

generate list L of entries in S in descending order;

count = 0;
while( count < p) {
    find next entry S[i][j] in L such that
        proc_unmap[i] > 0 and part_map[j] = 1;
    proc_unmap[i] = 0;
    part_map[j] = assigned;
    count++;
    map partition j to processor i;
}
```

我们在 PHG 中实现了三个子网格 – 进程映射函数, 接口分别如下:

```
int phgPartitionRemap(GRID *g, int nprocs, MPI_Comm comm),
```

```
int phgPartitionRemapV2(GRID *g, int nprocs, MPI_Comm comm),
```

```
int phgPartitionRemapV3(double ds[],int perm[],int np, MPI_Comm comm),
```

前两个函数中, `g` 为 PHG 的网格, 对其中每个单元 `e`, `e->mark` 进入函数时存储了它在新划分中所属的子网格编号, 而从函数返回时则存储着重映射后的新编号.

`phgPartitionRemapV3` 是一个通用的函数接口, `ds` 存储了当前进程迁往每个进程的数据迁移量, 映射结果在 `perm` 中返回, `np` 为新划分的个数, 数组 `ds` 与 `perm` 的长度均为 `np`. 该函数除可用于子网格 – 进程映射外, 还可用于计算其他一些应用中优化通信的映射.

函数 `phgPartitionRemap` 和 `phgPartitionRemapV2` 的功能是一样的, 后者仅存储相似矩阵的非零元, 有着更好的空间复杂度. 如果重映射后子网格编号发生了变化, 则它们返回一个大于零的值; 如果编号没有变化, 则返回零; 如果计算失败, 则它们返回小于零的值, 此时子网格编号保持不变.

### §4.2.5 划分质量度量

对并行自适应有限元计算而言, 网格划分质量直接影响总体计算时间. 最重要的划分质量度量指标是进程的计算负载的平衡. 此外, 还需要考虑极小化通信代价, 该目标对应着极小化子网格共享面的数目, 因为共享面上的数据近似等于计算中需要通信的数据. 在图划分中, 共享面与割边 (cut edge) 的概念是等价的. 这里, 我们采用两个度量指标来衡量划分质量: 表面比 (surface index) 和进程间连通系数 (interprocess connectivity). 表面比衡量的是通信量的大小, 进程间连通系数衡量的是需要通信的进程数 [62, 149].

表面比定义为子网格的边界面的数目与所有面的数目的比值 (在二维中为边的数目比值). 如果共有 $n$ 个子网格, $b_i$ 和 $f_i$ 分别代表子网格 $i$ 中边界面的数目和总面数, 那么最大局部表面比 (maximum local surface index) 定义为

$$r_M = \max_{0 \le i < n} \frac{b_i}{f_i}. \tag{4.5}$$

假如 $b_t$ 和 $f_t$ 分别表示所有子网格的边界面数目之和与整个网格中面的总数目, 那么可以定义全局表面比 (global surface index)

$$r_G = \frac{b_t}{f_t}. \tag{4.6}$$

而平均表面比 (average surface index) 则定义为

$$r_A = \frac{1}{n} \sum_{i=0}^{n-1} \frac{b_i}{f_i}. \tag{4.7}$$

进程间连通系数是指与给定子网格相互连通的子网格的数量. 它衡量与该进程进行通信的进程数目. 最大进程间连通系数 (maximum interprocess connectivity) 是程序可扩展性的一个重要衡量指标 [62].

## §4.3 PHG 的动态负载平衡模块

前面介绍了 PHG 中的网格划分算法, 本节将介绍 PHG 的动态负载平衡模块及相应的用户接口函数, 并给出一些实现细节.

### §4.3.1 动态负载平衡函数

PHG 中, 网格的负载不平衡因子保存在网格对象中, 并由 PHG 自动计算及维护. PHG 的动态负载平衡函数由用户调用, 一般在网格/基函数调整后立即调用, PHG 的动态负载平衡函数接口如下:

```
int phgBalanceGrid(GRID *g, FLOAT lif, INT sm, DOF *w, FLOAT p),
```

其中 `g` 为网格对象; `lif` 是动态负载平衡的阈值, 网格的负载不平衡因子大于该阈值则进行网格重划分; `sm` 用来控制子网格的数目, 使得子网格的平均单元数目不小于这个值, 用户可以将其设置负值或者零, 此时表示子网格的最小平均单元数目为 1; `w` 和 `p` 两个参数给出单元的计算负载, 即单元 $K$ 的计算负载为 $\mathtt{w}_K^p$, `w` 可以为 `NULL`, 表示所有单元的计算负载相等.

### §4.3.2 网格划分和数据迁移

PHG 的网格对象与一个 MPI 通信器相关联, 通信器中的进程数等于当前网格划分的子网格数, 当动态负载平衡函数被调用时, PHG 将首先根据 `sm` 参数、当前网格中的单元数、最大可用的进程数等判断是否需要调整通信器, 如果需要将创建一个包含更多进程的通信器. 接着调用划分程序划分网格, 划分结果直接保存在每个单元中, 单元 `e` 的成员 `mark` 存储了该单元所属的子网格编号.

网格划分函数结束时, 返回一个逻辑值说明网格分布是否被改变, `FALSE` 表明网格分布保持不变, `TRUE` 表明网格分布发生改变. 如果网格分布发生改变, 动态负载平衡函数将调用数据重分布函数重组子网格以及迁移自由度数据等. 其中, 重组子网格包括迁移网格单元、为新的子网格建立相应的分布式二叉树、建立子网格内部及子网格间的邻居关系等, 包含许多复杂的通信和处理, 在此不做详述.

## §4.4 数值算例

本节给出数值算例, 比较不同划分方法的性能、划分质量以及对有限元计算性能影响. 数值算例分为两个部分, 第一部分考察不同划分方法的划分质量, 衡量指标采用前面介绍的表面比 (surface index) 和进程间连通系数 (interprocess connectivity). 第二部分使用自适应有限元方法求解偏微分方程, 比较不同划分方

法对程序性能的影响.

### §4.4.1 划分质量比较

本节比较不同划分方法划分质量的优劣, 划分方法包括: ParMETIS, HSFC (PHG 的希尔伯特空间填充曲线划分方法), MSFC (PHG 的 Morton 空间填充曲线划分方法), RCB (Zoltan 的递归坐标二分划分方法) 和 Zoltan 的 HSFC. 其中第一个是图划分方法, 其余几个是几何划分方法.

本节使用了两个网格, 第一个网格 $\Omega_1$ 是一个圆柱体, 如图 4.6 所示, 直径小但长度较长, 有较大的长宽比, 网格的单元数目为 2,522,624. 第二个网格 $\Omega_2$ 是一个多孔薄板区域, 面积较大, 但很薄, 同时有很多的小孔, 如图 4.7 所示, 单元数目为 3,713,792. 两个网格均由 Netgen[81] 生成.

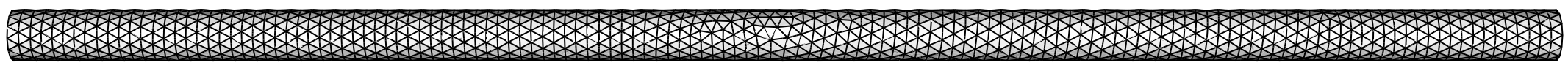

图 4.6 测试网格 $\Omega_1$ (圆柱体)

表 4.1 给出各种划分方法产生的子网格的最大进程间连通系数. 对第一个网格 $\Omega_1$, ParMETIS 的划分质量最好, 其次是 RCB, MSFC, PHG/HSFC, 最后是 Zoltan/HSFC, 当子网格数目小于或等于 64 时, PHG/HSFC 的最大进程间连通系数明显低于 Zoltan/HSFC. 对第二个网格结论是类似的. 需要说明的是这两个区域比较规则, 很适合 RCB 这类算法.

表 4.2 给出子网格的最大表面比和平均表面比, 这两个指标直接衡量了通信量的大小, 其中最大表面比衡量了单个进程最大的数据发送/接收量, 平均表面比衡量了全局的数据发送/接收量, 比值越小表明划分质量越好. 从表中可以看出 ParMETIS 的边界面所占的比例始终最低, 说明其划分质量最好; Zoltan/HSFC 的数值总是最大, 表明其划分质量在这几个划分方法里最差; MSFC 的划分质量比 PHG/HSFC 略差, 但同样优于 Zoltan/HSFC; 由于区域为规则区域, RCB 的划分质量优于 PHG/HSFC, MSFC 及 Zoltan/HSFC. 从表 4.1–4.2 可以看出, 图划分方法的划分质量始终好于几何方法的划分质量, 但前者计算划分的时间亦远大于后者.

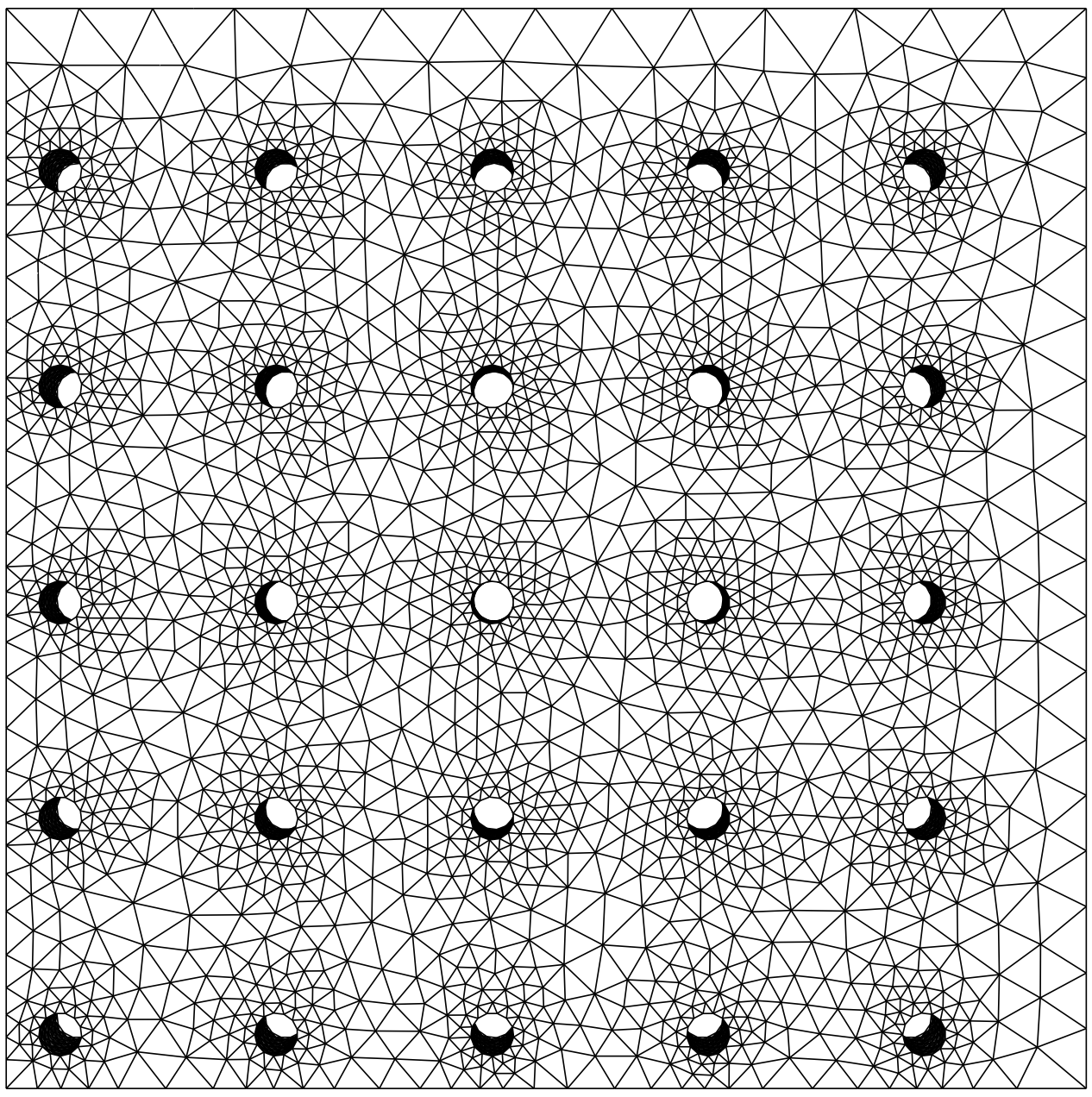

图 4.7 测试网格 $\Omega_2$ (多孔薄板)

### §4.4.2 划分质量对自适应有限元计算性能的影响

本节将 ParMETIS, Zoltan/HSFC, PHG/HSFC, PHG/RTK (PHG 的加密树划分方法), MSFC (PHG 中的 Morton 空间填充曲线划分方法), RCB (Zoltan 的递归坐标二分方法) 六种划分方法应用到自适应有限元计算中, 以研究不同划分方法对并行程序总体性能的影响. 我们将统计划分时间, 比较各个方法的划分速度; 统计动态负载平衡时间, 比较划分质量对数据迁移的影响; 统计有限元离散线性方程组求解时间, 比较划分质量对通信的影响; 统计每一个自适应步的总计算时间, 比较不同划分方法对并行程序总体性能影响. 本节将计算三个不同的算例, 分别求解三种不同类型的偏微分方程.

测试的硬件平台为科学与工程计算国家重点实验室的浪潮 TS10000 高性能集群 (LSSC-III), 计算结点为浪潮 NX7140N 刀片, 每个刀片包含两颗 Intel X5550 四核处理器和 24GB 内存, 其单核双精度浮点峰值性能为 10.68Gflops, 282 个计算结点的总浮点峰值性能为 24Tflops. 所有结点同时通过千兆以太网和 DDR Infiniband 网络互联.

表 4.1 $\Omega_1$ 与 $\Omega_2$ 的最大进程间连通系数

| $\Omega_1$ | | | | | | |
|---|---|---|---|---|---|---|
| # submeshes | 16 | 32 | 64 | 128 | 160 | 192 |
| ParMETIS | 2 | 2 | 5 | 8 | 8 | 9 |
| RCB | 2 | 2 | 5 | 11 | 12 | 13 |
| MSFC | 4 | 5 | 10 | 16 | 19 | 21 |
| PHG/HSFC | 3 | 6 | 13 | 23 | 24 | 23 |
| Zoltan/HSFC | 12 | 18 | 21 | 23 | 24 | 24 |
| $\Omega_2$ | | | | | | |
| # submeshes | 16 | 32 | 64 | 128 | 160 | 192 |
| ParMETIS | 7 | 7 | 7 | 10 | 11 | 12 |
| RCB | 7 | 7 | 8 | 11 | 12 | 12 |
| MSFC | 9 | 13 | 18 | 21 | 22 | 25 |
| PHG/HSFC | 8 | 10 | 13 | 18 | 20 | 21 |
| Zoltan/HSFC | 12 | 19 | 23 | 25 | 27 | 31 |

**例 4.4.1.** 本例求解如下 Dirichlet 边界条件 Helmholtz 方程,

$$\begin{cases} -\Delta u + u = f & (x,y,z) \in \Omega \\ u(x,y,z) = g & on\ \partial\Omega. \end{cases}$$

计算区域为圆柱体区域 $\Omega_1$, 取解析解如下:

$$u = \cos(2\pi x)\cos(2\pi y)\cos(2\pi z),$$

该算例中解是光滑的, 因此网格加密基本上是均匀的.

测试中使用了 32 个计算结点, 128 个进程. 计算区域如图 4.6 所示, 初始网格 $\mathcal{T}_1$ 由 Netgen [81] 生成, 含有 4,927 个四面体单元. 使用三阶协调拉格朗日元对方程进行离散, 调用数值代数软件包 Hypre 的 BoomerAMG 解法器 [22] 求解有限元离散产生的线性系统. 网格自适应使用一个后验型的误差指示子, 其定义如下

$$\eta_K^2 = h_K^2 \|f_h + \Delta u_h + u_h\|_{L^2(K)}^2 + \sum_{f\subset\partial K\cap\Omega} \frac{h_f}{2} \|[\frac{\partial u_h}{\partial n_f}]\|_{L^2(f)}^2.$$

在网格划分时, 单元的权重取为 1, 为了观察网格划分, 负载不平衡因子阈值设置得较小 (1.05), 划分次数较多, 有利于统计时间. 网格单元标记策略为最大标记策略 (maximum strategy), 参数 $\theta$ 取 0.5 (参看 §5.1.3.2).

**表 4.2 $\Omega_1$ 与 $\Omega_2$ 的表面比**

| maximum surface index (%, $\Omega_1$) | | | | | | |
|---|---|---|---|---|---|---|
| # submeshes | 16 | 32 | 64 | 128 | 160 | 192 |
| ParMETIS | 0.74 | 1.48 | 2.88 | 5.00 | 5.78 | 5.71 |
| RCB | 1.0 | 2.18 | 4.07 | 6.06 | 7.06 | 7.60 |
| MSFC | 4.2 | 6.51 | 10.6 | 16.2 | 18.9 | 20.7 |
| PHG/HSFC | 3.80 | 8.31 | 15.5 | 19.5 | 20.4 | 21.5 |
| Zoltan/HSFC | 13.9 | 19.1 | 25.4 | 28.1 | 30.9 | 35.8 |
| average surface index (%, $\Omega_1$) | | | | | | |
| # submeshes | 16 | 32 | 64 | 128 | 160 | 192 |
| ParMETIS | 0.64 | 1.28 | 2.39 | 3.62 | 4.05 | 4.42 |
| RCB | 0.85 | 1.68 | 3.43 | 5.17 | 6.03 | 6.58 |
| MSFC | 2.93 | 5.14 | 7.44 | 9.8 | 10.7 | 11.5 |
| PHG/HSFC | 2.78 | 5.03 | 7.18 | 9.34 | 10.2 | 10.8 |
| Zoltan/HSFC | 8.72 | 11.8 | 16.2 | 20.6 | 22.0 | 23.6 |
| maximum surface index (%, $\Omega_2$) | | | | | | |
| # submeshes | 16 | 32 | 64 | 128 | 160 | 192 |
| ParMETIS | 2.45 | 2.56 | 3.30 | 5.32 | 6.13 | 6.10 |
| RCB | 2.57 | 4.74 | 6.29 | 9.16 | 9.66 | 11.6 |
| MSFC | 3.0 | 5.11 | 7.06 | 9.74 | 11.1 | 11.8 |
| PHG/HSFC | 2.86 | 4.74 | 7.03 | 9.78 | 10.6 | 11.3 |
| Zoltan/HSFC | 8.89 | 12.3 | 16.5 | 19.5 | 21.2 | 22.8 |
| average surface index (%, $\Omega_2$) | | | | | | |
| # submeshes | 16 | 32 | 64 | 128 | 160 | 192 |
| ParMETIS | 1.12 | 1.45 | 2.18 | 3.43 | 3.87 | 4.21 |
| RCB | 1.89 | 3.04 | 3.93 | 5.90 | 6.58 | 7.2 |
| MSFC | 2.24 | 3.61 | 5.21 | 7.34 | 8.22 | 8.92 |
| PHG/HSFC | 2.19 | 3.29 | 4.88 | 6.92 | 7.81 | 8.27 |
| Zoltan/HSFC | 5.10 | 6.86 | 9.47 | 12.6 | 14.1 | 15.3 |

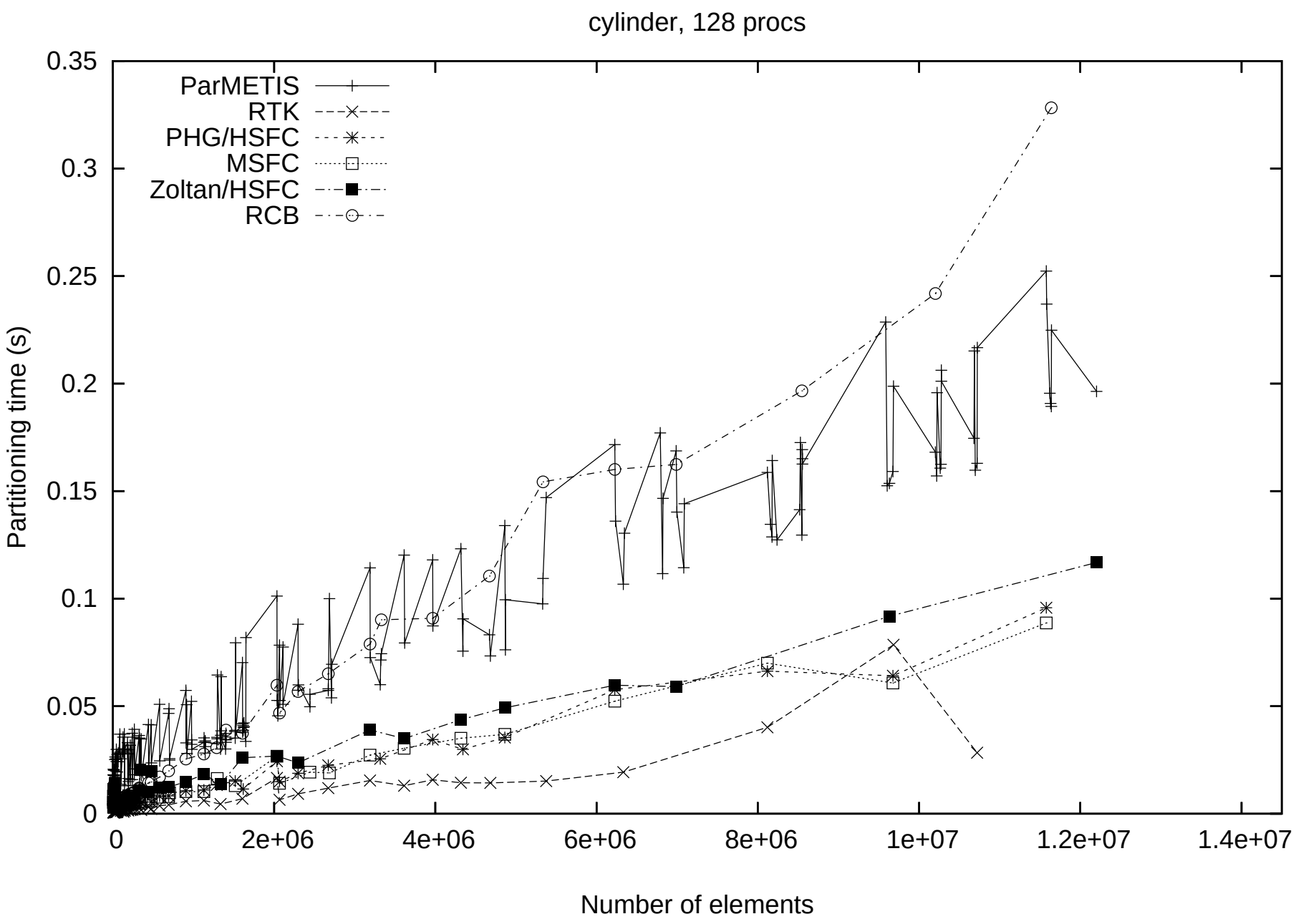

图 4.8 网格划分时间 (例 4.4.1)

图 4.8 给出划分时间, 从中可以看出 RTK 方法的划分速度最快, 其次是 MSFC, PHG/HSFC, Zoltan/HSFC. PHG 实现的 HSFC 及 MSFC 均比 Zoltan/HSFC 速度要快. ParMETIS 和 RCB 最慢. 从图上可以看出网格分布对 ParMETIS 的划分时间有较大影响, 随着网格调整, 其划分时间出现剧烈震荡. 总体来讲, 几何划分方法对网格分布不敏感, RCB, MSFC, PHG/HSFC, Zoltan/HSFC 的划分时间随着网格规模的增大以较稳定的速率增加.

图 4.9 给出动态负载平衡时间, 其中包含了网格划分和数据迁移时间, 其中数据迁移时间占的比重较大. 一个好的划分方法应该是增量的, 即网格的微小变化仅导致划分的微小变化, 这样会使得数据迁移量比较小. 从图上可以看出 RTK 的时间最短, 表明其数据迁移量较少, RTK 的时间增长也很稳定. 其次是 ParMETIS, ParMETIS 同样表现出震荡. 接着是 MSFC, PHG/HSFC, RCB, 用时最长的是 Zoltan/HSFC. 它们的时间增长趋势均比较稳定, 说明这些方法均有较好的可扩展性和比较低的计算复杂度.

一般情况下线性方程组的求解时间在自适应有限元计算中占统治地位, 图 4.10 和 4.11 分别给出线性方程组求解时间和每一个自适应步的时间. 从图中可以看出, RCB, ParMETIS, RTK 运行时间最短. 对 RCB 而言, 其划分质量一般要比 ParMETIS 及 RTK 差一些, 这个算例是特例, 因为计算区域是一个长圆柱体, 非常

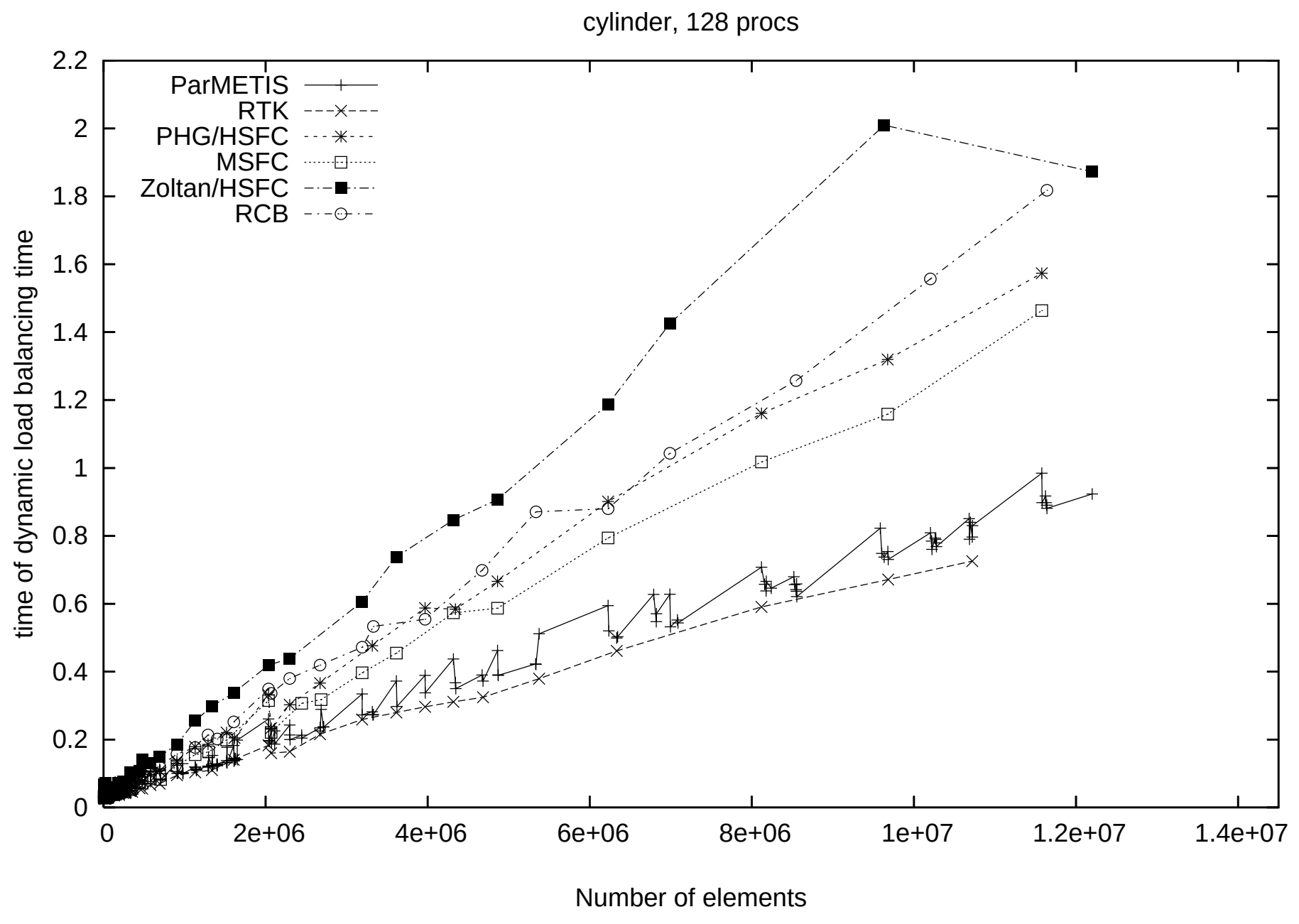

图 4.9 动态负载平衡时间 (例 4.4.1)

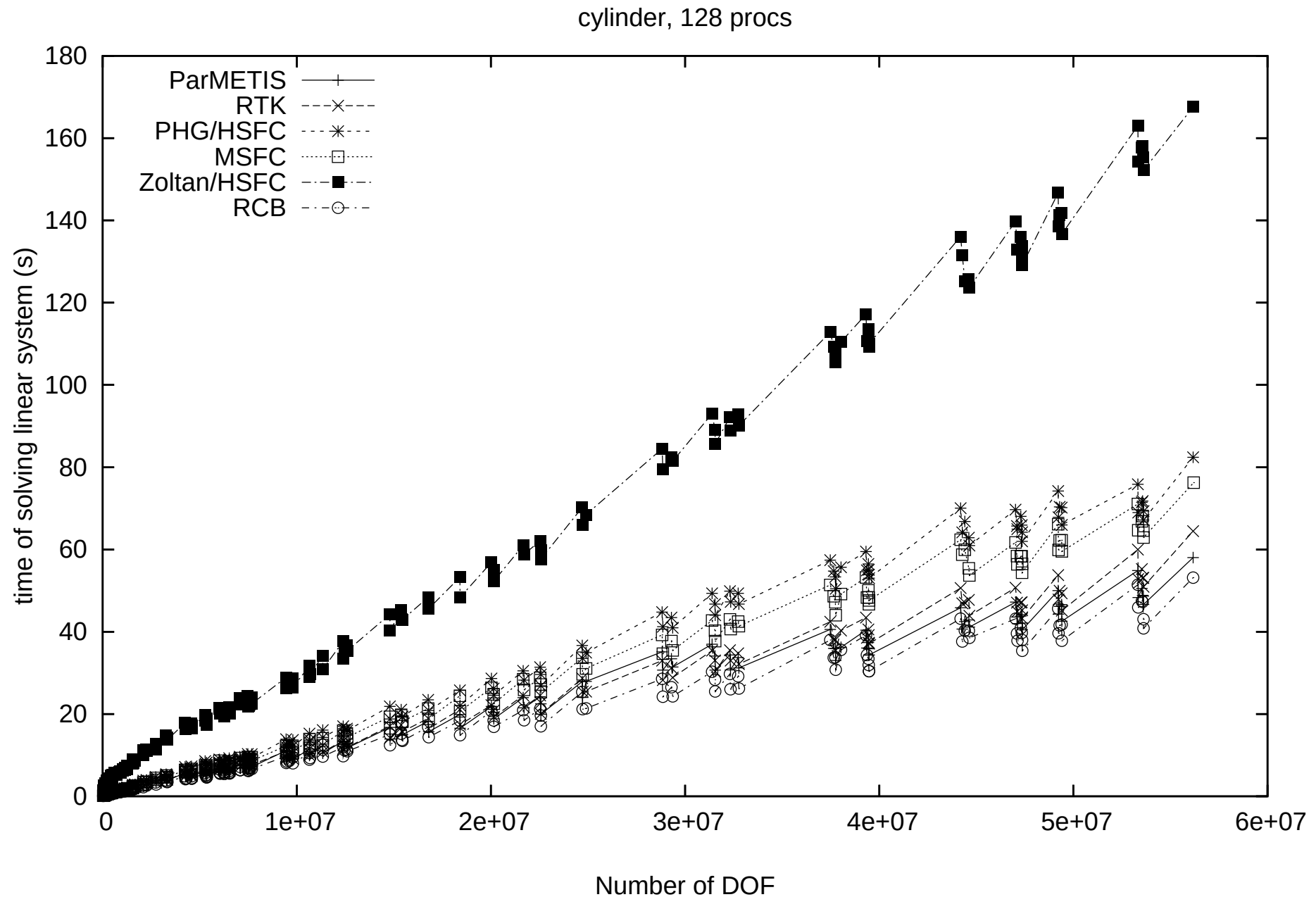

图 4.10 有限元离散系统求解时间, 横坐标为自由度数 (例 4.4.1)

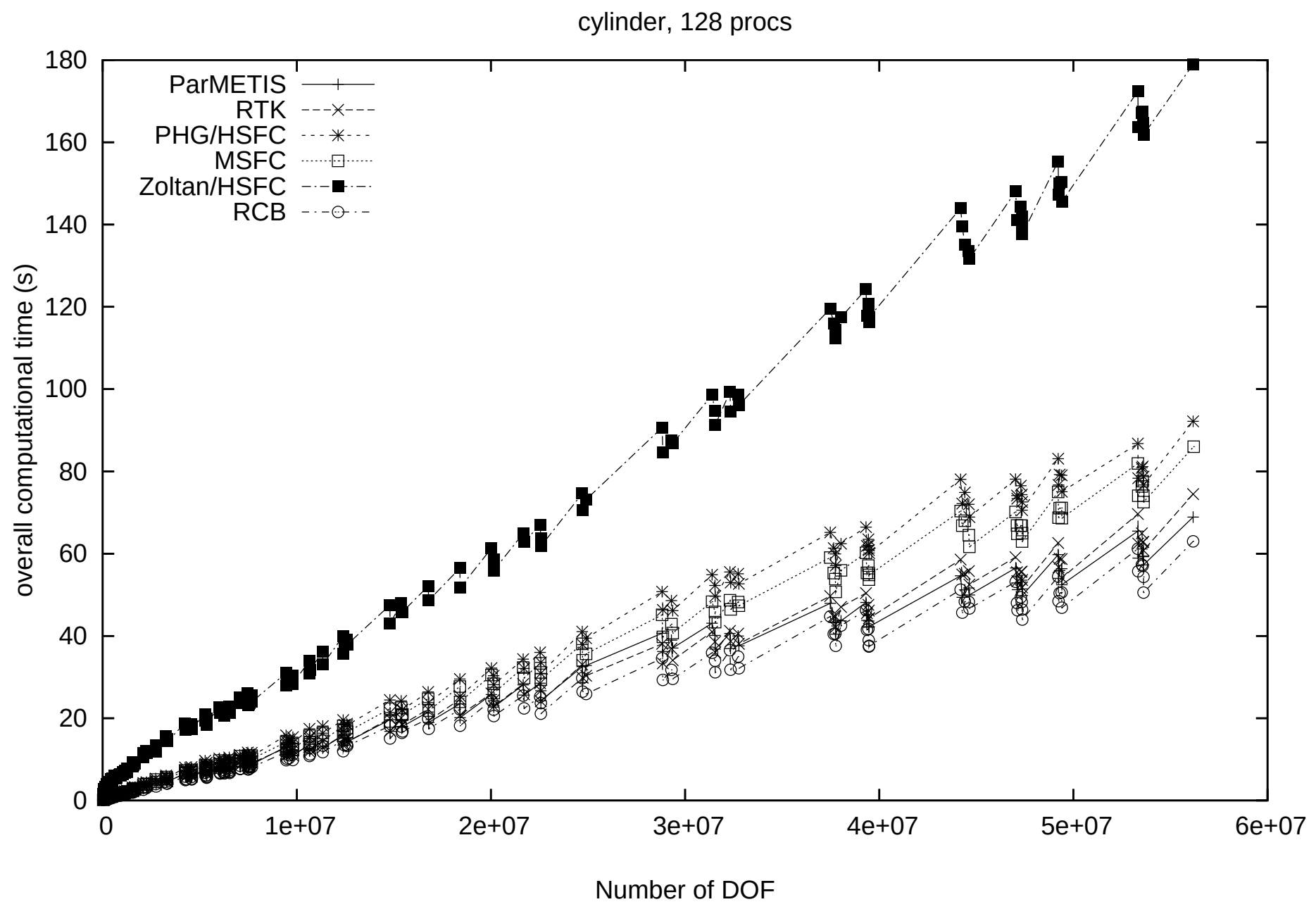

图 4.11 每一个自适应步时间 (例 4.4.1)

适合 RCB. 其次是 MSFC, PHG/HSFC, 用时最长的是 Zoltan/HSFC. Zoltan/HSFC 与 PHG/HSFC 是同一类型的划分方法, 由于 PHG 在映射时采取了保持区域长宽比的方式, 维持了区域的空间局部性, PHG/HSFC 的划分质量比 Zoltan/HSFC 高, 求解时间也较少.

**表 4.3 总体运行时间及划分次数 (例 4.4.1)**

| Method | total running time(s) | # of repartitionings |
|---|---|---|
| RCB | 3049.60 | 60 |
| ParMETIS | 3360.73 | 189 |
| RTK | 3465.63 | 59 |
| MSFC | 4088.01 | 58 |
| PHG/HSFC | 4493.43 | 59 |
| Zoltan/HSFC | 8954.21 | 50 |

表 4.3 给出该算例采用不同划分方法时的运行时间及划分次数, 算例包含 190 个自适应步. RCB 运行时间最短, 说明对计算区域比较规则的问题, 简单的几何划分方法是非常有效的. Zoltan/HSFC 的运行时间最长, 是别的方法的两倍以上,

表现最差. ParMETIS 的划分次数最多, 是别的方法的三倍. 划分次数过多会导致数据迁移频繁, 当网络带宽较低时, 就需要考虑数据迁移问题, 但由于其划分质量好, 线性方程组的求解时间较短.

**例 4.4.2.** 本例求解一个带有 Dirichlet 边界条件的 Poisson 方程

$$\begin{cases} -\Delta u = f & (x,y,z) \in \Omega \\ u(x,y,z) = g & on\ \partial\Omega. \end{cases}$$

其计算区域为多孔薄板区域 $\Omega_2$, 取解析解为

$$u = 4(1 - e^{-100x} - (1 - e^{-100})x)y(1-y)e^{-z},$$

它在 $x = 0$ 处呈现边界层现象.

计算区域如图 4.7 示, 初始网格是一致网格, 由 Netgen 生成, 含有 49,271 个四面体单元. 测试使用了 32 个计算节点, 128 个进程, 与前一个算例一样, 使用三阶拉格朗日元进行离散, 使用 PCG (Preconditioned Conjugate Gradient method) 求解离散得到的线性方程组, 采用残量型误差指示子, 形式如下

$$\eta_K^2 = h_K^2 \|f_h + \Delta u_h\|_{L^2(K)}^2 + \sum_{f \subset \partial K \cap \Omega} \frac{h_f}{2} \|[\frac{\partial u_h}{\partial n_f}]\|_{L^2(f)}^2.$$

标记策略为最大标记策略 (maximum strategy), 参数取为 0.5 (参看 §5.1.3.2).

图 4.12–4.15 分别给出了划分时间、动态负载平衡时间、线性方程组求解时间及每个自适应步计算时间. 划分时间如图 4.12 所示, 最快的仍然为 RTK, 其次为 MSFC, PHG/HSFC, Zoltan/HSFC, 最后是 ParMETIS 和 RCB. RCB 最慢, 同算例 4.4.1 一样, ParMETIS 又表现出震荡.

动态负载平衡时间如图 4.13 示, RTK 用时最少, 并且 RTK 的运行时间随着网格单元数量以线性增长; 其次是 ParMETIS, 它的运行略慢, 但其划分质量高, 通过减少数据迁移减少了负载平衡时间; 接着是 MSFC, PHG/HSFC; 运行最慢的是 Zoltan/HSFC 及 RCB.

线性方程组求解及每个自适应步的时间如图 4.14 和 4.15 所示, MSFC 运行时间最短, Zoltan/HSFC 运行时间最长, 其他几个划分方法的运行时间相差不大. RCB 在这个例子中的表现不如在例 4.4.1 中的表现, 因为在前者的计算区域是一个狭长的圆柱体, 较适合 RCB.

表 4.4 给出了总体运行时间及划分次数, 总的自适应次数为 238 次. 运行时间最长的为 Zoltan/HSFC, 几乎是别的方法的 2 倍. 算例 4.4.1 及 4.4.2 均表明: 在

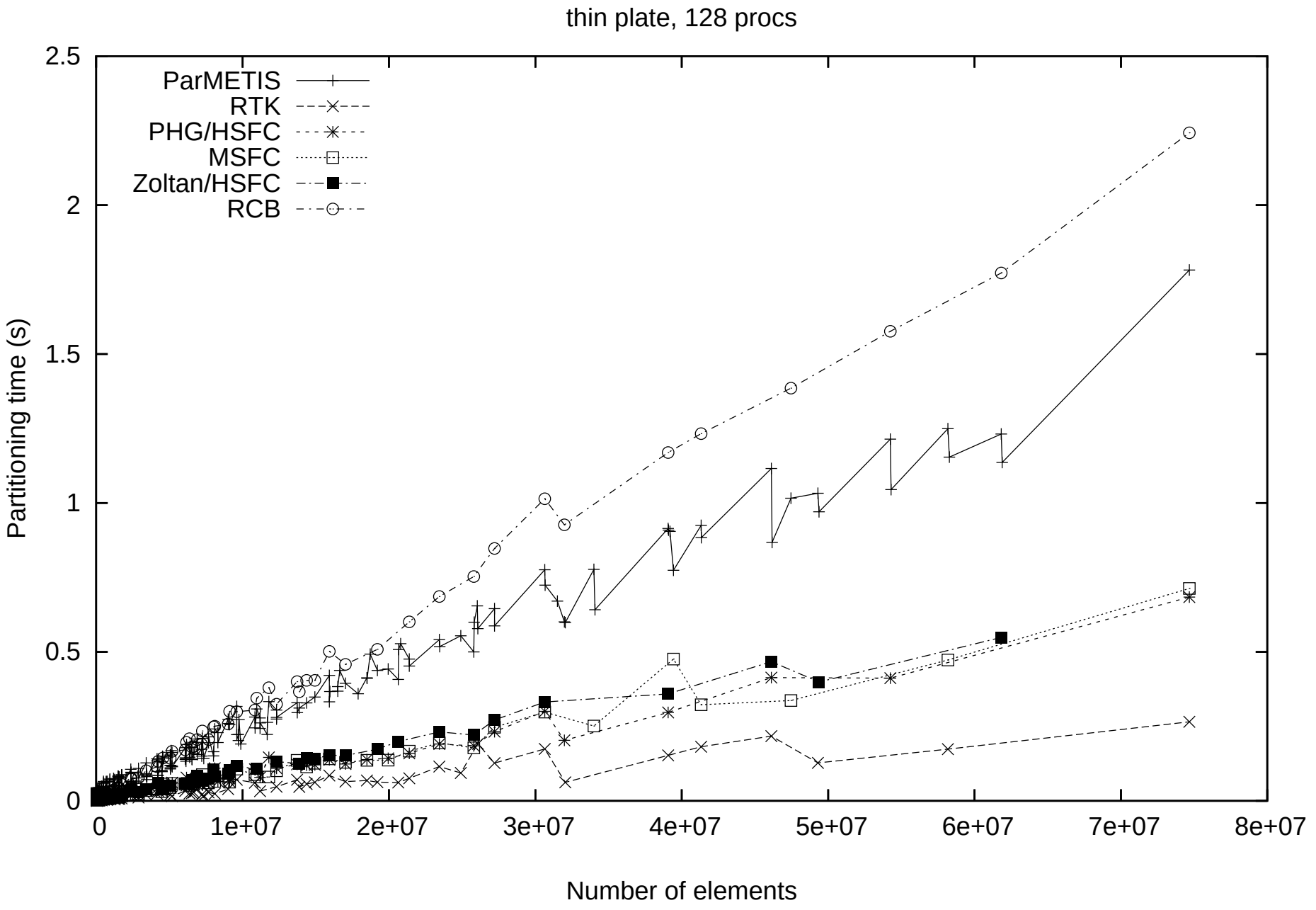

图 4.12 网格划分时间 (例 4.4.2)

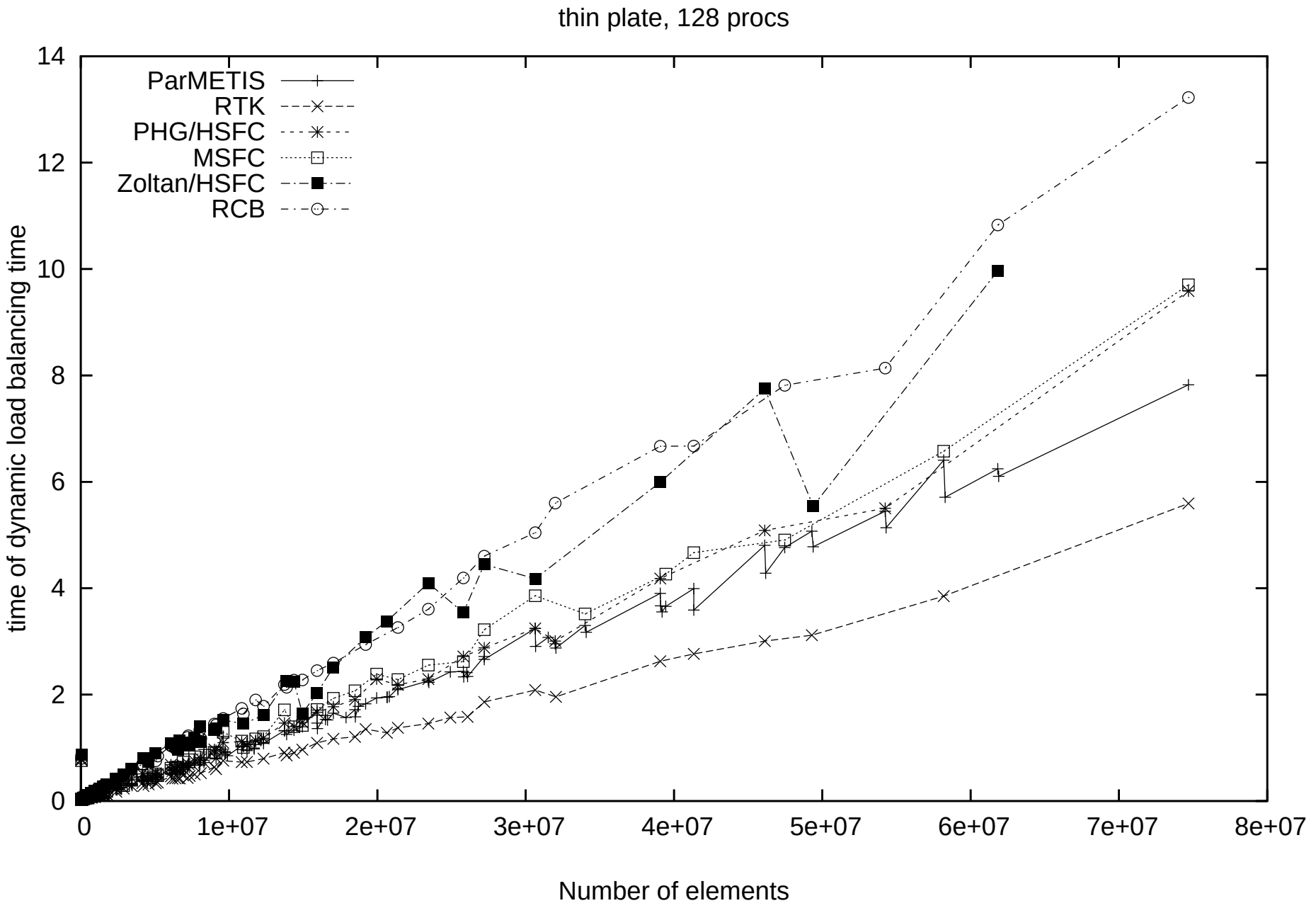

图 4.13 动态负载平衡时间 (例 4.4.2)

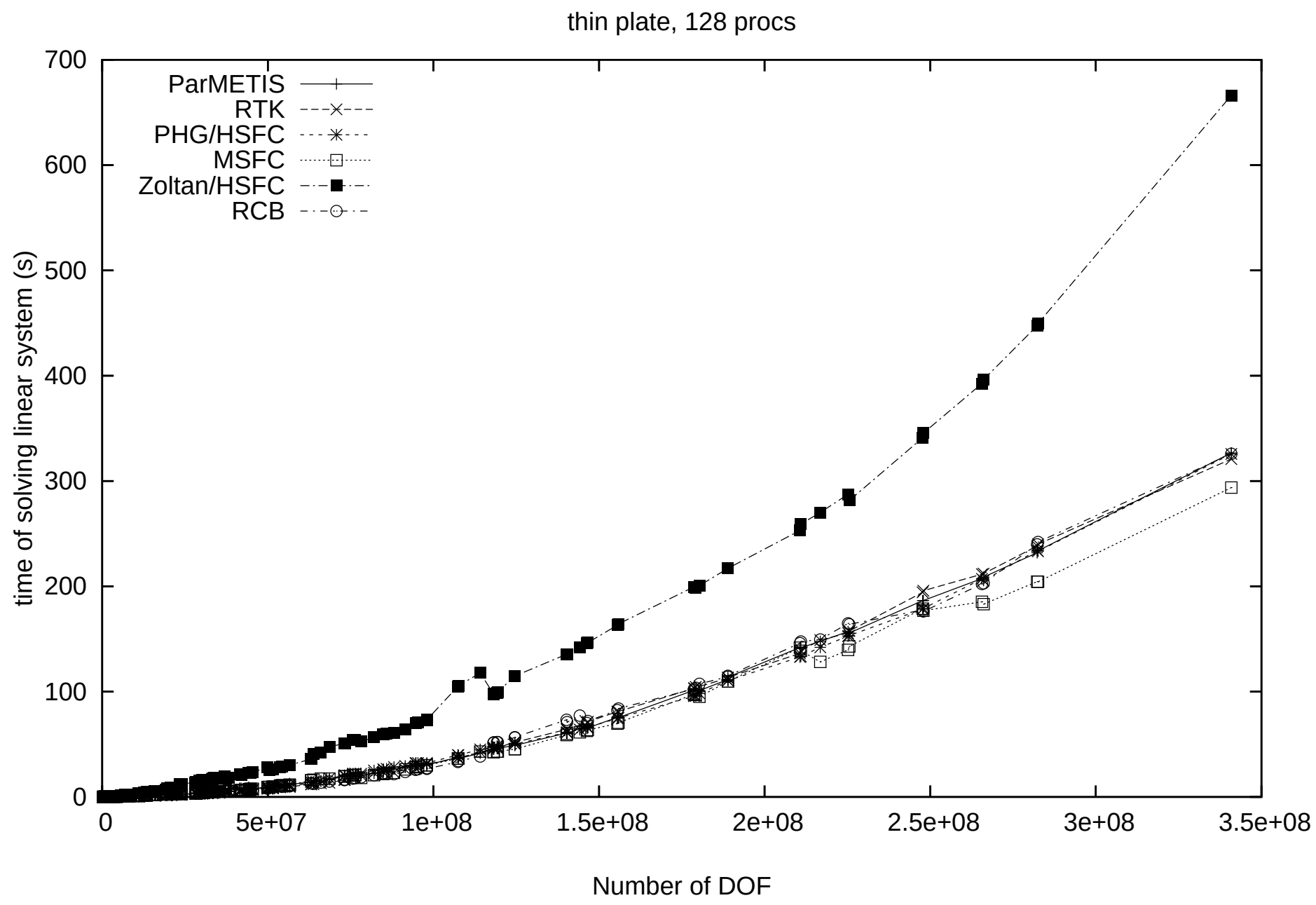

图 4.14 有限元离散系统求解时间, 横坐标为自由度数 (例 4.4.2)

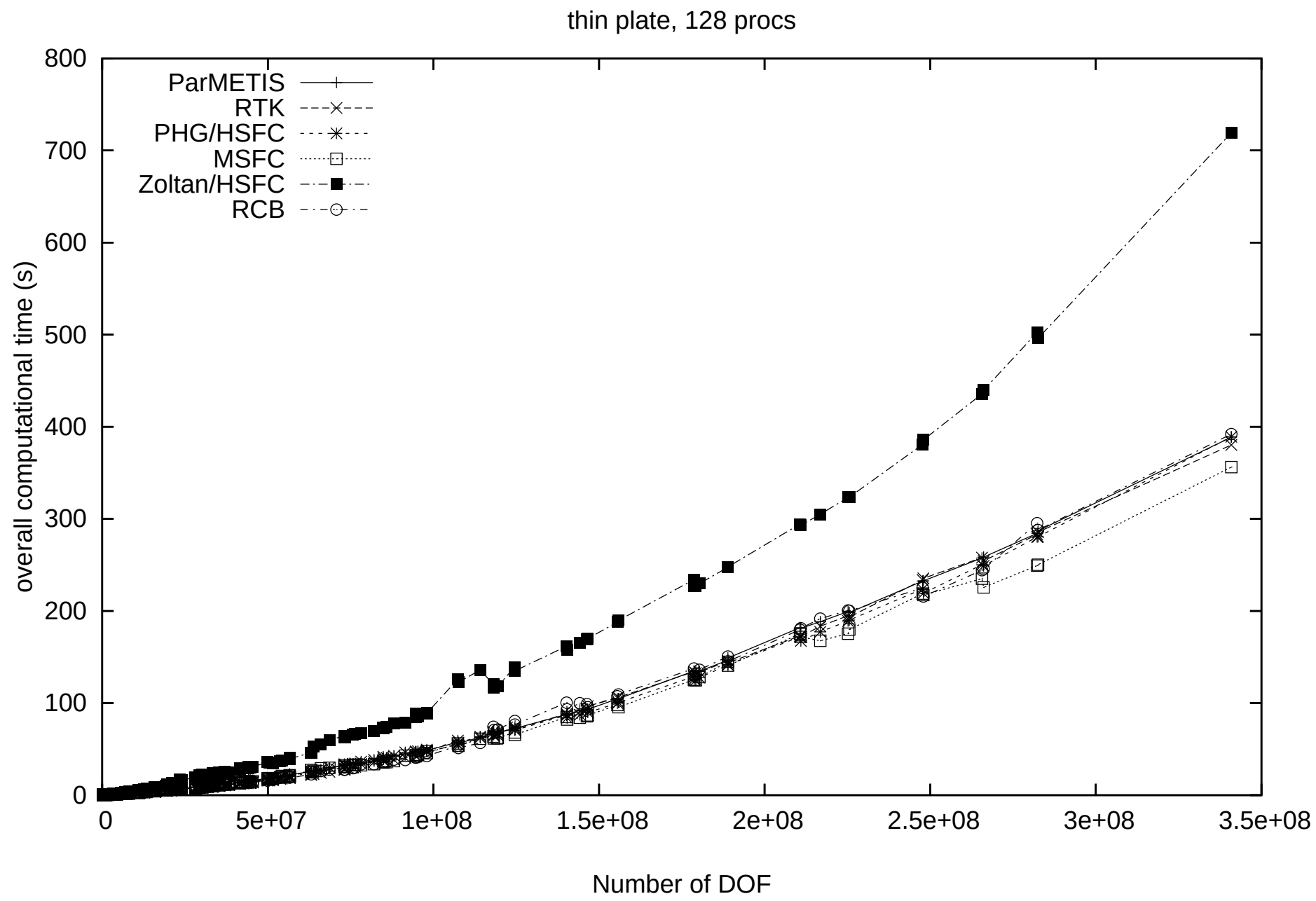

图 4.15 每一个自适应步时间 (例 4.4.2)

表 4.4 总体运行时间及划分次数 (例 4.4.2)

| Method | total running time(s) | # of repartitionings |
|---|---|---|
| MSFC | 6269.41 | 83 |
| RTK | 6514.84 | 89 |
| PHG/HSFC | 6570.44 | 83 |
| RCB | 6667.12 | 89 |
| ParMETIS | 6758.62 | 237 |
| Zoltan/HSFC | 12079.33 | 74 |

映射的过程中, 保持区域的空间局部性对划分质量是很重要的. 其他几个方法的运行时间比较接近. 值得一提的是 ParMETIS, 和上一个例子一样, 其划分次数最多, 这是由于问题的奇点是局部的, 网格的自适应局部加密会引起负载不平衡, 需要调整网格分布. 一般说来, Morton 空间填充曲线的空间局部性要弱于希尔伯特空间填充曲线, 但在这两个例子中, MSFC 的计算效果却始终优于希尔伯特空间填充曲线划分方法 (PHG/HSFC & Zoltan/HSFC), 所以在选择划分方法时应根据问题本身的特点、计算区域和划分质量等多个因素确定合适的划分策略, 以加快运算速度, 优化计算资源.

**例 4.4.3.** 本例求解的方程为线性抛物型方程

$$\begin{cases} u_t - \Delta u = f & in\ \Omega \times (0,T) \\ u = g & on\ \Gamma \times [0,T]. \end{cases} \tag{4.8}$$

取解析解为

$$u = \exp((25((x-\frac{1}{2}-\frac{2}{5}\sin(8\pi t))^2 + (y-\frac{1}{2}-\frac{2}{5}\cos(8\pi t))^2 + (z-1)^2) + 0.9)^{-1} - 2.5).$$

计算区域为 $\Omega_3 = (0,1)^3$, 时间区间为 $[0,1]$. 解的极大值在 $z=1$ 平面上运动, 网格在极值附近最密. 由于这是一个依赖于时间的问题, 网格自适应同时包含加密和放粗.

初始计算网格包含六个四面体, 时间离散采用向后欧拉格式, 空间离散采用二阶拉格朗日元, 使用 PCG (Preconditioned Conjugate Gradient method) 求解得到的线性方程组, 在时间和空间方向上同时采用自适应 [66, 156], 误差指示子为残量型的, 标记策略为最大标记策略 (maximum strategy), 加密和放粗的参数分别选为 0.6 和 0.1.

**表 4.5 程序运行时间 (例 4.4.3, 128 进程)**

| Method | Time TAL(s) | Time DLB(s) | Time SOL (s) | Time STP(s) |
|---|---|---|---|---|
| PHG/HSFC | 6525 | 0.0734 | 0.1886 | 0.9192 |
| Zoltan/HSFC | 6744 | 0.0917 | 0.1928 | 0.9501 |
| MSFC | 6902 | 0.0730 | 0.1966 | 0.9724 |
| RTK | 7015 | 0.0738 | 0.2050 | 0.9884 |
| RCB | 7131 | 0.1126 | 0.1938 | 1.0046 |
| ParMETIS | 7151 | 0.1421 | 0.2114 | 1.0075 |

**表 4.6 程序运行时间 (例 4.4.3, 192 进程)**

| Method | Time TAL(s) | Time DLB(s) | Time SOL(s) | Time STP(s) |
|---|---|---|---|---|
| PHG/HSFC | 6597 | 0.0932 | 0.1808 | 0.9294 |
| MSFC | 6601 | 0.0936 | 0.1863 | 0.9299 |
| Zoltan/HSFC | 6646 | 0.1046 | 0.1862 | 0.9362 |
| RCB | 7197 | 0.1176 | 0.1862 | 1.0139 |
| RTK | 7185 | 0.0799 | 0.2124 | 1.0123 |
| ParMETIS | 7218 | 0.1982 | 0.1942 | 1.0169 |

这里给出两组测试结果. 第一组测试使用 32 个计算结点、128 个进程, 共运行了 7098 个时间步, 每个时间步的平均单元数量为 663,151, 平均自由度数为 919,036. 结果见表 4.5, 其中第一项时间为总运行时间 (TAL), 第二项为单次动态负载平衡的平均时间 (DLB), 第三项为单次线性方程组的平均求解时间 (SOL), 第四项为单个时间步的平均计算时间 (STP). 从表 4.5 中可以看出, 几何划分方法的效果优于图划分方法, PHG/HSFC, Zoltan/HSFC, 及 MSFC 均优于 RTK 及 ParMETIS, RCB 除外. 其中 PHG/HSFC 运行时间最短, ParMETIS 运行时间最长. 可以看出, 当网格变化较剧烈时, 几何划分方法也有优势. 一般来说, 在静态划分中, 图划分方法划分质量最好, 最有优势; 在动态划分中, 情况则不一定.

第二组测试数据如表 4.6 所示, 其中使用了 32 结点、192 进程. 结果与第一组测试类似, 几何划分方法用时较少, 仍优于图划分方法. 在本算例的两个表格中, PHG/HSFC 均比 Zoltan/HSFC 用时略少, 说明 PHG 实现的版本的划分质量略高, 但没有前面两个算例的差别那么大, 因为在这个算例中, 计算区域为 $(0,1)^3$, Zoltan/HSFC 的映射结果和 PHG/HSFC 是类似的. 两个版本的差别在区域的长

宽比较大时体现得更为明显.

## §4.5 本章小结

本章介绍了 PHG 的层次网格结构, 网格划分方法及动态负载平衡模块, 并通过数值实验比较了各种网格划分方法在并行自适应有限元计算中的优劣.

通常来讲, 几何划分方法的速度快、实现简单, 但划分质量不高; 图划分方法的速度慢、算法复杂, 但划分质量非常高. 对静态问题或者网格变化较小的问题, 图划分方法是最好的选择; 对网格频繁调整的问题, 几何划分方法是很有竞争力的. 在实际应用中, 应综合考虑网格划分、数据迁移、问题求解等多个因素确定合适的划分方法.

# 第五章　$hp$ 自适应策略设计

$hp$ 自适应有限元方法同时调整网格和单元上的基函数, 它具有 $h$ 自适应有限元方法和 $p$ 自适应有限元方法的优点. 在适当的策略下, $hp$ 自适应有限元方法可以达到指数阶收敛. 关于 $hp$ 自适应有限元方法已有许多研究成果, 但自适应策略仍旧是其难点之一, 目前尚没有成熟的、广泛被接受的算法, 本章将对 $hp$ 自适应策略进行研究.

## §5.1　基础知识

在实际应用中, 很多科学和工程问题都可以使用偏微分方程或者常微分方程进行建模, 由于问题的解析解很难获得, 通常通过求解数值解来给出解析解的近似. 自适应有限元是求解偏微分方程的高效数值方法之一, 它根据数值解及已知数据对误差作出估计, 并且根据误差指示子调整网格/基函数, 利用最少的计算资源达到最好的计算结果. 自适应有限元方便处理奇性问题, 也方便处理复杂区域, 是当今科学与工程计算的热点研究领域之一 [157, 158, 159, 165, 199, 200, 201, 202].

### §5.1.1　模型问题

考虑椭圆形偏微分方程,

$$\begin{cases} -\nabla\cdot(A(x)\nabla u) = & f \text{ in } \Omega, \\ u|_{\Gamma_D} = & g_D \text{ on } \Gamma_D, \\ A(x)\dfrac{\partial u}{\partial \nu} = & g_N \text{ on } \Gamma_N, \end{cases} \tag{5.1}$$

其中 $A$ 为对称正定矩阵, $\Gamma_D, \Gamma_N \subset \partial\Omega$, $\Gamma_D \bigcap \Gamma_N = \emptyset$, $\Gamma_D \bigcup \Gamma_N = \partial\Omega$. 方程 (5.1) 的变分形式为: 求解 $u \in H^1(\Omega)$, 其中 $u = \bar{u} + G$, $\bar{u} \in V = \{v \in H^1(\Omega), v|_{\Gamma_D} = 0\}, G \in H^1(\Omega), G|_{\Gamma_D} = g_D$, 并且

$$\int_\Omega A\nabla\bar{u}\cdot\nabla v dx = \int_\Omega f v dx - \int_\Omega \nabla G\cdot\nabla v dx + \int_{\Gamma_N} g_N v dS,\ \forall v \in V. \tag{5.2}$$

引入如下符号:

$$a(u,v) = \int_\Omega A\nabla u\cdot\nabla v dx, \tag{5.3}$$

$$l(v) = \int_\Omega f v dx - \int_\Omega \nabla G\cdot\nabla v dx + \int_{\Gamma_N} g_N v dS. \tag{5.4}$$

则问题变为求解 $\bar{u} \in V$, 使得下述变分形式成立,

$$a(\bar{u}, v) = l(v), \quad \forall v \in V. \tag{5.5}$$

双线性型 $a(\cdot,\cdot)$ 是 $V$- 椭圆的并且有界, $l$ 是有界的线性泛函, 由 Lax-Milgram 定理可知, $u$ 存在唯一 [158, 207]. 为了简单起见, 后文将不再区分 $\bar{u}$ 与 $u$.

### §5.1.2 有限元离散

在变分问题 (5.5) 中, $V$ 是无限维 Banach 空间, 计算机无法处理无限维空间, 需要构造 $V$ 的有限维子空间 $V_h$ 进行近似, 并求解如下有限维离散问题: 求 $u \in V_h \subset V$, 满足如下变分形式

$$a(u_h, v_h) = l(v_h), \quad \forall v_h \in V_h. \tag{5.6}$$

同样由 Lax-Milgram 定理可知, 该问题的解存在唯一.

为了构造子空间 $V_h$, 首先对计算区域 $\Omega$ 做剖分.

**定义 5.1.1.** *(区域 $\Omega$ 的网格剖分) 将 $\bar{\Omega}$ 剖分成有限个闭子区域 $K_1, \cdots, K_m$, 每一个子区域 $K_i$ $(1 \le i \le m)$ 称为一个单元, 全部单元组成的集合为 $\mathcal{T}_h$ 称为 $\Omega$ 的一个网格剖分, 简称剖分. 单元满足如下性质:*

*(1) 对每个单元 $K \in \mathcal{T}_h$, $K$ 是闭集, 其内部 $\overset{o}{K}$ 是非空且连通的;*

*(2) 单元 $K$ 的边界 $\partial K$ 是 Lipschitz 连续的;*

*(3) $\bar{\Omega} = \bigcup_{K \in \mathcal{T}_h} K$;*

*(4) 任何两个不同的单元 $K_1, K_2 \in \mathcal{T}_h$, $\overset{o}{K}_1 \bigcap \overset{o}{K}_2 = \emptyset$;*

*(5) 每个 $K \in \mathcal{T}_h$, $\partial K$ 或者是 $\partial \Omega$ 的一部分, 或者是相邻单元的边界.*

剖分 $\mathcal{T}_h$ 称为区域 $\Omega$ 上的一个计算网格. 在二维区域中, 常用单元类型为三角形和四边形; 在三维区域中, 常用单元类型为四面体和六面体. 并行自适应有限元软件平台 PHG 采用的单元类型为四面体.

在网格 $\mathcal{T}_h$ 中, 任意单元 $K$, 记 $P_K$ 为某种多项式集合. 可以构造有限元空间 $V_h$ 使得其中的函数在网格 $\mathcal{T}_h$ 每一个单元上是一个多项式, 形式如下

$$X_h = \{v_h : v_h|_K \in P_K, \forall K \in \mathcal{T}_h\},$$

$$V_h = \{v_h \in X_h : v_h|_{\Gamma_D} = 0\},$$

当限定多项式次数时, $V_h$ 为有限维空间, 设其维数为 $n$, 则可以选取 $n$ 个基函数 $\{\varphi_1, \cdots, \varphi_n\}$. 令

$$u_h = \sum_{i=1}^{n} u_i \varphi_i, \tag{5.7}$$

则变分问题 (5.6) 等价于如下线性代数方程组

$$\sum_{i=1}^{n} a(\varphi_i, \varphi_j) u_i = l(\varphi_j), \quad j = 1, \cdots, n. \tag{5.8}$$

其系数矩阵为刚度矩阵 $\mathcal{A} = [a(\varphi_i, \varphi_j)]_{i,j=1}^{n}$, 未知量为 $u = (u_1, \cdots, u_n)^T$, 右端项为 $b = (l(\varphi_1), \cdots, l(\varphi_n))^T$, 问题最终可以写成

$$\mathcal{A}u = b. \tag{5.9}$$

至此便得到了变分问题 (5.6) 的有限元离散.

### §5.1.3 自适应有限元

误差估计和自适应加密由 Babuška 和 Rheinholdt 首先引入到有限元计算中 [160, 161]. 他们的方法基于估计数值解的残留, 使用残留得到一个局部更精确的新解, 同时定义了网格加密过程. 受他们的启发, 其他计算数学家深入地研究了这个方法, 设计了更有效的离散格式, 误差指示子等, 关于自适应有限元的理论结果及最新进展见文献 [162, 163, 164, 165, 166, 195, 199, 200, 202, 203].

#### §5.1.3.1 误差指示子和自适应策略

误差指示子是自适应有限元方法重要的组件, 它真实反映了实际误差. 一个好的误差指示子是真实误差的一个上界和下界, 上界表明真实误差被误差指示子控制, 即计算是可信的; 下界表明误差指示子被真实误差控制, 即误差指示子是有效的. 常用的误差指示子有如下几类:

- 残量型误差指示子 (residual estimate) [201, 202, 203]: 通过计算残量的范数来估计数值解的误差.

- 对偶权重残量误差指示子 (dual weighted residual estimate) [199]: 先计算数值解的残量, 然后乘以一个表示误差依赖关系的权重. 这种类型的误差指示子需要求解一个对偶变分问题.

- 多重基误差指示子 (hierarchical basis error estimate) [200]: 通过计算高阶元或者加密网格的有限元解, 估计两个解之间的残量.

- 平均化方法 (averaging methods) [200]: 使用局部外插或者数值解梯度平均估计误差子.
- 平衡残量误差指示子 (equilibrated residual error estimate) [200]: 将问题划分为自平衡的系统, 然后在每个点或者单元的一个 patch 中求解一个局部的 Dirichlet 或者 Neumann 问题, 通过局部问题给出误差指示子.

自适应过程中, 如果全局后验误差估计小于给定值, 则有限元解被认为达到了精度要求, 自适应循环结束; 否则, 局部自适应加密进程被调用, 通过调整单元分布、单元中的基函数及加密误差比较大的单元得到一个新的有限元空间, 计算在新空间上继续进行. 主要的自适应方法有如下几种:

- $h$ 自适应方法: 通过将误差大的网格单元加密成多个子单元实现, 它保持基函数的阶不变. 该方法的收敛阶是代数的, 是自适应有限元中最成熟最常用的方法.
- $p$ 自适应方法: 保持网格不变, 通过改变基函数的阶实现自适应. 该方法收敛速度快, 如果网格选取适当并且解足够光滑, 可以达到指数阶收敛. 但它对网格的要求比较高, 得到的线性方程组的条件数较高, 方程求解速度慢.
- $hp$ 自适应方法: $hp$ 自适应方法同时调整网格和基函数. 优点是收敛速度快, 在合适的策略下, 可以达到指数阶收敛, 缺点是理论不成熟, 缺乏稳健有效的自适应策略.
- $r$ 自适应方法: 移动网格方法, 保持网格规模不变, 根据数值解的误差分布修改网格点的位置. 在误差大的地方放置更多的网格点. 该方法在流体、燃烧、热传导等问题中应用较广泛.

#### §5.1.3.2 单元标记策略

本节仅限于讨论如何选取加密单元, 放粗单元的选取是类似的, 只需要改变一些不等式的方向即可. 许多研究人员对标记策略进行了探讨并提出了一些有效的策略. Dörfler 在文 [39] 中针对 Poisson 问题提出了保证误差下降策略, Morin 等人在文 [40] 中提出了 MNS 加密标记策略. 目前常用的单元加密标记策略主要有: 最大策略 (maximum strategy, MAX), 误差等分布策略 (equidistribution strategy, EQDIST), 保证误差下降策略 (guaranteed error reduction strategy, GERS), 和 MNS 加密策略 (MNS-refinement strategy, MNS). 本节先给出这些策略的数学描述.

首先介绍几个符号: $\mathcal{T}_h$ 表示当前网格的单元集合; $S$ 表示当前网格的边 (二维) 或面 (三维) 的集合; $\eta_K$ ($K \in \mathcal{T}_h$) 和 $\eta_s$ ($s \in S$) 表示 (定义在单元和面上的) 误差指示子; $\mathrm{osc}_A$ ($A \subset \mathcal{T}_h$) 是刻划数据振荡的量 (data oscillation); $\varepsilon$ 是停止标准, 即期望计算达到的精度 (tolerance); $p$ 表示误差指示子的指数, 与误差所使用的范数有关 (通常等于 2).

我们希望误差渐进相等地分布在各个单元上, 因而在标记加密单元时, 应尽量选择误差指示子较大的单元. 标记策略设计都遵循这一原理. 常用标记策略描述如下:

**最大策略 (maximum strategy, MAX)**

选取 $\theta \in (0,1)$, 将所有满足下述条件的单元标记为加密单元:

$$\eta_K \geq \theta \max_{K' \in \mathcal{T}_h} \eta_{K'} \tag{5.10}$$

**误差等分布策略 (equidistribution strategy, EQDIST)**

假定 $\mathcal{T}_h$ 中单元的数目为 $N$, 全局误差为 $\eta = (\sum\limits_{K \in \mathcal{T}_h} \eta_K^p)^{1/p}$, 假如误差指示子 $\eta_K$ 在所有单元中均相等, 则有 $\eta = N^{1/p}\eta_K$. 由此, 可以导出下述两个策略.

**EQDIST1**

选取 $\theta \in (0,1)$, 标记满足下述条件的单元为加密单元:

$$\eta_K \geq \theta \frac{\eta}{N^{1/p}} \tag{5.11}$$

**EQDIST2**

选取 $\theta \in (0,1)$, 标记满足下述条件的单元为加密单元:

$$\eta_K \geq \theta \frac{\varepsilon}{N^{1/p}} \tag{5.12}$$

**保证误差下降策略 (guaranteed error reduction strategy, GERS)**

Dörfler 在 [39] 中对于 Poisson 问题提出了能保证误差下降的策略, 该策略同样可以应用到其他问题. 给定 $\theta \in (0,1)$, 则加密单元的集合为满足下述条件的 $\mathcal{T}_h$ 的最小子集 $A$ ($\subseteq \mathcal{T}_h$):

$$\sum_{K \in A} \eta_K^p \geq \theta^p \sum_{K \in \mathcal{T}_h} \eta_K^p \tag{5.13}$$

**MNS 策略加密 (MNS-refinement strategy, MNS)**

该策略是 Morin 等人在证明自适应有限元的收敛性时提出的 [40]. 策略使用

定义在网格的边 (二维) 或面 (三维) 上的误差指示子 $\eta_s$ 和定义在单元上、刻划数据振荡的量 osc. 首先, 根据给定的参数 $\theta_1 \in (0,1)$, 构造满足下述条件的 $S$ 的最小子集 $S_1$:

$$\sum_{s\in S_1} \eta_s^p \geq \theta_1^p \sum_{s\in S} \eta_s^p \tag{5.14}$$

将所有至少有一条边 (二维) 或面 (三维) 属于 $S_1$ 的单元标记为加密单元, 并记加密单元的集合为 $A_1$. 对于 osc, 则根据给定的参数 $\theta_2 \in (0,1)$, 扩大 $A_1$, 使得加密单元集合是满足下述条件的 $\mathcal{T}_h$ 的最小子集 $A$ ($A_1 \subseteq A \subseteq \mathcal{T}_h$):

$$\mathrm{osc}_A \geq \theta_2 \ \mathrm{osc}_{\mathcal{T}_h}. \tag{5.15}$$

对 MAX 和 EQDIST 策略, 参数 $\theta$ 的值越大, 标记的单元的数目越少; 而对于 GERS 和 MNS 加密策略, 参数 $\theta$ 的值越大, 标记的单元则越多.

#### §5.1.3.3 标记策略的并行算法

本节讨论 GERS 和 MNS 加密策略在基于消息传递的分布式存储并行计算机上的实现问题, 同时介绍并行自适应有限元软件平台 PHG (Parallel Hierarchical Grid) 中实现这些标记策略的统一函数接口. 由于 MAX 策略和 EQUIDIST 策略很容易实现, 这里不作讨论.

根据上一节的介绍可知, GERS 策略和 MNS 加密策略的实现均可归结为如下问题: 给定集合 $G$, 定义在 $G$ 上的误差指示子 $\eta_K$ ($K \in G$), 及参数 $\theta \in (0,1)$, 求满足下式的 $G$ 的最小子集 $A$:

$$\sum_{K\in A} \eta_K \geq \theta \sum_{K\in G} \eta_K. \tag{5.16}$$

这里, 集合 $G$ 被划分成一些互不相交的子集, 分布在各个进程中.

上述问题最直接的算法是先将 $G$ 中的元素按照误差指示子 $\eta_K$ 的大小排序, 然后从大到小依次选出 $A$ 中的元素, 算法的计算复杂度为 $O(N\log N)$, $N = |G|$. PHG 没有采用这一算法, 因为它需要进行并行排序, 通信量较大.

给定 $\gamma \in [0,1]$, 定义集合 $A_\gamma$ 如下:

$$A_\gamma = \{K \mid K \in G, \eta_K \geq \gamma \max_{K'\in G} \eta_{K'}\}$$

记 $\eta_\gamma = \sum\limits_{K\in A_\gamma} \eta_K$ (它是集合 $A_\gamma$ 中误差指示子的和), $\eta = \sum\limits_{K\in G} \eta_K$, 则问题转化为给定 $\theta$ 求 $\gamma$ 使得 $\eta_\gamma = \theta\eta$. 易知 $\gamma$ 是 $\theta$ 的减函数. 我们用二分法计算满足要求的 $\gamma$.

上述算法很容易并行实现, 每次二分计算中的通信仅为一次规约操作, 算法的计算复杂度为 $O(N)$. 为了提高通信粒度, 也可以采用 $k$ 分 ($k > 2$) 的方法, 通过适当增加计算量来减少通信次数.

我们为 PHG 设计了统一的函数接口来完成单元标记, 该函数可以同时对加密和放粗进行标记, 函数接口定义如下:

```
void phgMarkElements(STRATEGY strategy_r, DOF *eta_r, FLOAT theta,
        DOF *osc, FLOAT zeta, int depth_r, STRATEGY strategy_c,
        DOF *eta_c, FLOAT gamma, int depth_c, FLOAT a),
```

其中, `strategy_r` 和 `strategy_c` 分别表示加密和放粗策略, 可以取值 `MARK_MAX`, `MARK_EQDIST`, `MARK_GERS`, `MARK_MNS`, `MARK_ALL`, `MARK_NONE`, 其中最后两个分别表示标记全部单元和不标记任何单元; `eta_r`, `osc` 和 `eta_c` 分别给出加密误差指示子, 数据振荡和放粗误差指示子; `theta`, `zeta` 和 `gamma` 给出标记参数; 而 `depth_r` 和 `depth_c` 则给出被标记的单元加密和放粗次数. 参数 `a` 仅被 EQDIST 策略使用, 对于加密而言, 标记的单元集合为 $\{K \mid \eta_K \geq \texttt{theta} \times \texttt{a}\}$. `strategy_r` 和 `strategy_c` 也可以取值为 `MARK_DEFAULT`, 它允许用户在运行程序时通过命令行选项 `-strategy`, `-refine-strategy` 和 `-coarsen-strategy` 指定标记策略. 默认策略为 `MARK_GERS`.

在这些策略中, GERS 是一个较好的选择, 通常情况下它表现出比其他策略更好的稳健性.

## §5.2 *hp* 自适应有限元

*hp* 自适应有限元由 Ivo Babuška 等人最早提出, 它收敛速度快, 受到研究人员的重视. *hp* 自适应策略仍然是 *hp* 自适应有限元方法的难点之一. 本节对自适应策略进行探讨.

### §5.2.1 有限元解空间和变分形式

本节接下来的内容中网格限定为协调四面体网格. 在经典 $h$ 自适应有限元中, 每个单元上的基函数的阶是相同的, 其有限元空间 $V_h$ 的构造相对简单. *hp* 自适应有限元中, 每个单元上的基函数的阶不一定相同. 用符号 $V_{h,p}$ 和 $\mathcal{T}_{h,p}$ 分别表示有限元空间和网格. 在协调四面体网格中, 单元上的基函数按照其所在的位置或者定义方式分为顶点基函数、边基函数、面基函数和体基函数, 它们分别定义在四面体的顶点、边、面和单元内部. 假设给定每个单元中体基函数的阶, 我们采用最小原则 (minimum rule) [167] 来定义各类基函数的阶, 它可使得有限元空间独

立于具体的基函数 [158]. 根据最小原则, 顶点、边或面基函数的阶取为包含该顶点、边或面的所有单元体基函数的阶的最小值.

所有这些基函数张成 $\mathcal{T}_{h,p}$ 上的协调有限元空间 $V_{h,p}$. $hp$ 自适应有限元的变分问题为: 求解 $u_{h,p} \in V_{h,p}$, 使得下式成立,

$$a(u_{h,p}, v_{h,p}) = l(v_{h,p}), \quad \forall v_{h,p} \in V_{h,p}. \tag{5.17}$$

有限元空间 $V_{h,p}$ 的基函数有很多选择, 但一般要求基函数满足两个条件:

- 基函数的支集 (support) 应该是局部的, 从而使得最终的刚度矩阵是稀疏的;
- 基函数应该是层次的 (hierarchic), 低阶基函数集合是高阶基函数集合的子集. 层次性用于保证有限元空间的协调性.

常用的层次基函数基于 Jacobi 多项式, Legendre 多项式和 Lobatto 函数构造, 详细内容参考 [168, 169, 170].

### §5.2.2 $hp$ 自适应加密策略

$hp$ 自适应有限元的难点之一是自适应加密策略, 即在单元上如何选取 $h$ 加密和 $p$ 加密. 原则是, 应该在方程的解光滑的区域选择 $p$ 加密, 在解有奇性的区域选择 $h$ 加密. 不幸的是, 解的性质在通常情况下是不知道的. 研究人员设计了不同的方法确定解的正则性. 下面介绍一些常见的 $hp$ 自适应加密策略:

(1) 利用解的正则性先验信息 (use of a priori knowledge of solution regularity).
$p$ 加密可以在解的光滑区域达到指数阶收敛, 但在有奇性的区域却没有 $h$ 加密有效. 由于这个原因, 许多 $hp$ 加密策略在解有奇性的区域采用 $h$ 加密, 别的区域采取 $p$ 加密. 最简单的策略是利用正则性的先验信息. 例如, 如果线性椭圆偏微分方程的系数是光滑的并且边界条件是分片解析的, 那么在区域的凹角处有点奇性 [172]. Ainsworth 和 Senior [172], Bernardi 等人 [184] 以及 Valenciano 等人 [185] 利用区域及已知数据的先验信息在有奇性的区域采用 $h$ 加密, 在其他区域采用 $p$ 加密.

(2) 类型参数 (type parameter).
Gui 和 Babuška [182] 及 Affia 等人 [186] 使用类型参数的概念, 设计了 $hp$ 自适应策略. 给定误差指示子 $\eta_{K,p_K}$ 和 $\eta_{K,p_K-1}$, 定义

$$R(K) = \begin{cases} \frac{\eta_{K,p_K}}{\eta_{K,p_K-1}}, & \eta_{K,p_K-1} \neq 0 \\ 0, & \eta_{K,p_K-1} = 0, \end{cases}$$

$R$ 用来评估解的光滑性. 给定参数 $\gamma$ $(0 \le \gamma < \infty)$, 如果 $R(K) \ge \gamma$, 单元 $K$ 将采取 $h$ 加密; 如果 $R(K) < \gamma$, 单元 $K$ 将采取 $p$ 加密. 另外, 如果 $\gamma = 0$, 单元 $K$ 将只采取 $h$ 加密, 而如果 $\gamma = \infty$, 则单元 $K$ 将只采取 $p$ 加密.

(3) 用低阶函数空间估计解的正则性 (estimate regularity using smaller p estimates). Süli, Houston 和 Schwab [173] 利用两个低阶基函数空间计算误差下降阶, 根据下降阶估计解的正则性. 设单元基函数的阶为 $p$, 如果在单元上计算基函数阶分别为 $p-1$, $p-2$ 时的误差 $\eta_{p-1}$ 和 $\eta_{p-2}$, 有下面近似表达式成立

$$\frac{\eta_{p-1}}{\eta_{p-2}} \approx (\frac{p-1}{p-2})^{1-m},$$

于是正则度 $m$ 可定义为:

$$m \approx 1 - \frac{\log(\eta_{p-1}/\eta_{p-2})}{\log((p-1)/(p-2))}.$$

当 $p \le m-1$ 时, 采取 $p$ 加密; 否则采取 $h$ 加密.

(4) 利用高阶函数空间估计解的正则性 (estimate regularity using larger p estimates). Ainsworth 和 Senior[177] 给出利用高阶基函数空间估计解正则性的策略. 设当前单元 $K$ 上基函数阶为 $p$, 分别计算基函数阶为 $p+1$, $p+2$, $p+3$ 时的误差. 该策略将问题 (5.17) 转化为满足平衡条件 (equilibrium condition) 的局部 Neumann 问题. 假设基函数阶为 $p+j$ 时的误差为 $e_j$, 那么有下面近似表达式成立

$$\|e_0\|_K^2 - \|e_j\|_K^2 \approx C^2(p+j)^{-2\alpha},$$

其中 $C$, $\alpha$ 为正的常数. 局部正则度 $m$ 定义为 $m = 1+\alpha$. 如果 $p \le m-1 = \alpha$, 则采取 $p$ 加密, 否则采取 $h$ 加密.

(5) 参考解 (reference solution).
Demkowicz 及其合作者 [190, 191, 192] 设计了一个 $hp$ 自适应策略. 假定当前计算网格及数值解分别为 $\mathcal{T}_{h,p}$ 和 $u_{h,p}$, 对网格和基函数做一致的 $h$ 和 $p$ 加密, 得到一个更细的网格 $\mathcal{T}_{h/2,p+1}$, 在这个网格上可以求得一个新的数值解 $u_{h/2,p+1}$. $u_{h,p}$ 与 $u_{h/2,p+1}$ 差值的某个范数作为全局误差, 如

$$\eta = \|u_{h,p} - u_{h/2,p+1}\|_{H^1(\Omega)}.$$

下一步是决定每个单元的最优加密方式, 考虑所有可能的 $h$ 加密方式和 $p$ 加密方式, 从中选择一个最优的加密方法. 该策略是有效的, 但非常复杂, 计算量特别大. Šolín 等人 [193] 给出了一个简化的版本.

(6) 误差下降预测 (predict error estimate on assumption of smoothness).

Melenk 等人 [174] 以及 Heuveline 等人 [176] 给出了启发式加密策略. 对二维四边形和三维六面体单元, 他们假设解是光滑的, 根据近似理论, 可以预测 $h$ 加密及 $p$ 加密的误差下降趋势, 而根据这个预测判断解是否是光滑的. 如果实际误差达到预测, 表明解是光滑的; 如果达不到预测, 则表明解不够光滑. 对光滑的区域采取 $p$ 加密, 对非光滑区域采取 $h$ 加密.

$hp$ 自适应有限元的加密策略没有 $h$ 自适应有限元那样成熟, 构造简单有效的自适应加密方法仍然是一个很大的挑战, $hp$ 自适应有限元基本理论也在发展之中. 上面给出了一些自适应策略的简单描述, 详细的过程可以参考相关文献, 还有一些其他的策略, 可以参考文献 [167, 171, 175, 177, 183, 187, 188, 189, 194, 196].

## §5.3 一种新的 $hp$ 自适应策略

本节介绍我们设计的 $hp$ 自适应加密策略, 该策略基于误差下降预测. 该策略中, 我们对不同的加密方式预测误差下降的速度, 如果实际误差与预测一致, 表明解是光滑的, 对单元采取 $p$ 加密; 如果误差下降没有达到预期, 表明解不够光滑, 对单元采取 $h$ 加密. 这个策略很大程度上受到 Melenk 等人 [174] 以及 Heuveline 等人 [176] 工作的启发. 本文对 $h$ 加密及 $p$ 加密的误差下降估计采用不同的假设, 得到不同的误差下降估计公式, 这个策略不仅适合二维四边形和三维六面体, 也适合二维三角形单元和三维四面体单元.

本节考虑线性椭圆型偏微分方程

$$\begin{cases} -\triangle u = & f \text{ in } \Omega, \\ u = & 0 \text{ on } \partial\Omega, \end{cases} \tag{5.18}$$

其中 $\Omega \subset R^d, d = 2, 3,\ f \in L^2(\Omega)$. 我们的目标是设计和定义计算网格 $\mathcal{T}_{h,p}$ 和一个有限元空间 $V_{h,p} \subset H_0^1(\Omega)$, 计算数值解 $u_{h,p} \in V_{h,p}$ 使得变分形式 (5.17) 成立, 并达到事先设定的停止标准.

下面介绍一些记号. 记 $\|\cdot\|$ 为能量模, $\|u\| = \sqrt{a(u,u)}$; $\varepsilon$ 为停止标准; $\eta_K$ 为定义在单元 $K$ 上的误差指示子; $\eta$ 定义为 $\eta = (\sum_{K\in\mathcal{T}_{h,p}} \eta_K)^{\frac{1}{2}}$. 用 $p_K$ 和 $h_K$ 分别表示单元 $K$ 上的基函数的阶及单元的尺寸. 如果一个单元被加密成多个子单元, 那么 $c_K$ 表示其子单元的个数. 最后引入符号 $N$ 表示自由度数目.

### §5.3.1 $hp$ 自适应策略

Melenk 等人 [174] 以及 Heuveline 等人 [176] 提出的 $hp$ 自适应策略基于如下近似理论: 如果对网格做一致剖分, 并且所有单元上基函数的阶均为 $p$, 记 $e = u - u_{h,p}$

为误差, 如果 $u \in H^m(\Omega)$, 那么当采用分片多项式基函数时有 [177, 178]

$$\|e\|_{H^1(\Omega)} \le C\frac{h^\mu}{p^{(m-1)}}\|u\|_{H^m(\Omega)}, \tag{5.19}$$

其中 $h$ 是网格的尺寸, $\mu = \min(p, m-1)$, $C$ 是独立于 $h$ 及 $p$ 的常数.

Melenk 及其合作者 [174] 针对二维四边形网格提出了一种基于误差下降预测的 $hp$ 自适应策略: 在四边形网格中, 假设当前单元的误差指示子为 $\eta_K$, 如果单元被加密成四个子单元, 那么每一个子单元上的误差预测为 $0.5\cdot 0.5^{p_K}\eta_K$; 如果单元上的基函数阶提高一阶, 那么误差预测 $\gamma_p\eta_K$, $\gamma_p$ ($\in (0,1)$) 为常数; 如果单元没有被调整 (包括网格及基函数), 则误差预测为 $\eta_K$, 即保持不变. 在自适应过程中, 一个单元被标记为加密单元, 则判断其误差下降是否达到预期: 如果误差下降达到预期, 表明解是光滑的, 下一个自适应步将采取 $p$ 加密; 否则将采取 $h$ 加密.

Heuveline 及其合作者 [176] 针对二维四边形和三维六面体也提出了基于误差下降预测的策略. 他们的策略是尽可能地多采用 $p$ 加密, 该策略分为几种情况: 1) 如果被标记加密的单元在上一个自适应步中没有被加密过, 则采取 $p$ 加密; 2) 如果单元 $K$ 是通过 $p$ 加密得到的, 那么误差预测为 $h_k\eta_K^{\text{old}}$, 如果误差小于这个预测, 则采取 $p$ 加密, 否则采取 $h$ 加密; 3) 如果单元是通过 $h$ 加密得到的, 那么误差预测为 $2^{-p_K}\eta_{K_M}^{\text{old}}$, 其中 $K_M$ 为其父单元, 如果误差小于这个预测, 则采取 $p$ 加密, 否则采取 $h$ 加密.

这两个策略适合二维四边形单元和三维六面体单元, 不能直接应用于三角形的局部自适应二分加密及四面体的局部自适应二分加密. 另外 Heuveline 及其合作者设计的策略存在对 $h_K$ 的依赖, 策略对单元尺寸敏感. 当单元尺寸较大的时候 ($h_K \ge 1$), 自适应过程中会有过多的 $p$ 加密, 不能适当地调整网格; 而当尺寸较小的时候 ($h_K \ll 1$), 则会有太多的 $h$ 加密, 收敛速度不一定能达到指数. 我们希望将这两个策略推广到网格二分加密的情况, 并且使得策略与网格尺寸无关.

这里给出一个 $hp$ 自适应策略, 该策略允许单元采取 $h$, $p$ 及 $hp$ 自适应加密. 对 $h$ 及 $p$ 自适应加密做不同的假设: 对 $h$ 自适应加密, 假设可以达到代数阶收敛; 对 $p$ 自适应加密, 假设可以达到指数阶收敛, 然后使用不同的误差估计来预测误差下降因子. 本文中一个单元允许同时采取 $h$ 及 $p$ 自适应加密, 并给出了这种情况的误差下降预测.

在 $h$ 自适应有限元方法中, 误差大的单元被加密成几个子单元, 其最优收敛速度是代数的, 表达式可以写成

$$\|e\| \le C_1 N^{\frac{-p}{d}}, \tag{5.20}$$

其中 $p$ 是单元上基函数的阶, $C_1$ 为常数. 将不等式 (5.20) 写成等式, 具有如下形式

$$\|e\| = C_1 N^{\frac{-p}{d}} = C_2 (\sum_{K\in T} \eta_K^2)^{\frac{1}{2}}, \tag{5.21}$$

其中 $C_2$ 是依赖于所采用的误差指示子的常数. 这是本文自适应策略的基本出发点.

如果对网格 $\mathcal{T}_{h,p}$ 采取一致 $h$ 加密, 每个单元被加密成 $c_K$ 个子单元, 得到新的网格 $\mathcal{T}_{h,p}^1$, 同时假设拥有相同父单元的子单元拥有相同的误差指示子. 现在自由度的数目近似为 $c_K N$. 如果解是光滑的, 并且达到最优代数收敛, 那么有下式近似成立

$$C_1 (c_K N)^{\frac{-p}{d}} = C_2 (\sum_{K\in \mathcal{T}_{h,p}^1} \eta_K^2)^{\frac{1}{2}} = C_2 (c_K \lambda_h^2 \sum_{K\in \mathcal{T}_{h,p}} \eta_K^2)^{\frac{1}{2}}, \tag{5.22}$$

其中 $\lambda_h$ 是误差下降因子. 综合式 (5.21) 及式 (5.22), 对 $h$ 自适应加密, 可以得到误差下降因子

$$\lambda_h^2 = \frac{1}{c_K} (\frac{1}{c_K})^{\frac{2p}{d}}.$$

我们采用和文章 [176] 一样的想法: 即尽可能多地采取 $p$ 加密. 对 $h$ 自适应加密, 本文使用一个放大的值做误差下降因子, 表达式如下

$$\lambda_h = (\frac{1}{c_K})^{\frac{p}{d}}. \tag{5.23}$$

下面推导 $p$ 加密的误差下降因子. 误差估计式 (5.19) 中需要知道解析解 $u$ 的正则性 $m$, 由于一般情况下 $m$ 是不知道的. 这里采用了一个启发式的技术. 假设 $p$ 大于等于 $m$, 并将不等式 (5.19) 写成等式, 则当基函数提高一阶时, 有

$$\|e\|_{H^1(\Omega)} = C (\frac{p}{p+1})^{m-1} \frac{h^\mu}{p^{(m-1)}} \|u\|_{H^m(\Omega)},$$

从而误差下降因子为

$$\lambda_p = (\frac{p}{p+1})^{m-1}.$$

对给定的问题, 需要使用合适的值近似 $m-1$. 在单元 $K$ 上, 我们使用单元 $K$ 上的基函数的阶近似解在单元上的局部正则性. 取 $m-1 = p/2$. 这个选择是合适的, 因为假设选取的值过低, 表明解的局部正则性比估计的好, 那么下一步自适应加密将采取 $p$ 加密, 否则下一个自适应步的加密将采取 $h$ 加密. 最终得到 $p$ 加密误差下降因子为

$$\lambda_p = (\frac{p}{p+1})^{\frac{p}{2}}. \tag{5.24}$$

在实际应用中, 当某一个单元采取 $h$ 加密时, 为了保证网格的协调性, 有可能使邻居单元也被 $h$ 加密, 如果这个邻居单元被标记了 $p$ 加密, 将允许这个单元同时进行 $h$ 自适应加密及 $p$ 自适应加密. 组合上面两个误差下降因子得到 $hp$ 加密的误差下降因子为,

$$\lambda_{hp} = (\frac{p}{p+1})^{\frac{p}{2}}(\frac{1}{c_K})^{\frac{p}{d}}. \tag{5.25}$$

误差下降估计式 (5.23), (5.24) 及 (5.25) 对网格的 $h$ 加密类型并没有限制, 仅需要知道每个单元 $h$ 加密后产生的子单元的个数 $c_K$. 该策略消除了 Heuveline 及其合作者提出的策略对 $h$ 的依赖, 使得我们的策略与网格尺寸无关.

接下来给出我们的 $hp$ 自适应加密策略, 此处使用 Heuveline 及其合作者 [176] 设计的一个框架, 并做了一些修改. 自适应过程如下:

**(1)** 对当前计算网格 $\mathcal{T}_{h,p}$, 如果误差指示子 $\eta$ 满足停止标准 ($\eta \le \varepsilon$), 则停止自适应循环.

**(2)** 使用最大策略标记加密单元.

**(3)** 对网格 $\mathcal{T}_{h,p}$ 中每一个单元 $K$, 如果 $K$ 被标记为加密单元:

- 如果 $K$ 是通过上一个自适应步的 $h$ 加密得到的, 并且基函数的阶没有改变. 记其父单元为 $K_M$, 检查是否有

  $$\eta_K^2 \le \lambda_h^2 \eta_{K_M}^2.$$

  如果上式成立, 将单元 $K$ 标记为 $p$ 加密, 将基函数的阶提高一阶; 否则, 将单元 $K$ 标记为 $h$ 加密;

- 如果 $K$ 是通过上一个自适应步的 $p$ 加密得到的, 记其父单元为 $K_M$, 检查是否有

  $$\eta_K^2 \le \lambda_p^2 \eta_{K_M}^2.$$

  如果上式成立, 将单元 $K$ 标记为 $p$ 加密, 将基函数的阶提高一阶; 否则, 将单元 $K$ 标记为 $h$ 加密;

- 如果 $K$ 是通过上一个自适应步的 $hp$ 加密得到的, 记其父单元为 $K_M$, 检查是否有

  $$\eta_K^2 \le \lambda_{hp}^2 \eta_{K_M}^2.$$

  如果上式成立, 将单元 $K$ 标记为 $p$ 加密, 将基函数的阶提高一阶; 否则, 将单元 $K$ 标记为 $h$ 加密;

- 如果单元 $K$ 在上一个自适应步没有被加密 (单元大小及基函数阶保持不变), 则将单元 $K$ 标记为 $p$ 加密, 将基函数的阶提高一阶.

**(4)** 根据单元的加密类型, 对单元实施 $p$ 加密及 $h$ 加密, 转至 **(1)**.

## §5.4 数值算例

本节用三个数值算例验证我们设计的 $hp$ 自适应策略. 数值算例中比较了三个不同的自适应策略: $h$ 自适应方法, 本文设计的 $hp$ 自适应策略及 Melenk 及其合作者设计的 $hp$ 自适应框架 [174]. 测试程序利用 PHG 编制, 测试环境同前一章.

由于问题是三维的, $d$ 为 3, 网格局部自适应加密算法为二分加密算法, 一个单元被加密成两个子单元, 因此 $c_K$ 为 2. 因而误差下降因子变为

$$\lambda_h = (\frac{1}{2})^{\frac{p}{3}}, \tag{5.26}$$

$$\lambda_p = (\frac{p}{p+1})^{\frac{p}{2}}, \tag{5.27}$$

$$\lambda_{hp} = (\frac{p}{p+1})^{\frac{p}{2}}(\frac{1}{2})^{\frac{p}{3}}. \tag{5.28}$$

这里使用的误差指示子由 Melenk 及其合作者设计 [174], 该误差指示子原为二维问题设计, 同样可以用在三维问题上, 其形式如下

$$\eta_K^2 = \frac{h_K^2}{p_K^2}\|f_{p_K} + \Delta u_N\|_{L^2(k)}^2 + \sum_{f\subset\partial K\cap\Omega}\frac{h_f}{2p_f}\|[\frac{\partial u_N}{\partial n_f}]\|_{L^2(f)}^2. \tag{5.29}$$

其中 $f_{p_K}$ 为 $f$ 在 $p-1$ 阶基函数空间上的 $L^2(K)$ 投影, $h_f$ 为单元 $K$ 上面 $f$ 的直径, $p_f = \max(p|_{K_1}, p|_{K_2})$, 其中 $f$ 为单元 $K_1$ 及 $K_2$ 的共享面.

最大策略中参数 $\theta$ 选取为 0.5, 基函数为 $H^1$ 层次基, 基函数初始阶数为 2, 最大阶数为 15, 解法器为预条件共轭梯度法 (PCG, Preconditioned Conjugate Gradient method), 预条件子为 block Jacobi.

在三维问题中, 合适的 $hp$ 自适应策略可以使误差达到指数阶收敛, 最优收敛有如下形式 [179]

$$\|e\| \le Ce^{-\gamma N^{1/5}}, \tag{5.30}$$

其中 $C$ 及 $\gamma$ 为常数, $N$ 为自由度数. 本节的图形中, 横轴以及纵轴的变量分别为自由度数和能量误差, 它们的刻度分别为 $N^{1/5}$ 和对数. 如果误差达到指数阶收敛, 那么误差曲线应该是一条直线.

图形中, `HAFEM` 表示仅使用 $h$ 加密的结果, `HP/PHG` 和 `HP/MK` 则分别表示我们的策略和 Melenk 及其合作者的策略, 二者的一个不同点在于: 如果一个被标记的单元在上一个自适应步中没有被调整 (包括网格和基函数), 那么 `HP/PHG` 将对这个单元采取 $p$ 自适应加密, 而 Melenk 等人的做法则是如果单元在这个自适应步中的误差指示子小于上一个自适应步中的误差指示子则采取 $p$ 自适应加密, 否则采取 $h$ 自适应加密.

**例 5.4.1.** *取解析解为* $u = e^{-z}(1+e^{100(x+y-1)})^{-1}$, *它在平面* $x+y-1=0$ *上呈现内层现象, 计算区域为* $\Omega=(0,1)^3$, *初始网格为一致网格, 单元数为 24.*

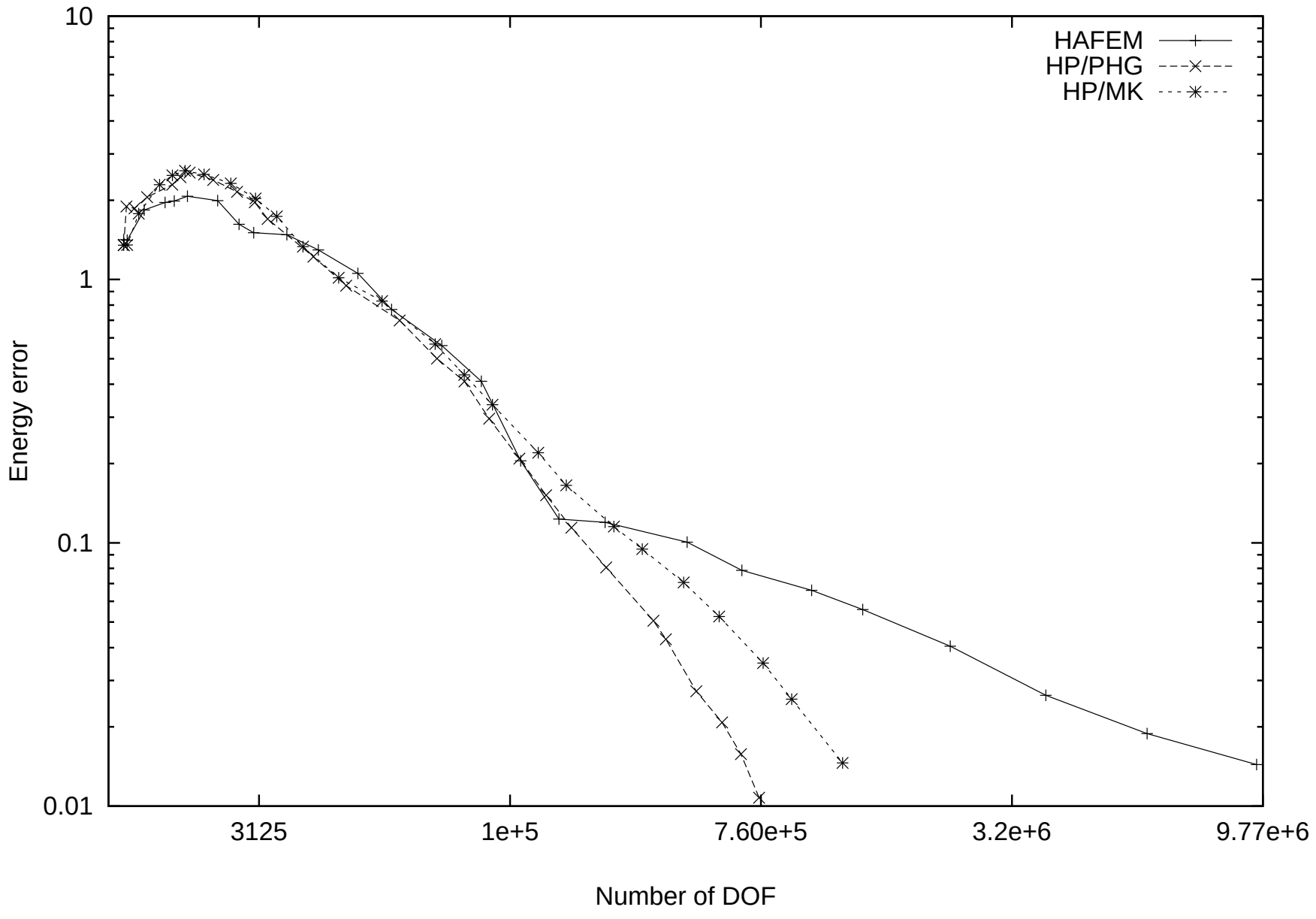

图 5.1 误差曲线 (例 5.4.1)

**表 5.1 不同策略的最终单元数、自由度数及能量误差 (例 5.4.1)**

| | # elements | # DOF | Energy error |
|---|---|---|---|
| HP/PHG | 23,920 | 750,009 | 1.08e-2 |
| HP/MK | 86,520 | 1,271,317 | 1.46e-2 |
| HAFEM | 6,969,762 | 9,528,910 | 1.44e-2 |

例 5.4.1 的计算结果如图 5.1 所示, 策略 `HP/PHG` 及 `HP/MK` 均达到了指数收敛, 但 `HP/PHG` 的收敛速度明显快于 `HP/MK`. 从表 5.1 也可以看出, `HP/MK` 的网格规模及

自由度规模远大于 HP/PHG 的规模, 说明 HP/MK 更多地做 $h$ 自适应加密, 而 HP/PHG 更多地做 $p$ 自适应加密, 效率更高. $hp$ 自适应有限元与 $h$ 自适应有限元相比, 前者的收敛速度远快于后者.

**例 5.4.2.** *取解析解为* $u=\cos(2\pi x)\cos(2\pi y)\cos(2\pi z)$, *该问题的难点在于解的高阶导数, 其奇数阶与偶数阶导数在每个点的行为不一样, $p$ 自适应加密不一定能改善数值解 [196]. 计算区域为 $L$- 形区域,* $\Omega=(-1,1)^3\setminus[0,1]\times[-1,0]\times[-1,1]$, *初始网格是一致的, 含 144 个单元.*

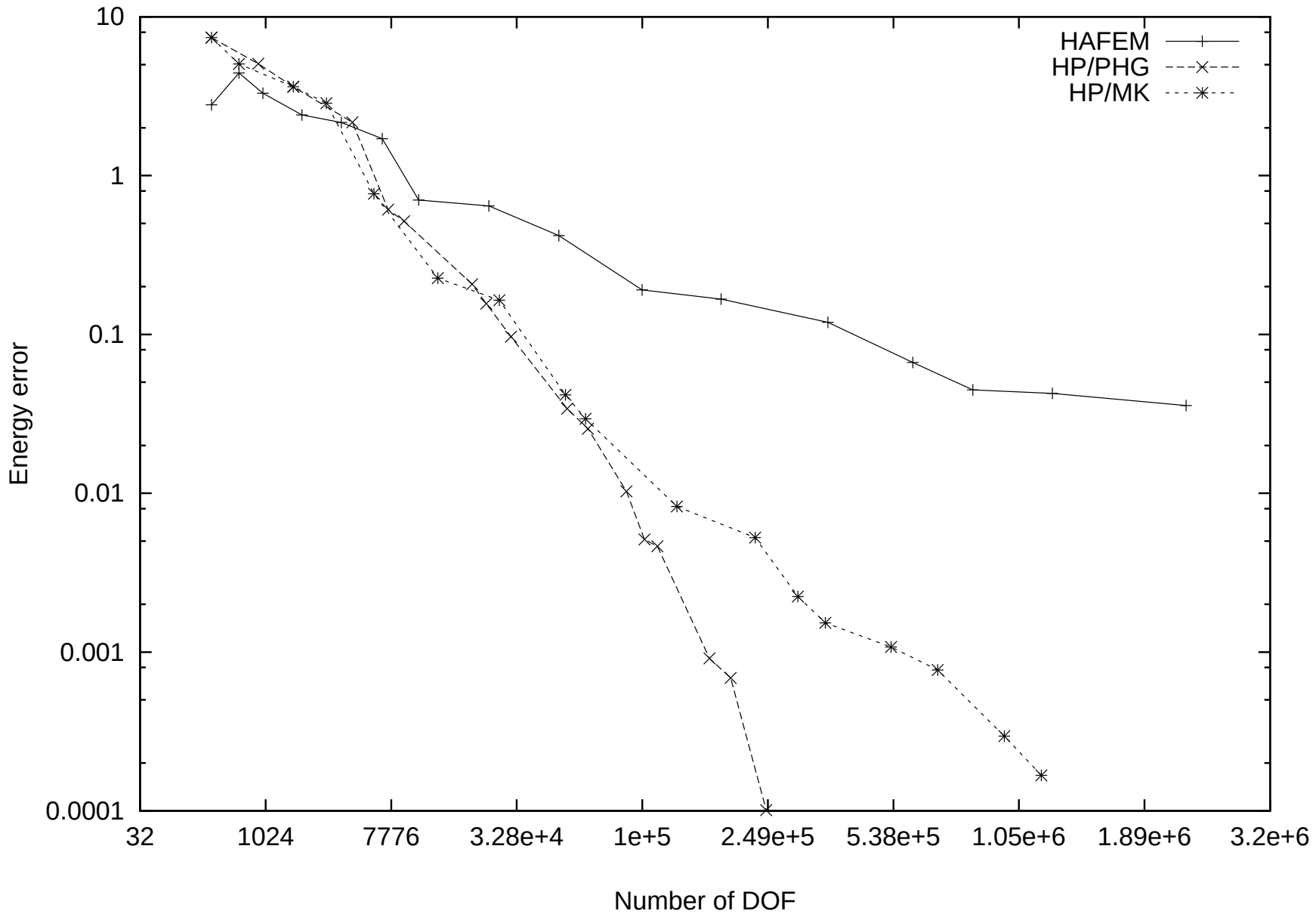

图 5.2 误差曲线 (例 5.4.2)

**表 5.2 不同策略的最终单元数、自由度数及能量误差 (例 5.4.2)**

| | # elements | # DOF | Energy error |
|---|---|---|---|
| HP/PHG | 3,772 | 246,046 | 1.01e-4 |
| HP/MK | 35,696 | 1,171,216 | 1.67e-4 |
| HAFEM | 1,663,068 | 2,263,137 | 3.57e-2 |

计算结果见图 5.2 和表 5.2, 两种 $hp$ 自适应策略均达到指数收敛, 但差别更明显, HP/MK 策略产生的最终网格单元数比 HP/PHG 高出近一个数量级, 自由度数目大概是 5 倍, 但最终的误差仍然大于 HP/PHG 的误差, 说明 HP/PHG 策略的效率

高于 HP/MK. 从表 5.2 可以看出, $h$ 自适应有限元产生的单元数量比 HP/PHG 高出两个数量级, 自由度数高一个数量级, 误差却大了两个数量级, $hp$ 自适应有限元方法相对于 $h$ 自适应有限元方法的优势明显.

**例 5.4.3.** *解析解为 $u=(x^2+y^2+z^2)^{\frac{1}{4}}$, 求解区域是 Fichera 区域, $\Omega=(-1,1)^3\setminus[0,1]^3$. 这个问题的计算区域有奇性, 同时解析解有一个点奇性. 初始网格为一致网格, 单元数目为 172.*

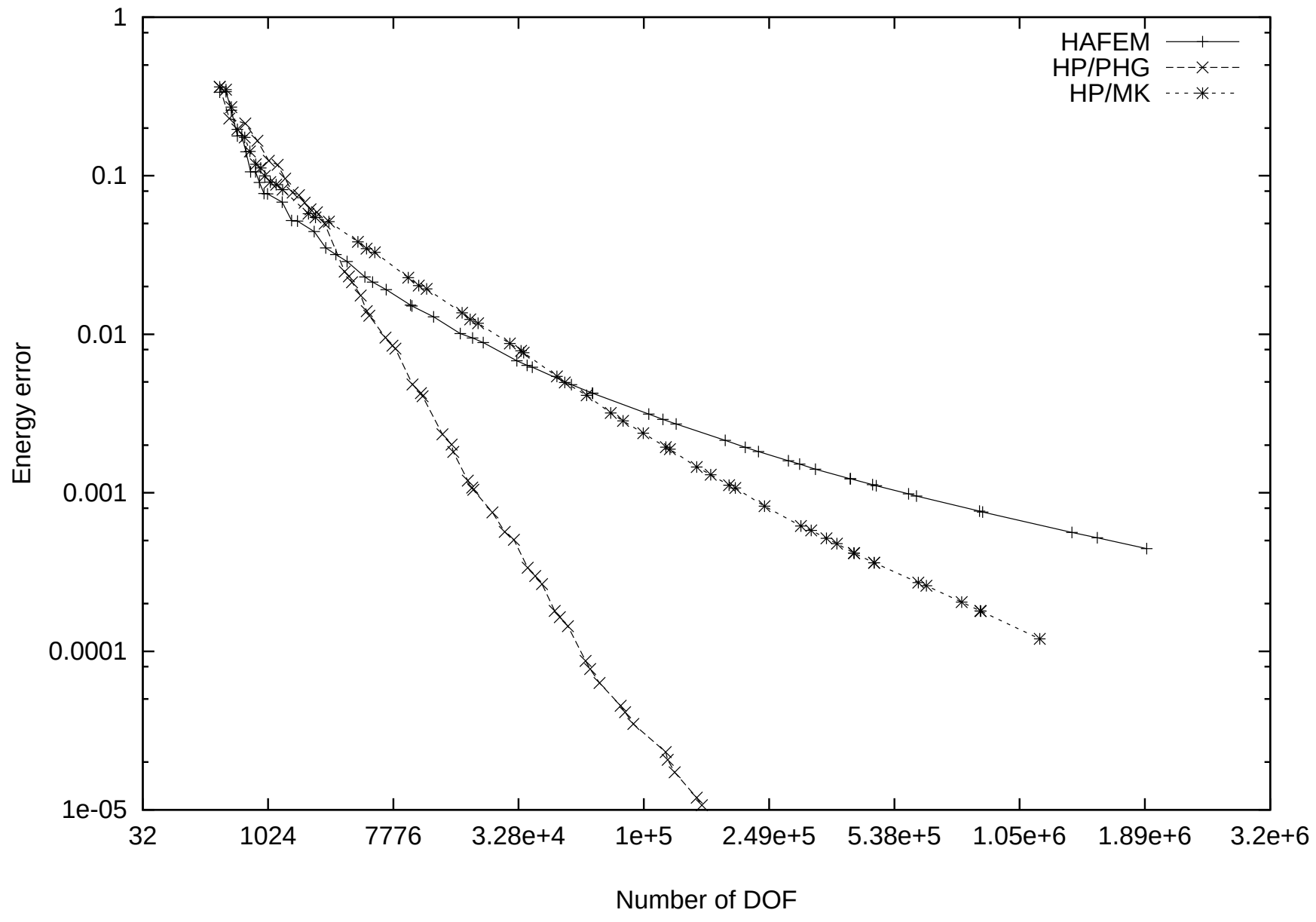

图 5.3 误差曲线 (例 5.4.3)

**表 5.3 不同策略的最终单元数、自由度数及能量误差 (例 5.4.3)**

| | # elements | # DOF | Energy error |
|---|---|---|---|
| HP/PHG | 3,429 | 155,812 | 1.07e-5 |
| HP/MK | 163,204 | 1,158,279 | 1.20e-4 |
| HAFEM | 1,377,588 | 1,904,054 | 4.44e-4 |

例 5.4.3 的计算结果见图 5.3 和表 5.3. 两种 $hp$ 自适应策略均达到指数收敛, 并且收敛曲线更光滑, 说明误差下降预测是准确的. 自适应策略 HP/MK 的单元数是策略 HP/PHG 的 50 倍, 自由度数目大约是 8 倍, 误差却大了一个数量级, 表明我

们的自适应策略更有效. $h$ 自适应有限元仅达到了代数收敛, 而 $hp$ 自适应有限元方法达到了指数收敛.

## §5.5 本章小结

本章介绍了 $hp$ 自适应有限元及常用的 $hp$ 自适应策略. 在研究现有 $hp$ 自适应策略的基础上提出了一个新的 $hp$ 自适应策略, 该策略基于误差下降预测. 我们给出了不同加密策略下误差下降速度的预测公式. 数值实验表明, 我们提出的 $hp$ 自适应策略对所测试的问题可以达到指数收敛, 并且优于 Melenk 及其合作者提出的策略.

# 第六章　总结与展望

本文对并行自适应有限元计算中的动态负载平衡及 $hp$ 自适应策略进行了研究. 对动态负载平衡, 我们研究了过顶点哈密尔顿圈及哈密尔顿路, 给出了存在性定理, 并设计了线性复杂度的构造算法; 设计了任意维希尔伯特序的编码解码算法, 并提出该曲线的一个变体; 改进并实现了加密树以及空间填充曲线两种网格划分方法. 对 $hp$ 自适应有限元方法, 我们提出了一个新的自适应加密策略.

动态负载平衡是并行自适应计算中的一个重要的、必不可少的环节. 在并行自适应有限元计算中，动态负载平衡通常通过网格重划分实现, 网格重划分的速度、网格重划分的质量以及网格重划分后数据迁移的开销是影响整个并行自适应有限元计算并行效率和并行可扩展性的重要因素. 此外, 网格重划分后, 新的子网格的重组和数据迁移涉及到复杂的数据处理和通信, 其并行程序设计亦是一项非常具有挑战性的工作. 本文介绍了作者在网格划分算法和 PHG 动态负载平衡模块的设计与实现方面的工作. 本文所实现的 PHG 动态负载平衡模块已经在实际应用中经受住了考验, 可以胜任数千 CPU 核、十亿以上单元规模的并行自适应有限元计算的要求 (该部分内容可以参见崔涛的博士论文), 并且在整个自适应有限元计算时间中只占很小的比例，展现出很高的效率和很好的稳健性, 基本能够满足当前应用的需求. 在未来, 我们将根据随着并行规模的增大及应用的复杂化而对动态负载平衡提出的新要求, 继续对相关算法及模块进行改进和完善.

$hp$ 自适应有限元方法能够达到很高的计算效率, 是一类非常有前景的算法, 是当前科学计算研究的热点问题之一. 当前, $hp$ 自适应有限元方法远未成熟, 离实际应用尚有很大的距离, 其中的难点之一便是 $hp$ 自适应策略的设计. 本文利用 PHG 平台对 $hp$ 自适应有限元的支持, 对 $hp$ 自适应策略进行了研究和数值实验. 文中提出了一种新的 $hp$ 自适应策略, 它在所适用的网格及网格加密方式方面比文献中提出的一种策略有所扩展, 并且数值实验的结果表明其在计算效果上亦优于文献中的策略. 此外, 这些数值实验亦验证了 PHG 平台中与 $hp$ 自适应相关的模块的正确性和适用性. 这些研究是初步的, 有待于未来从理论上和应用上进一步深入和展开.

# 参考文献

# 发表文章目录

# 致 谢

本文是在张林波研究员的指导下完成的, 在论文完成之际, 我衷心地感谢我的导师张林波研究员, 感谢张老师将我引入科学计算的殿堂, 感谢张老师对我的指导、关怀和支持. 张老师知识渊博, 治学态度严谨, 思维敏锐, 工作作风精益求精, 诲人不倦, 平易近人, 对我影响深远. 本论文从选题到完成, 每一步都是在导师的指导下完成的, 倾注了导师大量的心血. 我的每一点进步, 都离不开张老师的指导和帮助. 在此, 谨向张老师表示崇高的敬意和衷心的感谢.

感谢林群院士, 曹礼群老师, 郑伟英老师在学习和工作上的支持和帮助, 同时, 和你们的讨论也使我获益匪浅. 谢谢你们.

感谢计算数学与科学工程计算研究所, 研究所学术氛围浓厚, 学习环境舒适. 本文的工作是在科学与工程计算国家重点实验室完成的, 感谢实验室提供的优越的科研环境和高水平的计算环境. 感谢白英老师、吴继萍老师、丁如娟老师、张纪平老师、朱琳爱义老师的支持和帮助. 感谢数学与系统科学研究院的老师及图书馆等部门老师.

感谢我们课题组的成员. 感谢钱莹师姐, 崔涛师兄, 冷伟, 成杰, 王昆, 谢妍, 感谢你们在学习和工作上给予的帮助和建设性的意见. 感谢钱莹师姐和崔涛师兄在论文上给予的帮助.

在这五年中, 结识了很多的朋友, 王辛师姐、肖源明师兄、刘歆师兄、阴小波师兄、翟方曼师姐、初晴、张黎、丁晓东、张静静、杜锐、唐明筠、付云姗、胡娟、赵旭鹰、陈华杰、刘伟、田霞、陈景润、滕飞、聂宁明、程明厚、张娅、翟淑星、杨熙、李明、李金, 谢谢你们在学习和生活上给予的支持、鼓励和帮助. 感谢初晴和她老公王震在工作上给予的宝贵的建议和无私的帮助. 谢谢你们.

感谢家人的养育和支持, 感谢家人无私的奉献. 谢谢你们.